\documentclass[preprint, showkeys, showpacs, amsmath, amssymb]{revtex4}

\usepackage{graphicx}
\begin{document}

\title{Diploidy and the selective advantage for sexual reproduction in unicellular organisms}

\author{Maya Kleiman}
\affiliation{Department of Chemistry, Ben-Gurion University of the Negev, Be'er-Sheva, Israel}
\author{Emmanuel Tannenbaum}
\email{emanuelt@bgu.ac.il}
\affiliation{Department of Chemistry, Ben-Gurion University of the Negev, Be'er-Sheva, Israel}

\begin{abstract}

This paper develops mathematical models describing the evolutionary dynamics of both asexually and sexually reproducing populations of diploid unicellular organisms.  The asexual and sexual life cycles are based on the asexual and sexual life cycles in {\it Saccharomyces cerevisiae}, or Baker's yeast, which normally reproduces by asexual budding, but switches to sexual reproduction when stressed.  The mathematical models consider three reproduction pathways:  (1)  Asexual reproduction.  (2) Self-fertilization (3)  Sexual reproduction.  We also consider two forms of genome organization.  In one case, we assume that the genome consists of two multi-gene chromosomes, while in the second case we consider the opposite extreme and assume that each gene defines a separate chromosome, which we call the multi-chromosome genome.  These two cases are considered in order to explore the role that recombination has on the mutation-selection balance and the selective advantage of the various reproduction strategies.  We assume that the purpose of diploidy is to provide redundancy, so that damage to a gene may be repaired using the other, presumably undamaged copy (a process known as homologous recombination repair).  As a result, we assume that the fitness of the organism only depends on the number of homologous gene pairs that contain at least one functional copy of a given gene.  If the organism has at least one functional copy of every gene in the genome, we assume a fitness of $ 1 $.  In general, if the organism has $ l $ homologous pairs that lack a functional copy of the given gene, then the fitness of the organism is $ \kappa_l $.  The $ \kappa_l $ are assumed to be monotonically decreasing, so that $ \kappa_0 = 1 > \kappa_1 > \kappa_2 > \dots > \kappa_{\infty} = 0 $.  For nearly all of the reproduction strategies we consider, we find, in the limit of large $ N $, that the mean fitness at mutation-selection balance is $ \max\{2 e^{-\mu} - 1, 0\} $, where $ N $ is the number of genes in the haploid set of the genome, $ \epsilon $ is the probability that a given DNA template strand of a given gene produces a mutated daughter during replication, and $ \mu = N \epsilon $.  The only exception is the sexual reproduction pathway for the multi-chromosomed genome.  Assuming a multiplicative fitness landscape where $ \kappa_l = \alpha^{l} $ for $ \alpha \in (0, 1) $, this strategy is found to have a mean fitness that exceeds the mean fitness of all of the other strategies.  Furthermore, while the other reproduction strategies experience a total loss of viability due to the steady accumulation of deleterious mutations once $ \mu $ exceeds $ \ln 2 $, no such transition occurs in the sexual pathway.  Indeed, in the limit as $ \alpha \rightarrow 1 $ for the multiplicative landscape, we can show that the mean fitness for the sexual pathway with the multi-chromosomed genome converges to $ e^{-2 \mu} $, which is always positive.  We explicitly allow for mitotic recombination in this work, which, in contrast to previous studies using different models, does not have any advantage over other asexual reproduction strategies.  The results of this paper provide a basis for understanding the selective advantage of the specific meiotic pathway that is employed by sexually reproducing organisms.  The results of this paper also suggest an explanation for why unicellular organisms such as {\it Saccharomyces cerevisiae} (Baker's yeast) switch to a sexual mode of reproduction when stressed.  While the results of this paper are based on modeling mutation-propagation in unicellular organisms, they nevertheless suggest that, in more complex organisms with significantly larger genomes, sex is necessary to prevent the loss of viability of a population due to genetic drift.  Finally, and perhaps most importantly, the results of this paper demonstrate a selective advantage for sexual reproduction with fewer and much less restrictive assumptions than previous work.

\end{abstract}


\maketitle

\section{Introduction}

The evolution and maintenance of sexual reproduction is regarded as one of the central problems of evolutionary biology (Bell 1982; Williams 1975; Maynard-Smith 1978; Michod 1995; Hurst and Peck 1996; Agrawal 2006; Visser and Elena 2007).  The various theories for the selective advantage for sex fall into one of two general categories:  The first category of theories argues that sex provides a mechanism to purge deleterious mutations from a genome (Kondrashov 1988; Muller 1964; Bruggeman et al. 2003; Paland and Lynch 2006; Bernstein et al. 1984; Michod 1995, Nedelcu et al. 2004; Barton and Otto 2005; Keightley and Otto 2006), while the second category of theories argues that sex provides greater genetic variability that allows populations to adapt more quickly to changing environments (Bell 1982; Hamilton et al. 1990; Howard and Lively 1994).

The first category of theories has two versions:  The first version, called the {\it Deterministic Mutation Hypothesis}, argues simply that sex provides a mechanism for purging deleterious mutations from a population, and thereby repairing the germ line (Kondrashov 1988).  The problem with this theory is that it requires what appears to be an overly restrictive assumption regarding the dependence of organismal fitness on the number of deleterious mutations in the genome:  In order for the Deterministic Mutation Hypothesis to hold, the organismal fitness must decrease increasingly rapidly with the number of deleterious mutations.  This is a phenomenon known as {\it synergistic epistasis}, and the problem with this assumption is that it is not at all clear whether or not it is correct.  Furthermore, the theory only works if mutation rates are at least one per genome per replication cycle, which is not the case for many simpler organisms that are capable of reproducing sexually.

The second version of the first category of theories argues that sex prevents the accumulation of mutations in a finite population.  The argument is that a finite, asexually reproducing population will steadily accumulate deleterious mutations over time.  This phenomenon has been termed {\it Muller's Ratchet} (Muller 1964).  An alternative view holds that, in a finite population, random mutations will lead to the elimination of organisms with the wild-type genome.  Instead, random associations will be formed between functional and non-functional copies of genes at different locations in the genome.  This is termed the {\it Hill-Robertson} effect, and leads to a reduction in fitness.  In both interpretations of the consequence of finite populations, sexual reproduction breaks up associations between genes and thereby provides a mechanism for restoring mutation-free genomes.  This process can slow down or even stop Muller's Ratchet, or alternatively, it may greatly mitigate the fitness reduction due to the Hill-Robertson effect (Keightley and Otto 2006).  The problem with this theory is that it relies on the assumption of a finite population, which is often interpreted as meaning that the population must be taken to be ``small" in some sense.  This is an ill-defined term, since it is not clear what the cutoff for a ``small" population should be (generally this means that the population is sufficiently small that there are measurable deviations from infinite population behavior, due to significant reductions in genetic variation when compared with the infinite population at mutation-selection balance).  

The second category of theories also has two versions:  The first version argues that sexual reproduction allows a population to adapt more quickly to changing environments (Bell 1982).  The idea is that sexual reproduction allows for recombination among different organisms, and thereby increases the genetic variation of a population.  In a dynamic environment, this increased variation will increase the chances that some organism has a fit genome, thereby leading to faster adaptation (Bell 1982).  This theory is sometimes called the {\it Vicar of Bray Hypothesis}, named after an English cleric who was known for changing his opinion as political circumstances dictated (Bell 1982).

The second version of this category of theories is known as the {\it Red Queen Hypothesis}, and states that sexual reproduction evolved as a way for relatively slowly reproducing host organisms to survive in a co-evolutionary ``genetic arms race" with quickly reproducing parasites.  This theory derives its name from a character named the Red Queen in Lewis Carroll's {\it In the Looking Glass}, who states, ``It takes all the running you can do to stay in one place" (Hamilton et al. 1990).

While this second category of theories is not necessarily incorrect, it is not clear that it offers a single, universal explanation for the evolution and maintenance of sexual reproduction.  The reason for this is that there are sexually reproducing organisms that have remained essentially unevolved for millions of years in what appear to fairly static environments (e.g. sharks and crocodiles).  As a result, while sexual reproduction may indeed have a selective advantage over asexual reproduction in dynamic environments, it is not clear that either a dynamic environment or co-evolutionary dynamics are necessary conditions for sexual reproduction to be advantageous over asexual reproduction.

The question of the evolution and maintenance of sexual reproduction is actually composed of several questions.  These are:  (1)  How did sex evolve, and what were the evolutionary pressures leading to its emergence?  (2)  Once sex emerged, what were/are the selective advantages leading to its maintenance and ubiquity?  (3)  Why is there such a large variety in the specific implementation of sexual reproduction strategies among different organisms?  For example, in some organisms, sexual reproduction is merely used as a stress response.  Many other organisms, insects for example, can either reproduce asexually (parthenogenesis) or sexually.  Still other organisms reproduce almost exclusively sexually, but can reproduce asexually if there is no other option.  In some organisms there is no sex differentiation, that is, each individual is a hermaphrodite capable of producing both sperm and eggs.  Other organisms have male/female differentiation, however in all female environments some of the females can transform into males.  Furthermore, the males play widely varying roles in organisms with male/female differentiation.  In some organisms, the males compete intensively for females, so that only a small percentage of males ever succeed in mating, however those who do generally control a relatively large group of females.  These males invest very little energy in the raising of their offspring.  This may be contrasted with organisms where males take an active role in the raising of the offspring.  In these circumstances, typically a male only mates with a single female, and a higher percentage of males are able to find female mates.

Clearly, there must be different regimes where the various implementations of asexual and sexual reproduction strategies are respectively advantageous.  A cost-benefit analysis that could identify these parameter regimes in a manner that is consistent with observation is a central aspect of the overall question of the evolution and maintenance of sexual reproduction.

It is therefore clear that the question of the evolution and maintenance of sexual reproduction is in fact a complex issue that cannot be addressed in a single study.  Rather, this issue can only be resolved within the context of a concerted research program that addresses a relatively broad array of questions.  

Nevertheless, research on the evolution and maintenance of sexual reproduction must first begin by understanding the basic advantage that this reproduction strategy provides.  Once this basic advantage is understood, it is then possible to study why specific implementations of asexual and sexual reproduction strategies are observed in different regimes, and it is also possible to attempt to reconstruct the evolutionary pathways for the emergence of sexual reproduction.

As has been discussed in the preceding paragraphs, the various theories for the selective advantage for sexual reproduction all suffer from one or more deficiencies.  As a result, even though much progress has been made in our understanding of the maintenance of sexual reproduction in many classes of organisms, the most fundamental question regarding the evolution and maintenance of sexual reproduction is still regarded as an open problem in evolutionary biology.  

Unicellular organisms are the ideal systems for studying the basic advantage for sexual reproduction over asexual reproduction.  There are two reasons for this:  First of all, because sexual reproduction already occurs in unicellular organisms, it makes sense to first study the selective advantage for sexual reproduction in these organisms, since their relative simplicity compared with multicellular organisms suggests that it will be possible to uncover the basic advantage for sexual reproduction without having to deal with additional complications.  Second, because unicellular organisms that are capable of reproducing sexually can also reproduce asexually, understanding the selective advantage for sexual reproduction in unicellular organisms will also help to delineate parameter regimes where asexual or sexual reproduction strategies are respectively advantageous.

{\it Saccharomyces cerevisiae}, or Baker's yeast, is a model diploid unicellular organism that engages in a form of sexual reproduction when stressed.  Thus, in this paper, we develop mathematical models describing asexual and sexual reproduction in unicellular organisms, where we take life cycles that are based on the asexual and sexual life cycles in {\it S. cerevisiae} (Herskowitz 1988; Mable and Otto 1998; De Massy et al. 1994; Roeder 1995).  We assume multi-gene genomes comprised of semiconservatively replicating, double-stranded DNA molecules.  While we still make a number of simplifying assumptions, we nevertheless believe that the models considered in this paper are sufficiently realistic to be relevant for actual biological systems.  Consequently, we believe that the results we obtain in this paper may be used to draw definite conclusions about the relative selective advantage of various reproduction strategies in unicellular organisms.

We consider three distinct reproduction mechanisms:  Asexual reproduction, self-fertilization, and sexual reproduction.  Furthermore, for each reproduction mechanism we consider two extremes of genome organization, in order to explore the effect of recombination on the selective advantage for the various reproduction strategies:  A two-chromosomed, multi-gene genome, and a multi-chromosomed genome where each chromosome consists of a single gene.

The mathematical models considered here assume that the only purpose of diploidy is to provide genetic redundancy, or more specifically, a mechanism to repair double-stranded genetic damage on one gene using the other, presumably undamaged, corresponding region in the homologous gene.  This process is known as {\it homologous recombination repair}.  As a result, we assume that all organisms whose genomes contain at least one functional copy of every gene have the wild-type fitness, taken to be $ 1 $.  While it is possible that loss of function in one of the genes in a homologous pair can lead to a loss of fitness, if a cell has at least one functional copy of every gene in the genome, then it should remain viable.  As a result, for a genome with a large number of genes, the fitness penalty for having an additional homologous pair with one non-functional copy of a gene should become steadily smaller as the number of homologous pairs with one non-functional copy of a gene increases.  Thus, for the purposes of simplicity, we consider an initial fitness landscape where there is no fitness penalty for having homologous pairs with only one non-functional copy of a gene.  In any event, it makes sense that the overall purpose of diploidy is to provide a mechanism for repair and does not in general increase fitness.  For if the latter was the case, then it is not clear why two should be some kind of ``magic number", in the sense that fitness is optimized when an organism has two functional copies of every gene.  If fitness could be significantly increased by increasing the number of copies of a given gene, then it seems that the optimal number of copies of a gene should be highly gene-dependent (for example, highly expressed genes may be present in numerous copies, while one copy may suffice for genes that are only expressed from time to time).

While the fitness of the organism remains the wild-type fitness of $ 1 $ as long as the genome has at least one functional copy of every gene, we assume that the fitness of the organism is reduced for every homologous gene pair that lacks a functional copy of a given gene.  Thus, if $ l $ is the number of homologous gene pairs in the genome lacking a functional copy of the given gene, then the fitness of the organism is $ \kappa_l $, where we assume that the $ \kappa_l $ are monotonically decreasing, so that $ \kappa_0 = 1 > \kappa_1 > \kappa_2 > \dots > \kappa_{\infty} = 0 $.  

Based on the analysis that follows, we obtain, in the limit of large $ N $, that the mean fitnesses at mutation-selection balance for nearly all reproduction pathways is $ \max\{2 e^{-\mu} - 1, 0\} $, where $ N $ is the number of genes in the haploid set of the genome, $ \epsilon $ is the probability that a given template DNA strand of a given gene produces a mutant daughter as a result of replication, and $ \mu = N \epsilon $.  The only exception is for the case of sexual reproduction in the multi-chromosomed genome.  Here, the mean fitness can significantly exceed $ \max\{2 e^{-\mu} - 1, 0\} $.

Furthermore, except for sexual reproduction in the multi-chromosomed genome, all of the other reproduction strategies experience a total loss of viability once $ \mu $ exceeds $ \ln 2 $.  Here, the evolutionary dynamics of the population is characterized by the steady accumulation of deleterious mutations, and a steady decrease in fitness, eventually leading to a steady-state mean fitness of $ 0 $.  In the quasispecies model of evolutionary dynamics, this is known as the {\it error catastrophe}, which is characterized by a localization to delocalization transition of the population over the genome space (Eigen 1971; Tannenbaum and Shakhnovich 2005).  Because the population fitness drops to zero in this case, the population also undergoes what is known as {\it lethal mutagenesis}.  While the error catastrophe and lethal mutagenesis are formally distinct phenomena, they can often be associated with one another, as is the case with the models being considered here (Bull and Wilke 2008).  

However, for sexual reproduction in the multi-chromosomed genome, the error catastrophe does not occur as long as $ \kappa_l > 0 $ for each $ l $.  This result is interesting, for, although it is based on an analysis of unicellular organisms, it nevertheless suggests that sexual reproduction is necessary to prevent genetic drift and population extinction in more complex organisms that have long genomes.  For example, for {\it S. cerevisiae}, $ \mu $ is on the order of $ 0.01 $, which is well below $ \ln 2 \approx 0.69 $, while for humans ({\it H. sapiens}), $ \mu $ is on the order of $ 3 $, which is considerably larger than $ \ln 2 $.  Thus, {\it S. cerevisiae} may not need to reproduce sexually in order to remain viable (though sexual reproduction provides a selective advantage under stressful conditions), but humans may simply die out if they were to reproduce asexually.

It must be emphasized that this paper assumes a static fitness landscape, and assumes an infinite population, so that the selective advantage for sex does not arise due to a dynamic environment or a small population.  Furthermore, in contrast to the Deterministic Mutation Hypothesis, we believe that our fitness landscape is a more ``generic" one.  In particular, synergistic epistasis is not necessary for sex to have a selective advantage.  It is also not necessary for $ \mu $ to be larger than $ 1 $ for sex to have an advantage.  In fact, as long as each of the $ \kappa_l > 0 $ for $ l < \infty $, then sexual reproduction in the multi-chromosomed genome has a selective advantage over the other reproduction strategies for all values of $ \mu $.  

Thus, in this paper, we have developed a model that yields a selective advantage for sex under fewer and far less restrictive assumptions than previous models.  Interestingly, our model essentially does this by explicitly incorporating the role of diploidy, which is a level of realism that was not considered in many previous studies.  

\section{Description of the Organismal Genomes and Fitness Landscapes}

In this section, we describe the two modes of genome organization that we will consider in this paper.  Figure 1 may be useful for what follows.

\subsection{Two-chromosomed genome}

We begin with the two-chromosomed genome.  Here, we assume that a unicellular organism has a diploid genome consisting of two chromosomes, where each chromosome has $ N $ genes, labelled $ 1, \dots, N $.  We also assume that with each gene is associated a ``master" sequence (actually a pair of complementary sequences, since we are dealing with double-stranded DNA), corresponding to a functional copy of the gene, while any mutation to the master sequence renders the gene non-functional.  This is the multi-gene generalization of the single-fitness-peak approximation often made in quasispecies models of evolutionary dynamics (Bull et al. 2005; Wilke 2005; Tannenbaum and Shakhnovich 2005).  While this assumption is obviously oversimplified (indeed, recent research suggests that genes may, on average, sustain up to six mutations before losing functionality (Zeldovich et al. 2007)), it is the simplest non-trivial landscape that allows for mutation and selection (as opposed to random genetic drift).  Furthermore, the single-fitness-peak landscape reflects the fact that only a small fraction of all gene sequences will encode a gene carrying out a specific function, which is why the single-fitness-peak approximation has been known to provide correct order-of-magnitude estimates of various biological parameters (Kamp and Bornholdt 2002).

We may denote a given chromosome by $ \sigma = s_1 s_2 \dots s_N $, where each $ s_i = 1 $ if gene $ i $ is functional, and 
$ s_i = 0 $ if gene $ i $ is non-functional.  This means that the genome of a given organism may be represented by 
$ \{\sigma_1, \sigma_2\} $, where $ \sigma_1, \sigma_2 $ represent each of the two chromosomes in the genome.

During replication, the two DNA strands of each chromosome separate, and each strand forms the template for the synthesis of a complementary daughter strand (Tannenbaum and Shakhnovich 2005).  Because mutations can occur during each daughter strand synthesis, both daughter genes of a given parent gene may contain mutations.  We let $ p $ denote the probability that a template strand from a master copy of a gene forms a mutation-free daughter, so that $ 1 - p $ is the probability that the template strand forms a mutated daughter.  If the template strand already has a mutation, then we assume that sequence lengths are sufficiently long that any new mutations occur in a previously unmutated portion of the strand, so that a mutated template strand forms a non-functional daughter gene with probability $ 1 $.  This assumption is known as the {\it neglect of backmutations} (Tannenbaum and Shakhnovich 2005).  Mutation gives rise to a transition probability $ p(\sigma', \sigma) $, which is defined as the probability that a given template strand from chromosome $ \sigma' $ produces the daughter chromosome $ \sigma $.

We also define $ \epsilon = 1 - p $, and we define $ \mu = N \epsilon $.  $ \mu $ is the average number of mutated genes produced from $ N $ template gene strands per replication cycle.  In what follows in this paper, we will consider the limit of $ N \rightarrow \infty $ with $ \mu $ held constant, which is equivalent to holding the per genome replication fidelity constant in the limit of large genomes.

It should be noted that we are not necessarily assuming that the only source of mutations in the genome is due to point-mutations during replication.  The model allows for mutations that accumulate in the genome in between replications, due to base modifications and damage that occurs as a result of free radicals, radiation, and spontaneous chemical alterations.  During the growth phase of the cell, repair mechanisms are constantly at work repairing this genetic damage.  However, these genetic repair mechanisms are not infinitely fast, and so cannot completely eliminate all genetic damage.  As a result, at the time of replication, there will always be some bases that are damaged, which can then lead to the fixation of mutations in the daughter genome as a consequence of daughter strand synthesis.  This leads to an effective per genome, per replication cycle point mutation rate that is somewhat larger than would be expected if one considered daughter strand synthesis errors alone.

We let $ r_i $ denote the probability of mitotic recombination in this model (Mandegar and Otto 2007), which is the probability that the two daughter chromosomes of a given parent co-segregate into the identical daughter cell.  Mitotic recombination generally refers to individual genes.  However, in this model, we assume that the genes on a given chromosome all co-segregate together, so that $ r_i $ in this case refers to co-segregation of chromosomes.  In the multi-chromosome model to be discussed below, individual genes may segregate independently of one another, so that $ r_i $ then more accurately reflects the biological definition of mitotic recombination.

We assume that cells replicate with first-order growth kinetics.  We let $ \kappa_{\{\sigma_1, \sigma_2\}} $ denote the first-order growth rate constant of cells with genome $ \{\sigma_1, \sigma_2\} $, and we let $ n_{\{\sigma_1, \sigma_2\}} $ denote the number of organisms in the population with genome $ \{\sigma_1, \sigma_2\} $.  

We define an ordered strand-pair representation of the population, by defining $ n_{(\sigma_1, \sigma_2)} = (1/2) n_{\{\sigma_1, \sigma_2\}} $ if $ \sigma_1 \neq \sigma_2 $, and $ n_{(\sigma, \sigma)} = n_{\{\sigma, \sigma\}} $.  We also define $ \kappa_{(\sigma_1, \sigma_2)} = \kappa_{\{\sigma_1, \sigma_2\}} $.  

The ordered strand-pair representation leads to a method for characterizing a given ordered strand-pair by three parameters, denoted $ l_{10}, l_{01}, l_{00} $.  $ l_{10} $ denotes the number of homologous gene pairs for which the allele in $ \sigma_1 $ is functional (i.e. a ``1'' gene) and the allele in $ \sigma_2 $ is non-functional (i.e. a ``0'' gene).  $ l_{01} $ denotes the number of homologous gene pairs for which the allele in $ \sigma_1 $ is non-functional, and the allele in $ \sigma_2 $ is functional.  $ l_{00} $ denotes the number of homologous gene pairs where both alleles in $ \sigma_1 $ and $ \sigma_2 $ are non-functional.  We may also define $ l_{11} $ to be the number of homologous gene pairs where both alleles in $ \sigma_1 $ and $ \sigma_2 $ are functional.  Note that $ l_{11} = N - l_{10} - l_{01} - l_{00} $.  Also note that, by definition of the fitness landscape given in the Introduction, we have that $ \kappa_{(\sigma_1, \sigma_2)} = \kappa_{l_{00}} $.

\subsection{Multi-chromosomed genome}

For the multi-chromosomed genome, we assume a diploid genome consisting of $ N $ homologous gene-pairs, where each gene defines a separate chromosome, giving rise to a genome consisting of $ 2 N $ genes.  We assume that the homologous pairs segregate independently of one another, though for each homologous pair we may assume a mitotic recombination probability $ r_i $, defined as in the previous subsection.  Indeed, unless otherwise specified, all of the definitions in the multi-gene, two-chromosome model are the same for the multi-chromosome model being considered here.

Because the genes all lie on separate chromosomes, a diploid genome may be characterized by the two parameters 
$ l_{10}, l_{00} $, as opposed to the three parameters $ l_{10}, l_{01}, l_{00} $ as in the previous subsection.  Here, a diploid genome characterized by the parameters $ l_{10}, l_{00} $ has exactly $ l_{10} $ homologous pairs with one functional gene and one non-functional gene (i.e. a ``1'' and a ``0''), and $ l_{00} $ homologous pairs with two non-functional genes.  As before, we have $ l_{11} = N - l_{10} - l_{00} $.

Although both the two-chromosomed and multi-chromosomed genomes represent extremes of genome organization, we argue that, due to the {\it Law of Independent Assortment of Alleles} in classical genetics, the dynamics arising from the multi-chromosomed genome more closely approximates the true segregation dynamics of genes in actual organisms.

\section{Asexual reproduction}

\subsection{Description of the reproduction pathway}

In the asexual reproduction pathway, each chromosome replicates, and then the daughter chromosomes segregate into one of the two daughter cells.  Each daughter cell receives two of the daughter chromosomes from a given homologous pair, and it is assumed that daughter chromosomes from distinct homologous pairs segregate independently of one another.

If there is no mitotic recombination, then the two daughters of a given parent segregate into distinct daughter cells.  With mitotic recombination, the two daughter chromosomes (or genes, in the case of the multi-chromosomed genome) of a given parent chromosome co-segregate into the same daughter cell.  As mentioned previously, mitotic recombination for each homologous pair occurs with probability $ r_i $.

Figure 2 illustrates the asexual reproduction pathway.

\subsection{Two-chromosomed genome}

\subsubsection{Evolutionary dynamics equations}

In Appendix A.1, we show that the evolutionary dynamics of a population of asexually reproducing organisms with two-chromosomed genomes is given by,
\begin{widetext}
\begin{eqnarray}
&   &
\frac{d z_{l_1, l_2, l_3}}{dt} = 
-(\kappa_{l_3} + \bar{\kappa}) z_{l_1, l_2, l_3}
+ 2 r_i 
\sum_{l_1' = 0}^{N - l_1 - l_2 - l_3} 
\sum_{l_2' = 0}^{l_1}
\sum_{l_3' = 0}^{l_2} 
\sum_{l_4' = 0}^{l_3} 
\sum_{l_5' = 0}^{l_3 - l_4'}
\sum_{l_6' = 0}^{l_3 - l_4' - l_5'} 
\kappa_{l_6'} 
z_{l_1' + l_2' + l_3' + l_4', l_5', l_6'}
\times \nonumber \\
&   &
\frac{(l_1' + l_2' + l_3' + l_4')!}{l_1'! l_2'! l_3'! l_4'!} [(1 - \epsilon)^2]^{l_1'} [\epsilon (1 - \epsilon)]^{l_2'} [\epsilon (1 - \epsilon)]^{l_3'} (\epsilon^2)^{l_4'}
\times \nonumber \\
&    &
\frac{(N - l_1' - l_2' - l_3' - l_4' - l_5' - l_6')!}{(l_1 - l_2')! (l_2 - l_3')! (l_3 - l_4' - l_5' - l_6')! (N - l_1 - l_2 - l_3 - l_1')!}
\times \nonumber \\
&    &
[\epsilon (1 - \epsilon)]^{l_1 - l_2'} [\epsilon (1 - \epsilon)]^{l_2 - l_3'} (\epsilon^2)^{l_3 - l_4' - l_5' - l_6'}  [(1 - \epsilon)^2]^{N - l_1 - l_2 - l_3 - l_1'}
\nonumber \\
&   &
+ 2 (1 - r_i) 
\sum_{l_1' = 0}^{l_1} 
\sum_{l_2' = 0}^{l_2} 
\sum_{l_3' = 0}^{l_3} 
\sum_{l_4' = 0}^{l_3 - l_3'} 
\sum_{l_5' = 0}^{l_3 - l_3' - l_4'}
\kappa_{l_5'} z_{l_1' + l_3', l_2' + l_4', l_5'}
\frac{(l_1' + l_3')!}{l_1'! l_3'!} (1 - \epsilon)^{l_1'} \epsilon^{l_3'}
\frac{(l_2' + l_4')!}{l_2'! l_4'!} (1 - \epsilon)^{l_2'} \epsilon^{l_4'}
\times \nonumber \\
&   &
\frac{(N - l_1' - l_2' - l_3' - l_4' - l_5')!}{(l_1 - l_1')! (l_2 - l_2')! (l_3 - l_3' - l_4' - l_5')! (N - l_1 - l_2 - l_3)!}
\times \nonumber \\
&   &
[\epsilon (1 - \epsilon)]^{l_1 - l_1'} [\epsilon (1 - \epsilon)]^{l_2 - l_2'} (\epsilon^2)^{l_3 - l_3' - l_4' - l_5'}
[(1 - \epsilon)^2]^{N - l_1 - l_2 - l_3}
\end{eqnarray}
\end{widetext}
Here, $ z_{l_1, l_2, l_3} $ defines the total fraction of the ordered strand-pair population characterized by the parameters $ l_{10} = l_1, l_{01} = l_2, l_{00} = l_3 $, and $ \bar{\kappa}(t) $ is the average first-order growth rate constant of the entire population, a quantity known as the {\it mean fitness}.  We have that $ \bar{\kappa}(t) = \sum_{l_1 = 0}^{N} \sum_{l_2 = 0}^{N - l_1} \sum_{l_3 = 0}^{N - l_1 - l_2} \kappa_{l_3} z_{l_1, l_2, l_3} $.

\subsubsection{Mean fitness at mutation-selection balance}

For all of the reproduction strategies being considered in this paper, the central object of interest is the mean fitness of the population at mutation-selection balance (or equivalently, at steady-state).  The reason for this is that the mean fitness, by measuring the first-order growth rate constant of the population as a whole, determines which population will drive the other to extinction when two or more populations are mixed together.  Due to the nature of exponential growth, the population with the largest mean fitness will drive the others to extinction, which means that the reproduction strategy that the winning population employs is the reproduction strategy that has the selective advantage over the others for the given set of parameters.

This approach to determining which reproduction strategy is optimal for a given set of parameters is known as the {\it group selection} approach.  The group selection approach may be criticized in that it does not take into account the fact that selection acts on individuals, rather than populations.  An individual organism whose genes code for an optimal survival strategy in the given environment will out-reproduce the other organisms in the population.  This survival strategy may not necessarily coincide with the optimal survival strategy for the population as a whole.  Indeed, it is well-known that the group selection approach is inadequate for taking into account effects such as co-evolutionary dynamics, parasitism, and defection from cooperative strategies.

Despite the deficiencies of the group selection approach in general, it can give correct results under certain circumstances.  In cases where different populations or individuals do not directly interact with one another, so that one organism does not increase its fitness at the expense of the other, the group selection approach is a valid method for determining which genes will be selected for in a given environment.

In this paper, we make the simplifying assumption that populations with distinct reproduction strategies do not mix with one another (that is, sexuals interact with sexuals, asexual with asexuals, etc.), so that in our case the group selection approach is valid.  The group selection approach, however, would be problematic if we wished to consider not the maintenance of sexual reproduction, but rather the evolution and emergence of sexual reproduction from an asexual population.  Indeed, in recent work we found that pure sexual replicators could not arise from an asexual population, because their initial population density would be so low as to lead to large mating times that would completely eliminate any benefit for sex (Tannenbaum and Fontanari 2008).
 
At mutation-selection balance, the mean fitness is given by,
\begin{equation}
\bar{\kappa} = \max\{\kappa_l [2 (1 - \epsilon)^{N - l} - 1]| l = 0, \dots, N\}
\end{equation}

It must be emphasized that this result is the exact finite $ N $ solution for the steady-state mean fitness, and does not depend on the value of $ r_i $.  

In the limit as $ N \rightarrow \infty $ with $ \mu $ held constant, we have,
\begin{equation}
\bar{\kappa} \rightarrow \max\{2 e^{-\mu} - 1, 0\}
\end{equation}
where this result is both independent of $ r_i $ and the specific nature of the fitness function $ \{\kappa_l\} $ (assuming that the fitness function satisfies the monotonicity condition given in the Introduction).

The transition between the two functional forms for $ \bar{\kappa} $ at $ \mu = \ln 2 $ corresponds to a localization to delocalization transition known as the error catastrophe.  Beyond this value of $ \mu $, the mutation rate is sufficiently high that natural selection can no longer localize the population to a given region of the genome space, and the result is the loss of viability due to genetic drift.  If we include decay terms into our model (e.g. death or loss of organisms due to flow out of a chemostat), then this loss of viability can lead to the extinction of the population (a phenomenon known as lethal mutagenesis).

To avoid encumbering the biologically relevant results of our model (i.e. the steady-state mean fitness) with the detailed mathematical derivations, we have placed the mathematical derivations in the following subsubsection.  We believe that the mathematical analysis is sufficiently interesting that it should not be relegated to an Appendix.  However, we place it in a separate section from the main results so that the reader can choose to simply skip over the mathematical details.

\subsubsection{Mathematical derivation of the mean fitness at mutation-selection balance}

To determine the mean fitness at mutation-selection balance, denoted by $ \bar{\kappa} $, we proceed as follows:  We define a {\it generating function} (Wilf 2006) $ w_{l}(\beta_1, \beta_2, t) $, defined over the population distribution $ \{z_{l_1, l_2, l_3}\} $, via,
\begin{equation}
w_l(\beta_1, \beta_2, t) = \sum_{l_1 = 0}^{N - l} \sum_{l_2 = 0}^{N - l - l_1} \beta_1^{l_1} \beta_2^{l_2} z_{l_1, l_2, l}
\end{equation}
and we also let $ w_l(\beta_1, \beta_2) $ denote the steady-state value of $ w_l(\beta_1, \beta_2, t) $.

In Appendix D we show that, at mutation-selection balance, the following equation holds for $ \beta_1 = \beta, \beta_2 = 1 - \beta $:
\begin{widetext}
\begin{eqnarray}
&   &
\frac{\partial w_l(\beta, 1 - \beta, t)}{\partial t} \geq
\kappa_l [2 (1 - \epsilon)^{N - l} (r_i w_l(1, 0, t) + (1 - r_i) w_l(\beta, 1 - \beta, t)) - w_l(\beta, 1 - \beta, t)] 
\nonumber \\
&   &
- \bar{\kappa}(t) w_l(\beta, 1 - \beta, t)
\end{eqnarray}
\end{widetext}
where equality holds if $ l = 0 $, or if $ z_{l_1, l_2, l_3} = 0 $ for $ l_3 < l $.  Setting $ \beta = 1 $ we obtain,
\begin{eqnarray}
&   &
\frac{\partial w_l(1, 0, t)}{\partial t} \geq [\kappa_l (2 (1 - \epsilon)^{N - l} - 1) - \bar{\kappa}(t)] w_l(1, 0, t)
\nonumber \\
\end{eqnarray}
and so, if we assume that the system converges to a stable steady-state, then we must have that $ \bar{\kappa} \geq \kappa_l [2 (1 - \epsilon)^{N - l} - 1] $ for all $ l = 0, \dots, N $, and so $ \bar{\kappa} \geq \max\{\kappa_l [2 (1 - \epsilon)^{N - l} - 1]| l = 0, \dots, N\} $.

Let $ l^{*} $ denote the smallest value of $ l_3 $ such that there exist $ l_1, l_2 $ for which $ z_{l_1, l_2, l_3} > 0 $ at steady-state.  Because the $ z_{l_1, l_2, l_3} $ sum to $ 1 $, it follows that some of them must be positive, and hence such an $ l^{*} $ must exist.

We have that $ w_{l^{*}}(1/2, 1/2) > 0 $.  If we also have that $ w_{l^{*}}(1, 0) > 0 $, then,
\begin{equation}
0 = [\kappa_{l^{*}} (2 (1 - \epsilon)^{N - l^{*}} - 1) - \bar{\kappa}] w_{l^{*}}(1, 0)
\end{equation}
which implies that $ \bar{\kappa} = \kappa_{l^{*}} [2 (1 - \epsilon)^{N - l^{*}} - 1] $.

If, on the other hand, we have that $ w_{l^{*}}(1, 0) = 0 $, then we obtain,
\begin{equation}
0 = [\kappa_{l^{*}} (2 (1 - r_i) (1 - \epsilon)^{N - l^{*}} - 1) - \bar{\kappa}] w_{l^{*}}(\frac{1}{2}, \frac{1}{2})
\end{equation}
which implies that $ \bar{\kappa} = \kappa_{l^{*}} [2 (1 - r_i) (1 - \epsilon)^{N - l^{*}} - 1] $.  

If $ r_i = 0 $ then the two expressions for $ \bar{\kappa} $ are identical.  If $ r_i > 0 $, however, then the second expression is smaller than the first, which is impossible, given the inequality that $ \bar{\kappa} $ must satisfy.  Therefore, for $ r_i > 0 $, we must have that $ w_{l^{*}}(1, 0) > 0 $ and so in any case we have $ \bar{\kappa} = \kappa_{l^{*}} [2 (1 - \epsilon)^{N - l^{*}} - 1] $.
However, given that $ \bar{\kappa} \geq \max\{\kappa_l [2 (1 - \epsilon)^{N - l} - 1]| l = 0, \dots, N\} $, we must have that 
$ \bar{\kappa} = \kappa_{l^{*}} [2 (1 - \epsilon)^{N - l^{*}} - 1] = \max\{\kappa_l [2 (1 - \epsilon)^{N - l} - 1]| l = 0, \dots, N\} $.

Now, let us consider the limit as $ N \rightarrow \infty $ while holding $ \mu $ fixed, and let us consider two different regimes, the first where $ 2 e^{-\mu} - 1 > 0 $, and the second where $ 2 e^{-\mu} - 1 \leq 0 $.  The first regime corresponds to the interval $ 0 \leq \mu < \ln 2 $, while the second corresponds to the interval $ \mu \geq \ln 2 $.

Given that the $ \kappa_l $ are monotonically decreasing, and given that $ \lim_{l \rightarrow \infty} \kappa_l = 0 $, it follows that, given any $ \epsilon' > 0 $, there exists some $ l_{\epsilon'} > 0 $ such that $ \kappa_l  < \epsilon' $ whenever $ l > l_{\epsilon'} $.  We may relax this condition somewhat, in order to allow for the possibility that finite genome sizes affect the fitness landscape, but that the fitness landscape nevertheless converges as $ N \rightarrow \infty $ to a landscape that satisfies the property given above.

Thus, we assume that the fitness landscape has the following property:  For every $ \epsilon' > 0 $, there exists an $ l_{\epsilon'} > 0 $ and an $ N_{\epsilon'} > 0 $ such that $ \kappa_l < \epsilon' $ whenever $ l > l_{\epsilon'} $ and $ N > N_{\epsilon'} $.

So, suppose that $ \mu \in [0, \ln 2) $, so that $ 2 e^{-\mu} - 1 > 0 $.  Then let us assume that $ l $, $ N $ are sufficiently large so that $ \kappa_{l'} < 2 e^{-\mu} - 1 $ for all $ l' \geq l $.  Then, given $ \epsilon' > 0 $, choose $ N_{\epsilon'} $ to be such that 
$ |(1 - \epsilon)^{n} - e^{-\mu}| < \epsilon' $ for all $ n \geq N_{\epsilon'} $.  Then, for $ l' < l $ we have, for $ N \geq N_{\epsilon'} + l $, that,
\begin{eqnarray}
&   &
\kappa_{l'} [2 (1 - \epsilon)^{N - l'} - 1] < \kappa_{l'} [2 (e^{-\mu} + \epsilon') - 1] 
\nonumber \\
&   &
= \kappa_{l'} (2 e^{-\mu} - 1) + 2 \kappa_{l'} \epsilon'
\leq 2 e^{-\mu} - 1 + 2 \epsilon' 
\end{eqnarray}
Now, for $ l' \geq l $ we have that,
\begin{equation}
\kappa_{l'} [2 (1 - \epsilon)^{N - l'} - 1] \leq \kappa_{l'} < 2 e^{-\mu} - 1 < 2 e^{-\mu} - 1 + 2 \epsilon'
\end{equation}
and so we have that $ \bar{\kappa} < 2 e^{-\mu} - 1 + 2 \epsilon' $.  However, we also have, for $ N \geq N_{\epsilon'} + l $, that,
\begin{eqnarray}
&   &
\bar{\kappa} \geq 2 (1 - \epsilon)^{N} - 1 > 2 (e^{-\mu} - \epsilon') - 1 = 2 e^{-\mu} - 1 - 2 \epsilon'
\nonumber \\
\end{eqnarray}
and so we have that $ 2 e^{-\mu} - 1 - 2 \epsilon' < \bar{\kappa} < 2 e^{-\mu} - 1 + 2 \epsilon' $.  Since $ \epsilon' > 0 $ is arbitrary, it follows that, for $ \mu \in [0, \ln 2) $, we have that $ \bar{\kappa} \rightarrow 2 e^{-\mu} - 1 $ as $ N \rightarrow \infty $.

Now suppose that $ \mu \in [\ln 2, \infty) $, so that $ 2 e^{-\mu} - 1 \leq 0 $.  Then given some $ \epsilon' > 0 $, choose $ l $, $ N $ to be sufficiently large so that $ \kappa_{l'} < \epsilon' $ for all $ l' \geq l $.  Then, choose $ N_{\epsilon'} $ to be such that 
$ |(1 - \epsilon)^{n} - e^{-\mu}| < \epsilon'/2 $ for all $ n \geq N_{\epsilon'} $.  Then, for $ l' < l $ we have, for $ N \geq N_{\epsilon'} + l $, that,
\begin{eqnarray}
&   &
\kappa_{l'} [2 (1 - \epsilon)^{N - l'} - 1] < \kappa_{l'} [2 (e^{-\mu} + \frac{\epsilon'}{2}) - 1]
\nonumber \\
&   &
= \kappa_{l'} (2 e^{-\mu} - 1) + \kappa_{l'} \epsilon' \leq \epsilon'
\end{eqnarray}
while for $ l' \geq l $ we have that,
\begin{equation}
\kappa_{l'} [2 (1 - \epsilon)^{N - l'} - 1] \leq \kappa_{l'} < \epsilon'
\end{equation}
and so we have that $ \bar{\kappa} < \epsilon' $.  However, we also have that $ \bar{\kappa} \geq 0 $, so since $ \epsilon' > 0 $ is arbitrary, it follows that, for $ \mu \in [\ln 2, \infty) $, $ \bar{\kappa} \rightarrow 0 $ as $ N \rightarrow \infty $.

The result of our analysis is that $ \bar{\kappa} = \max\{2 e^{-\mu} - 1, 0\} $ in the $ N \rightarrow \infty $ limit.  

When $ r_i = 0 $, we may prove that $ z_{l_1, l_2, l_3} = 0 $ at steady-state whenever $ l_1 + l_2 + l_3 < N $.  We will prove this by contradiction.  So, suppose that there exist $ l_1, l_2, l_3 $ where $ l_1 + l_2 + l_3 < N $ such that $ z_{l_1, l_2, l_3} > 0 $.  Then let us choose $ l_3^* $ to be the smallest value of $ l_3 $ for which there exist $ l_1, l_2 $ with $ l_1 + l_2 + l_3 < N $ and $ z_{l_1, l_2, l_3} > 0 $.  This means that, whenever $ l_3 < l_3^* $, then $ z_{l_1, l_2, l_3} > 0 \Rightarrow l_1 + l_2 + l_3 = N $.

Now, given $ l_3^* $, choose $ l_1^*, l_2^* $ so that $ l_1^* + l_2^* $ is the smallest value of $ l_1 + l_2 $ for which $ z_{l_1, l_2, l_3^*} > 0 $.  Then, in Eq. (1), setting $ l_1 = l_1^*, l_2 = l_2^*, l_3 = l_3^* $, we have that $ l_1' + l_2' + l_3' + l_4' + l_5' \leq
l_1^* + l_2^* + l_3^* < N $, so by definition of $ l_3^* $, we must have $ z_{l_1' + l_3', l_2' + l_4', l_5'} = 0 $ for $ l_5' < l_3^* $.  Therefore, in Eq. (1), we need only consider $ l_5' = l_3^* $, which implies that $ l_3' = l_4' = 0 $, and so $ z_{l_1' + l_3', l_2' + l_4', l_5'} = z_{l_1', l_2', l_3^*} $.  Furthermore, because $ l_1' \leq l_1^* $, $ l_2' \leq l_2^* $, we have $ l_1' + l_2' \leq l_1^* + l_2^* $, with equality if and only if $ l_1' = l_1^* $, $ l_2' = l_2^* $.  By definition of $ l_1^*, l_2^* $, it follows that $ z_{l_1', l_2', l_3^*} = 0 $ unless $ l_1' = l_1^* $, $ l_2' = l_2^* $.  Putting everything together, we obtain that, at steady-state, Eq. (1) becomes, for $ r_i = 0 $,
\begin{equation}
0 = [\kappa_{l_3^*} (2 (1 - \epsilon)^{N - l_3^*} (1 - \epsilon)^{N - l_1^* - l_2^* - l_3^*} - 1) - \bar{\kappa}] z_{l_1^*, l_2^*, l_3^*}
\end{equation}
which implies that $ \bar{\kappa} = \kappa_{l_3^*} (2 (1 - \epsilon)^{N - l_3^*} (1 - \epsilon)^{N - l_1^* - l_2^* - l_3^*} - 1) $.  However, because $ l_1^* + l_2^* + l_3^* < N $, it follows that $ (1 - \epsilon)^{N - l_1^* - l_2^* - l_3^*} < 1 \Rightarrow
\bar{\kappa} < \kappa_{l_3^*}(2 (1 - \epsilon)^{N - l_3^*} - 1) \Rightarrow\Leftarrow $, from the result for $ \bar{\kappa} $.  With this contradiction, our claim is proved.

\subsection{Multi-chromosomed genome}

\subsubsection{Evolutionary dynamics equations}

In Appendix A.2, we show that the evolutionary dynamics of a population of asexually reproducing organisms with multi-chromosomed genomes is given by,
\begin{widetext}
\begin{eqnarray}
&   &
\frac{d z_{l_1, l_2}}{dt} = -(\kappa_{l_2} + \bar{\kappa}(t)) z_{l_1, l_2} 
+ 2 \sum_{l_1' = 0}^{N - l_1 - l_2} \sum_{l_2' = 0}^{l_1} \sum_{l_3' = 0}^{l_2} 
\sum_{l_4' = 0}^{l_2 - l_3'} \kappa_{l_4'} z_{l_1' + l_2' + l_3', l_4'}
\times \nonumber \\
&   &
\frac{(l_1' + l_2' + l_3')!}{l_1'! l_2'! l_3'!} [\frac{r_i}{2} (1 - \epsilon)^2]^{l_1'} [(1 - \epsilon) (1 - r_i (1 - \epsilon))]^{l_2'} 
[\epsilon + \frac{r_i}{2} (1 - \epsilon)^2]^{l_3'}
\times \nonumber \\
&   &
\frac{(N - l_1' - l_2' - l_3' - l_4')!}{(l_1 - l_2')! (l_2 - l_3' - l_4')! (N - l_1 - l_2 - l_1')!}
[2 \epsilon (1 - \epsilon)]^{l_1 - l_2'} (\epsilon^2)^{l_2 - l_3' - l_4'} [(1 - \epsilon)^2]^{N - l_1 - l_2 - l_1'} 
\nonumber \\
\end{eqnarray}
\end{widetext}
where $ z_{l_1, l_2} $ is the total fraction of the population whose genomes are characterized by the parameters $ l_{10} = l_1, l_{00} = l_2 $, and the mean fitness $ \bar{\kappa}(t) $ is given by $ \bar{\kappa}(t) = \sum_{l_1 = 0}^{N} \sum_{l_2 = 0}^{N - l_1} 
\kappa_{l_2} z_{l_1, l_2} $.

\subsubsection{Mean fitness at mutation-selection balance}
 
As with the two-chromosomed genome, the mean fitness for the multi-chromosomed genome at mutation-selection balance is given by,
\begin{equation}
\bar{\kappa} = \max\{\kappa_l [2 (1 - \epsilon)^{N - l} - 1]| l = 0, \dots, N\}
\end{equation} 
where this result is independent of the value of $ r_i $.  

In the limit as $ N \rightarrow \infty $ with $ \mu $ held fixed, we obtain that,
\begin{equation}
\bar{\kappa} \rightarrow \max\{2 e^{-\mu} - 1, 0\}
\end{equation}
 
\subsubsection{Mathematical derivation of the mean fitness at mutation-selection balance}

To determine the mean fitness at mutation-selection balance, we proceed analogously to the two-chromosomed case:  We define a generating function $ w_{l}(\beta, t) $, defined over the population distribution $ \{z_{l_1, l_2}\} $, via,
\begin{equation}
w_l(\beta, t) = \sum_{k = 0}^{N - l} \beta^{k} z_{k, l}
\end{equation}
and we also let $ w_l(\beta) $ denote the steady-state value of $ w_l(\beta, t) $.

Following a similar procedure to the derivation in Appendix D, we may show that,
\begin{widetext}
\begin{equation}
\frac{\partial w_l(\beta, t)}{\partial t} \geq -(\kappa_{l} + \bar{\kappa}(t)) w_l(\beta, t) 
+ 2 (1 - \epsilon)^{N - l} [1 + (2 \beta - 1) \epsilon]^{N - l} 
\kappa_l w_l(\frac{(\frac{1}{2} - \beta) r_i (1 - \epsilon) + \beta}{1 + (2 \beta - 1) \epsilon}, t)
\end{equation}
\end{widetext}
with equality if $ l = 0 $ or if $ z_{l_1, l_2} = 0 $ for $ l_2 < l $.  

Setting $ \beta = 1/2 $ gives,
\begin{equation}
\frac{\partial w_l(\frac{1}{2}, t)}{\partial t} \geq 
[\kappa_{l} (2 (1 - \epsilon)^{N - l} - 1) - \bar{\kappa}(t)] w_l(\frac{1}{2}, t)
\end{equation}
with equality if $ l = 0 $ or if $ z_{l_1, l_2} = 0 $ for $ l_2 < l $.  As with the two-chromosomed model, this implies that $ \bar{\kappa} \geq \max\{\kappa_l [2 (1 - \epsilon)^{N - l} - 1]| l = 0, \dots, N\} $.

Let $ l^{*} $ denote the smallest value of $ l_2 $ such that there exists an $ l_1 $ for which $ z_{l_1, l_2} > 0 $ at steady-state.  Then since $ z_{l_1, l_2} = 0 $ for $ l_2 < l^{*} $, we have, at steady-state, that,
\begin{equation}
0 = [\kappa_{l^{*}} (2 (1 - \epsilon)^{N - l^{*}} - 1) - \bar{\kappa}] w_{l^{*}}(\frac{1}{2})
\end{equation} 

Because there exists an $ l_1 $ for which $ z_{l_1, l^{*}} > 0 $, it follows that $ w_{l^{*}}(1/2) > 0 $, and so 
$ \bar{\kappa} = \kappa_{l^{*}} [2 (1 - \epsilon)^{N - l^{*}} - 1] \Rightarrow \bar{\kappa} = \max\{\kappa_{l} [2 (1 - \epsilon)^{N - l} - 1]| l = 0, \dots, N\} $.  

As is the case for the two-chromosomed model, it follows that $ \bar{\kappa} \rightarrow \max\{2 e^{-\mu} - 1, 0\} $ as $ N \rightarrow \infty $.

For $ r_i = 0 $, suppose that there exist $ l_1, l_2 $ with $ l_1 + l_2 < N $ such that $ z_{l_1, l_2} > 0 $ at steady-state.  Then let $ l_2^* $ be the smallest value of $ l_2 $ for which there exists an $ l_1 $ with $ l_1 + l_2 < N $ such that $ z_{l_1, l_2} > 0 $ at steady-state.  Then let $ l_1^* $ be the smallest value of $ l_1 $ such that $ z_{l_1, l_2^*} > 0 $ at steady-state.

In Eq. (15), when $ r_i = 0 $ we have that $ l_1' = 0 $.  We also have, for $ l_1 = l_1^*, l_2 = l_2^* $, that $ l_2' + l_3' + l_4' \leq
l_1^* + l_2^* < N $, and so $ z_{l_2' + l_3', l_4'} = 0 $ for $ l_4' < l_2^* $, and so we may take $ l_4' = l_2^* $, $ l_3' = 0 $.  Now, by definition of $ l_1^* $, we have that $ z_{l_2', l_2^*} = 0 $ whenever $ l_2' < l_1^* $, and so we may take $ l_2' = l_1^* $. 

At steady-state, Eq. (15) then becomes,
\begin{equation}
0 = [\kappa_{l_2^*} (2 (1 - \epsilon)^{N - l_2^*} (1 - \epsilon)^{N - l_1^* - l_2^*} - 1) - \bar{\kappa}] z_{l_1^*, l_2^*}
\end{equation}
and so $ \bar{\kappa} = \kappa_{l_2^*} (2 (1 - \epsilon)^{N - l_2^*} (1 - \epsilon)^{N - l_1^* - l_2^*} - 1) $.  Since $ l_1^* + l_2^* < N $, it follows that $ (1 - \epsilon)^{N - l_1^* - l_2^*} < 1 \Rightarrow \bar{\kappa} < \kappa_{l_2^*} (2 (1 - \epsilon)^{N - l_2^*} - 1) \Rightarrow\Leftarrow $, thereby proving our claim.

\section{Self-Fertilization}

\subsection{Description of the reproduction pathway}

In the self-fertilization reproduction pathway, a diploid cell first divides via the asexual pathway into two diploid daughter cells.  Each of the diploid daughter cells then divide into two haploids, where each haploid receives exactly one chromosome from each homologous pair.  The result is four haploids, which then pair at random with one another and fuse to form two diploid cells.  

This pathway is illustrated in Figure 3 for a two-chromosomed genome.  As with the case for asexual reproduction, the multi-chromosomed case is similar, except that distinct homologous pairs segregate independently of one another.

\subsection{Two-chromosomed genome}

For the two-chromosomed genome, the equations for self-fertilization are identical to the equations for asexual replication, where $ r_i = 1/3 $.  The reason for this is that a given parent diploid cell produces four haploids containing four chromosomes.  Because mating is random, a given chromosome has a probability of $ 1/3 $ of pairing with any other chromosome, which gives $ r_i = 1/3 $.

\subsection{Multi-chromosomed genome}

\subsubsection{Evolutionary dynamics equations}

In Appendix B, we show that the evolutionary dynamics of a population of organisms reproducing via the self-fertilization pathway are, for the multi-chromosome case, given by,
\begin{widetext}
\begin{eqnarray}
&   &
\frac{d z_{l_1, l_2}}{dt} = -(\kappa_{l_2} + \bar{\kappa}(t)) z_{l_1, l_2} +
\frac{2}{3} \sum_{l_1' = 0}^{N - l_1 - l_2} \sum_{l_2' = 0}^{l_1} 
\sum_{l_3' = 0}^{l_2} \sum_{l_4' = 0}^{l_2 - l_3'} 
\kappa_{l_4'} z_{l_1' + l_2' + l_3', l_4'}
\frac{(l_1' + l_2' + l_3')!}{l_1'! l_2'! l_3'!} 
\times \nonumber \\
&   &
[(\frac{r_i}{2} (1 - \epsilon)^2)^{l_1'} ((1 - \epsilon) (1 - r_i (1 - \epsilon)))^{l_2'} (\epsilon + \frac{r_i}{2} (1 - \epsilon)^2)^{l_3'}
\nonumber \\
&   &
+ 2 (\frac{1 - r_i}{4} (1 - \epsilon)^2)^{l_1'} ((1 - \epsilon) (1 - \frac{1 - r_i}{2} (1 - \epsilon)))^{l_2'} (\epsilon + \frac{1 - r_i}{4} (1 - \epsilon)^2)^{l_3'}]
\times \nonumber \\
&   &
\frac{(N - l_1' - l_2' - l_3' - l_4')!}{(l_1 - l_2')! (l_2 - l_3' - l_4')! (N - l_1 - l_2 - l_1')!}
[2 \epsilon (1 - \epsilon)]^{l_1 - l_2'} (\epsilon^2)^{l_2 - l_3' - l_4'} [(1 - \epsilon)^2]^{N - l_1 - l_2 - l_1'}
\nonumber \\
\end{eqnarray}
\end{widetext}

\subsubsection{Mean fitness at mutation-selection balance}

The mean fitness for the multi-chromosomed genome with the self-fertilization pathway at mutation-selection balance is given by,
\begin{equation}
\bar{\kappa} = \max\{\kappa_l [2 (1 - \epsilon)^{N - l} - 1]| l = 0, \dots, N\}
\end{equation} 
where this result is independent of the value of $ r_i $.  

In the limit as $ N \rightarrow \infty $ with $ \mu $ held fixed, we obtain that,
\begin{equation}
\bar{\kappa} \rightarrow \max\{2 e^{-\mu} - 1, 0\}
\end{equation}

\subsubsection{Mean fitness at mutation-selection balance in the limit where $ N \rightarrow \infty $}

Defining $ w_l(\beta, t) $ as for the case of asexual reproduction in the multi-chromosomed genome, we obtain,
\begin{widetext}
\begin{eqnarray}
&   &
\frac{\partial w_l(\beta, t)}{\partial t} \geq -(\kappa_{l} + \bar{\kappa}(t)) w_l(\beta, t)
+ \frac{2}{3} [2 \beta \epsilon (1 - \epsilon) + (1 - \epsilon)^2]^{N - l} 
\kappa_{l} 
\times \nonumber \\
&   &
[w_l(\frac{(\frac{1}{2} - \beta) r_i (1 - \epsilon) + \beta}{1 + (2 \beta - 1) \epsilon}, t) + 
2 w_l(\frac{(\frac{1}{2} - \beta) \frac{1 - r_i}{2} (1 - \epsilon) + \beta}{1 + (2 \beta - 1) \epsilon}, t)]
\end{eqnarray}
\end{widetext}
with equality if $ l = 0 $ or if $ z_{l_1, l_2} = 0 $ for $ l_2 < l $.

Setting $ \beta = 1/2 $ gives,
\begin{equation}
\frac{\partial w_l(\frac{1}{2}, t)}{\partial t} \geq [\kappa_{l} (2 (1 - \epsilon)^{N - l} - 1) - \bar{\kappa}(t)] w_l(\frac{1}{2}, t)
\end{equation}

Following a similar analysis to the one performed for the asexual, multi-chromosomed case, we obtain that $ \bar{\kappa} = \max\{\kappa_{l} [2 (1 - \epsilon)^{N - l} - 1]| l = 0, \dots, N\} $, and that $ \bar{\kappa} \rightarrow \max\{2 e^{-\mu} - 1, 0\} $ in the limit where $ N \rightarrow \infty $.

\section{Sexual Reproduction}

\subsection{Description of the reproduction pathway}

In the sexual reproduction pathway, we assume that a diploid cell produces four haploids in the same manner as for the self-fertilization pathway.  However, instead of the four haploids fusing with one another, the haploids enter a haploid pool, where they fuse at random with haploids produced by other diploid parent cells.  This reproduction pathway is illustrated in Figure 4.

In contrast to self-fertilization, where we assume that the haploid fusion is fast (since the haploids are in close proximity to one another, having been produced by the same parent), with sexual reproduction we must take into consideration the haploid population.

A given haploid genome, whether it is derived from the two-chromosomed or multi-chromosomed diploid genome, may be characterized by the parameter $ l_0 $, which is the number of non-functional genes in the cell.  We may then let $ n_{l_0} $ denote the number of haploids in the population whose genomes are characterized by the parameter $ l_0 $.  Now, because a diploid cell contains twice the number of chromosomes as the corresponding haploid, we define the total population $ n $ to be $ n_D + n_H/2 $, where $ n_D $ is the total population of diploids, and $ n_H $ is the total population of haploids.  We then define the haploid population fractions $ z_l $ via $ z_l = (1/2) n_l/n $.  We define the total haploid population fraction $ z_H = \sum_{l = 0}^{N} z_l = (1/2) n_H/n $.

We assume that haploid fusion is a second-order process characterized by a second-order rate constant $ \gamma $.  If $ V $ denotes the system volume, then we assume that, as the population grows, the volume increases so as to maintain a constant population density $ \rho \equiv n/V $.

\subsection{Two-chromosomed genome}

\subsubsection{Evolutionary dynamics equations}

In Appendix C.1, we show that the evolutionary dynamics of a population of sexually reproducing organisms with two-chromosomed genomes is given by, 
\begin{widetext}
\begin{eqnarray}
&   &
\frac{d z_{l_1, l_2, l_3}}{d t} = 
-(\kappa_{l_3} + \bar{\kappa}(t)) z_{l_1, l_2, l_3} +
2 \gamma \rho \frac{(l_1 + l_3)! (l_2 + l_3)!}{l_1! l_2! l_3!}
\times \nonumber \\
&   &
(\prod_{k = 1}^{l_1} \frac{N - l_1 - l_2 - l_3 + k}{N - l_1 - l_3 + k})
(\prod_{k = 1}^{l_3} \frac{1}{N - l_3 + k})
z_{l_1 + l_3} z_{l_2 + l_3}
\nonumber \\
&   &
\frac{d z_l}{dt} = -\bar{\kappa}(t) z_l - 2 \gamma \rho z_H z_l +
2 \sum_{l_1 = 0}^{N - l} \sum_{l_2 = 0}^{l} \sum_{l_3 = 0}^{l - l_2} \sum_{l_4 = 0}^{l - l_2 - l_3}
\kappa_{l_4} z_{l_1 + l_2, l_3, l_4} 
\frac{(l_1 + l_2)!}{l_1! l_2!} (1 - \epsilon)^{l_1} \epsilon^{l_2}
\times \nonumber \\
&   &
\frac{(N - l_1 - l_2 - l_3 - l_4)!}{(l - l_2 - l_3 - l_4)! (N - l - l_1)!} \epsilon^{l - l_2 - l_3 - l_4} (1 - \epsilon)^{N - l - l_1}
\end{eqnarray}
\end{widetext}

In this paper, we will consider for simplicity the limit as $ \gamma \rho \rightarrow \infty $, so that the characteristic haploid fusion time is negligible.  At this stage, we are neglecting the time cost for sex associated with the characteristic haploid fusion time, in order to see if we can first identify a basic advantage for sex before considering costs that can reduce or eliminate this advantage. 

In the $ \gamma \rho \rightarrow \infty $ limit, we have that the evolutionary dynamics equations are given by,
\begin{widetext}
\begin{eqnarray}
&   &
\frac{d z_{l_1, l_2, l_3}}{dt} = -(\kappa_{l_3} + \bar{\kappa}(t)) z_{l_1, l_2, l_3} +
\frac{2}{\bar{\kappa}(t)} \frac{(l_1 + l_3)! (l_2 + l_3)!}{l_1! l_2! l_3!}
\times \nonumber \\
&   &
(\prod_{k = 1}^{l_1} \frac{N - l_1 - l_2 - l_3 + k}{N - l_1 - l_3 + k})
(\prod_{k = 1}^{l_3} \frac{1}{N - l_3 + k})
f_{l_1 + l_3} f_{l_2 + l_3}
\nonumber \\
\end{eqnarray}
where,
\begin{eqnarray}
&   &
f_l \equiv \sum_{l_1 = 0}^{N - l} \sum_{l_2 = 0}^{l} \sum_{l_3 = 0}^{l - l_2} \sum_{l_4 = 0}^{l - l_2 - l_3}
\kappa_{l_4} z_{l_1 + l_2, l_3, l_4} 
\frac{(l_1 + l_2)!}{l_1! l_2!} (1 - \epsilon)^{l_1} \epsilon^{l_2}
\times \nonumber \\
&   &
\frac{(N - l_1 - l_2 - l_3 - l_4)!}{(l - l_2 - l_3 - l_4)! (N - l - l_1)!} \epsilon^{l - l_2 - l_3 - l_4} (1 - \epsilon)^{N - l - l_1}
\end{eqnarray}
\end{widetext}

\subsubsection{Mean fitness at mutation-selection balance in the limit where $ N \rightarrow \infty $}

The generating function approach that was successfully used to obtain the mean fitnesses of the non-sexual reproduction pathways does not work for the sexual reproduction pathway.  Nevertheless, in the $ N \rightarrow \infty $ limit, it is possible to derive an analytical expression for the mean fitness of the two-chromosomed, sexual reproduction pathway, at mutation-selection balance.  Interestingly, in the limit as $ N \rightarrow \infty $ at fixed $ \mu $, we obtain that,

\begin{equation}
\bar{\kappa} = \max\{2 e^{-\mu} - 1, 0\}
\end{equation}
which is identical to the $ N \rightarrow \infty $ limit of the other reproduction strategies.

Figure 5 shows a plot of $ \bar{\kappa} $ versus $ \mu $ for $ N = 50 $.  We assume a multiplicative fitness landscape, defined by $ \kappa_l = \alpha^l $, with $ \alpha = 0.8 $.  We present plots of $ \bar{\kappa} $ using both the analytical, $ N \rightarrow \infty $ result, and results obtained by solving for the steady-state of the evolutionary dynamics equations using fixed-point iteration.  Note the good agreement between the analytical result and the results obtained by fixed-point iteration.  Due to finite size effects, the numerically computed values of $ \bar{\kappa} $ near $ \mu = \ln 2 $ are slightly larger than the analytical, $ N \rightarrow \infty $ result. 

\subsubsection{Mathematical derivation of the mean fitness at mutation-selection balance in the limit where 
$ N \rightarrow \infty $}

In the limit as $ \gamma \rho \rightarrow \infty $, we obtain that $ z_l \rightarrow 0 $ for $ l = 0, \dots, N $, so that $ \bar{\kappa}(t) z_l \rightarrow 0 $.  However, it is possible that $ \gamma \rho z_H z_l $ converges to some finite and possibly non-zero value.  Assuming a steady-state for the haploid population (because the $ z_l = 0 $) we obtain, 

\begin{widetext}
\begin{eqnarray}
&   &
\gamma \rho z_H z_l = \sum_{l_1 = 0}^{N - l} \sum_{l_2 = 0}^{l} \sum_{l_3 = 0}^{l - l_2} \sum_{l_4 = 0}^{l - l_2 - l_3}
\kappa_{l_4} z_{l_1 + l_2, l_3, l_4} 
\frac{(l_1 + l_2)!}{l_1! l_2!} (1 - \epsilon)^{l_1} \epsilon^{l_2}
\times \nonumber \\
&   &
\frac{(N - l_1 - l_2 - l_3 - l_4)!}{(l - l_2 - l_3 - l_4)! (N - l - l_1)!} \epsilon^{l - l_2 - l_3 - l_4} (1 - \epsilon)^{N - l - l_1}
\end{eqnarray}
\end{widetext}

Summing $ l $ from $ 0 $ to $ N $ gives $ \gamma \rho z_H^2 = \bar{\kappa}(t) $.  Therefore, defining $ \tilde{z}_l = z_l/z_H $, we may solve for $ \tilde{z}_l $ in terms of $ \bar{\kappa}(t) $ and the diploid population fractions.  Substituting the results into the dynamical equations for the diploid population, we obtain Eq. (29).

In the limit as $ N \rightarrow \infty $, one possible solution is simply that the population is completely delocalized over the sequence space, and so $ \bar{\kappa} = 0 $.  So, we first consider the regime where $ \bar{\kappa} > 0 $.  In Appendix E.1, we show, in the limit as $ N \rightarrow \infty $, that,
\begin{widetext}
\begin{equation}
\frac{(l_1 + l_3)! (l_2 + l_3)!}{l_1! l_2! l_3!} (\prod_{k = 1}^{l_1} \frac{N - l_1 - l_2 - l_3 + k}{N - l_1 - l_3 + k})
(\prod_{k = 1}^{l_3} \frac{1}{N - l_3 + k}) \rightarrow \frac{1}{l_3!} (\frac{l_1 l_2}{N})^{l_3} e^{-\frac{l_1 l_2}{N}}
\end{equation}
\end{widetext}

So, at mutation-selection balance where $ \bar{\kappa} > 0 $, we have that,
\begin{eqnarray}
&    &
z_{l_1, l_2, l_3} = \frac{2}{\bar{\kappa} (\bar{\kappa} + \kappa_{l_3})}
\frac{1}{l_3!} (\frac{l_1 l_2}{N})^{l_3} e^{-\frac{l_1 l_2}{N}} f_{l_1 + l_3} f_{l_2 + l_3}
\nonumber \\
\end{eqnarray}

Noting that $ \tilde{z}_l = f_l/\bar{\kappa} $, we may substitute the expression for $ z_{l_1, l_2, l_3} $ into the definition of the $ f_l $ to obtain, in the limit of large $ N $, that,
\begin{widetext}
\begin{equation}
\tilde{z}_l = 2 e^{-\mu} \sum_{k = 0}^{l} \frac{\mu^k}{k!} \tilde{z}_{l - k}
\sum_{l_4 = 0}^{l - k} \frac{\kappa_{l_4}}{\bar{\kappa} + \kappa_{l_4}}
\sum_{l_1 = 0}^{N - l}
\frac{1}{l_4!} (\frac{l_1 (l - l_4 - k)}{N})^{l_4} e^{-\frac{l_1 (l - l_4 - k)}{N}} 
\tilde{z}_{l_1 + l_4} 
\end{equation}
\end{widetext}

Now, let $ l^{*} $ denote the smallest value of $ l $ for which $ \tilde{z}_l > 0 $.  Since $ \tilde{z}_l = 0 $ for all 
$ l < l^{*} $, we have,
\begin{equation}
1 = 2 e^{-\mu} \sum_{l_4 = 0}^{l^{*}}
\frac{\kappa_{l_4}}{\bar{\kappa} + \kappa_{l_4}}
\sum_{l_1 = 0}^{N - l^{*}}
\frac{1}{l_4!} (\frac{l_1 (l^{*} - l_4)}{N})^{l_4} e^{-\frac{l_1 (l^{*} - l_4)}{N}}
\tilde{z}_{l_1 + l_4}
\end{equation}

In Appendix E.1, we show that the distribution for the $ \tilde{z}_l $ approaches a Gaussian with mean that scales as $ \sqrt{N} $ and a standard-deviation that scales as $ N^{1/4} $.  If we then define a probability density function $ p(x) $ via $ \sqrt{N} \tilde{z}_l = p(l/\sqrt{N}) $, then in the limit of large $ N $ we may write,
\begin{equation}
1 = 2 e^{-\mu} \int_{0}^{\infty} d x_1 e^{-x x_1} p(x_1) 
\sum_{l_4 = 0}^{\infty} \frac{\kappa_{l_4}}{\bar{\kappa} + \kappa_{l_4}}\frac{1}{l_4!} (x x_1)^{l_4} 
\end{equation}
where $ x \equiv l^{*}/\sqrt{N} $.  Now, using the inequality,
\begin{equation}
\frac{\kappa_{l_4}}{\bar{\kappa} + \kappa_{l_4}} \leq \frac{1}{\bar{\kappa} + 1}
\end{equation}
we have,
\begin{equation}
1 \leq \frac{2 e^{-\mu}}{\bar{\kappa} + 1} \int_{0}^{\infty} d x_1 p(x_1)
= \frac{2 e^{-\mu}}{\bar{\kappa} + 1}
\end{equation}
and so $ \bar{\kappa} \leq 2 e^{-\mu} - 1 $.

Now, if we define $ w_{100} = \sum_{l = 0}^{N} z_{l, 0, 0} $, then it is possible to show, for finite $ N $, that,
\begin{equation}
\frac{d w_{100}}{dt} = [2 (1 - \epsilon)^{N} - 1 - \bar{\kappa}(t)] w_{100}
\end{equation}
from which it follows that $ \bar{\kappa} \geq 2 (1 - \epsilon)^{N} - 1 $ in order for the steady-state to be stable.  In particular, as $ N \rightarrow \infty $, we obtain that $ \bar{\kappa} \geq 2 e^{-\mu} - 1 $.  Combined with the previous analysis giving that $ \bar{\kappa} \leq 2 e^{-\mu} - 1 $, we obtain that $ \bar{\kappa} = 2 e^{-\mu} - 1 $.  However, since $ \bar{\kappa} \geq 0 $, we have that $ \bar{\kappa} = \max\{2 e^{-\mu} - 1, 0\} $.

\subsection{Multi-chromosomed genome}

\subsubsection{Evolutionary dynamics equations}

In Appendix C.2, we show that the evolutionary dynamics of a population of sexually reproducing organisms with multi-chromosomed genomes is given by,
\begin{widetext}
\begin{eqnarray}
&   &
\frac{d z_{l_1, l_2}}{dt} = -(\kappa_{l_2} + \bar{\kappa}(t)) z_{l_1, l_2} 
+ 2 \gamma \rho \sum_{l = 0}^{l_1} 
\frac{(l + l_2)! (l_1 - l + l_2)!}{l! (l_1 - l)! l_2!}
\times \nonumber \\
&   &
(\prod_{k = 1}^{l} \frac{N - l_1 - l_2 + k}{N - l - l_2 + k})
(\prod_{k = 1}^{l_2} \frac{1}{N - l_2 + k})
z_{l + l_2} z_{l_1 - l + l_2} 
\nonumber \\
&   &
\frac{d z_l}{dt} = -\bar{\kappa}(t) z_{l} - 2 \gamma \rho z_{l} z_H 
+ 2 \sum_{l_1 = 0}^{N - l} \sum_{l_2 = 0}^{l} \sum_{l_3 = 0}^{l - l_2}
\kappa_{l_3} z_{l_1 + l_2, l_3} 
\frac{(l_1 + l_2)!}{l_1! l_2!} (\frac{1 - \epsilon}{2})^{l_1} (\frac{1 + \epsilon}{2})^{l_2}
\times \nonumber \\
&   &
\frac{(N - l_1 - l_2 - l_3)!}{(l - l_2 - l_3)! (N - l - l_1)!} 
\epsilon^{l - l_2 - l_3} (1 - \epsilon)^{N - l - l_1} 
\end{eqnarray}
\end{widetext}

Following a similar procedure to the two-chromosomed case, we obtain, in the limit as $ \gamma \rho \rightarrow \infty $, that,
\begin{widetext}
\begin{eqnarray}
&   &
\frac{d z_{l_1, l_2}}{dt} = -(\kappa_{l_2} + \bar{\kappa}(t)) z_{l_1, l_2} 
+ \frac{2}{\bar{\kappa}(t)} \sum_{l = 0}^{l_1} 
\frac{(l + l_2)! (l_1 - l + l_2)!}{l! (l_1 - l)! l_2!}
\times \nonumber \\
&   &
(\prod_{k = 1}^{l} \frac{N - l_1 - l_2 + k}{N - l - l_2 + k})
(\prod_{k = 1}^{l_2} \frac{1}{N - l_2 + k})
f_{l + l_2} f_{l_1 - l + l_2} 
\nonumber \\
\end{eqnarray}
where,
\begin{eqnarray}
&   &
f_l \equiv \sum_{l_1 = 0}^{N - l} \sum_{l_2 = 0}^{l} \sum_{l_3 = 0}^{l - l_2}
\kappa_{l_3} z_{l_1 + l_2, l_3} 
\frac{(l_1 + l_2)!}{l_1! l_2!} (\frac{1 - \epsilon}{2})^{l_1} (\frac{1 + \epsilon}{2})^{l_2}
\times \nonumber \\
&   &
\frac{(N - l_1 - l_2 - l_3)!}{(l - l_2 - l_3)! (N - l - l_1)!} 
\epsilon^{l - l_2 - l_3} (1 - \epsilon)^{N - l - l_1} 
\end{eqnarray}
\end{widetext}

\subsubsection{Mean fitness at mutation-selection balance in the limit where $ N \rightarrow \infty $}

From Appendix E.2, we have that, whenever $ \bar{\kappa} > 0 $, then it is obtained by solving the pair of equations,
\begin{eqnarray}
&   &
1 = 2 e^{-\lambda^2} \sum_{l = 0}^{\infty} \frac{1}{l!} \frac{\lambda^{2 l} \kappa_l}{\bar{\kappa} + \kappa_l}
\nonumber \\
&   &
\mu = \lambda^2 (1 - 2 e^{-\lambda^2} \sum_{l = 0}^{\infty} \frac{1}{l!} \frac{\lambda^{2 l} \kappa_{l + 1}}
{\bar{\kappa} + \kappa_{l + 1}})
\end{eqnarray}

When $ \kappa_l = \delta_{l0} $, where $ \delta_{ij} $ denotes the Kronecker delta function, we have, from the second equation, that $ \lambda^2 = \mu $.  Substituting into the first equation,
we obtain,
\begin{equation}
1 = \frac{2 e^{-\mu}}{\bar{\kappa} + 1} \Rightarrow \bar{\kappa} = 2 e^{-\mu} - 1 
\end{equation}
and so $ \bar{\kappa} = \max \{2 e^{-\mu} - 1, 0\} $.

However, if $ \kappa_l > 0 $ for finite $ l $, then we find that the steady-state mean fitness for the sexual reproduction pathway for the multi-chromosomed genome exceeds the mean fitness of $ \max\{2 e^{-\mu} - 1, 0\} $ for the other pathways.  Admittedly, we have only checked this for multiplicative fitness landscapes for which $ \kappa_l = \alpha^{l} $, where $ \alpha \in (0, 1) $.  However, we conjecture that this result will hold more generally, since the multiplicative fitness landscape seems to be a reasonable first approximation for how $ \kappa_l $ will vary with $ l $.  Essentially, what we are doing with the multiplicative landscape is averaging over the various fitness penalties induced by knocking out a given homologous pair in the genome.  To be more precise, we are making an optimal curve fit of the form $ \alpha^l $ to the fitness values $ \kappa_0 = 1, \kappa_1, \kappa_2, \dots, \kappa_{\infty} = 0 $.  This can be done by taking the natural logarithm of the fitness functions, and finding the optimal linear fit $ l \ln \alpha $ for the points $ \ln \kappa_0 = 0, \ln \kappa_1, \ln \kappa_2, \dots, \ln \kappa_{\infty} = -\infty $.

The fitness values $ \{\kappa_l\} $ are themselves averages of the true fitness landscape of the organism:  For a given value of $ l $, $ \kappa_l $ is taken to be the average of all fitnesses obtained from all possible genomes having $ l $ homologous pairs lacking a functional copy of their respective genes.      

The fitness increase of the multi-chromosomed sexual pathway over the other pathways becomes larger as $ \alpha $ increases from $ 0 $ to $ 1 $.  Crucially, the multi-chromosomed sexual pathway does not appear to exhibit any kind of change in the functional form of $ \bar{\kappa} $ at some critical $ \mu $, signaling the onset of an error threshold.  Thus, it appears that the multi-chromosomed sexual reproduction pathway considered in this paper does not have an error threshold, so that a sexual population can survive at mutation rates where a non-sexual population would lose viability and presumably go extinct.

Figure 6 shows a plot of $ \bar{\kappa} $ versus $ \mu $ for $ N = 50 $, assuming a multiplicative landscape with $ \alpha = 0.8 $.  We present plots obtained by numerically solving for $ \bar{\kappa} $ using the $ N \rightarrow \infty $ equations given in Eq. (44) (using a combination of fixed-point iteration and binary search), by numerically solving the evolutionary dynamics equations themselves using fixed-point iteration, and by stochastic simulations of finite populations of reproducing organisms.  For comparison, we also include a plot of the function $ \max \{2 e^{-\mu} - 1, 0\} $.  Note the good agreement that is obtained between the stochastic simulations, the fixed-point iteration of the evolutionary dynamics equations, and the numerical solution of the $ N \rightarrow \infty $ equations.

We can obtain an analytical, closed form expression for $ \bar{\kappa} $ in the limit that $ \alpha \rightarrow 1 $.  We find that $ \lim_{\alpha \rightarrow 1} \bar{\kappa} = e^{-2 \mu} $, a result that will be derived in the following subsubsection.  Because $ e^{-2 \mu} > 0 $, this result is consistent with our claim that there is no error threshold for sexual reproduction with the multi-chromosomed genome.  Furthermore, because $ e^{-2 \mu} \geq 2 e^{-\mu} - 1 $, with equality only occurring for $ \mu = 0 $, we obtain that this result is also consistent with our observations that the sexual, multi-chromosomed mean fitness exceeds the mean fitness of the other reproduction pathways as long as $ \alpha > 0 $.

Figure 7 shows a plot of $ \bar{\kappa} $ versus $ \mu $ for a multiplicative landscape with $ \alpha = 0.99 $.  Because $ \alpha $ is so close to $ 1 $ here, we were unable to show results from either fixed-point iteration of the evolutionary dynamics equations themselves, nor results from stochastic simulations, since the required value of $ N $ in both cases, and the required population size in the latter case, would be so large as to make computation times prohibitive.  However, we may readily obtain numerical expressions for $ \bar{\kappa} $ by solving the $ N \rightarrow \infty $ equations given by Eq. (44), and comparing the result with the analytical expression of $ e^{-2 \mu} $.  As can be seen in Figure 7, the results are indistinguishable.

\subsubsection{Mathematical derivation of the mean fitness at mutation-selection balance in the limit as $ N \rightarrow \infty $}

Following a similar procedure for the two-chromosomed case, we have, in the limit of large $ N $, that the steady-state distribution $ z_{l_1, l_2} $ satisfies,
\begin{equation}
z_{l_1, l_2} = \frac{2}{\bar{\kappa} (\bar{\kappa} + \kappa_{l_2})} \sum_{l = 0}^{l_1} \frac{1}{l_2!} (\frac{l (l_1 - l)}{N})^{l_2}
e^{-\frac{l (l_1 - l)}{N}} f_{l + l_2} f_{l_1 - l + l_2}
\end{equation}
where this analysis of course assumes that $ \bar{\kappa} > 0 $.

Substituting into the definition for $ f_l $, and following a similar procedure as was done for the two-chromosomed genome, we have, in the large $ N $ limit,
\begin{eqnarray}
&   &
\tilde{z}_l = 2 e^{-\mu} \sum_{l_3 = 0}^{l} \frac{\kappa_{l_3}}{\bar{\kappa} + \kappa_{l_3}} 
\sum_{k = 0}^{l - l_3} \frac{\mu^{k}}{k!} 
\sum_{l_1 = 0}^{N - l}
\frac{(l_1 + l - l_3 - k)!}{l_1! (l - l_3 - k)!} (\frac{1}{2})^{l_1 + l - l_3 - k} 
\times \nonumber \\
&   &
\sum_{l_4 = 0}^{l_1 + l - l_3 - k} \frac{1}{l_3!} (\frac{l_4 (l_1 + l - l_3 - k - l_4)}{N})^{l_3}
e^{-\frac{l_4 (l_1 + l - l_3 - k - l_4)}{N}} \tilde{z}_{l_4 + l_3} \tilde{z}_{l_1 + l - k - l_4}
\end{eqnarray}

As $ N $ becomes large, we have observed that the $ \tilde{z}_l $ approach a Gaussian distribution with a mean that scales as $ \sqrt{N} $ and a standard deviation that scales as $ N^{1/4} $.  As a result, if we define a variable $ x = l/\sqrt{N} $, then in the limit as $ N \rightarrow \infty $ we can transform from a discrete representation in terms of the $ \tilde{z}_l $ into a continuous representation in terms of a probability density $ p(x) $, where conservation of probability implies that $ p(x) (1/\sqrt{N}) = \tilde{z}_l \Rightarrow p(x) = \tilde{z}_l \sqrt{N} $.

The transformation from a discrete to a continuous representation allows us to re-write Eq. (47) as an integral equation.  We then take the Laplace transform of both sides of the equation.  Since we are dealing with the large $ N $ limit, we expand the Laplace transforms on both sides of the equation out to $ 1/\sqrt{N} $ and equate the two first-order expansions.  This leads to a set of equalities that must hold in the limit of large $ N $, which gives us the pair of equations in Eq. (44) that must be solved in order to obtain $ \bar{\kappa} $.  The details of this derivation may be found in Appendix E.2.

Now, let us analyze the behavior of Eq. (44) in the limit as $ \alpha \rightarrow 1 $.  In this limit, we expect that $ \lambda^2 \rightarrow \infty $.  The reason for this is as follows:  In Appendix E.2, we define $ \lambda $ to be such that $ \lambda \sqrt{N} $ is the average number of defective genes in a given haploid.  We also have that, as $ N \rightarrow \infty $, the $ \tilde{z}_l $ converge to a Gaussian distribution that in fact approaches a $ \delta $-function centered at $ \lambda \sqrt{N} $.  In Appendix E.2, we show that the probability that two haploids, each having $ \lambda \sqrt{N} $ defective genes, will fuse to form a diploid with exactly $ l $ homologous gene pairs lacking a functional copy of the given gene is given by,

\begin{equation}
\frac{1}{l!} \lambda^{2 l} e^{-\lambda^2}
\end{equation}

Therefore, on average, the overlap of two haploids at mutation-selection balance will produce a diploid with a fitness of 
\begin{equation}
\sum_{l = 0}^{\infty} \frac{1}{l!} (\lambda^2 \alpha)^{l} e^{-\lambda^2} = e^{-\lambda^2 (1 - \alpha)}
\end{equation} 

In order for the steady-state distribution to be localized, we expect, for a given $ \mu > 0 $, that this quantity is below some value that is less than the wild-type fitness of $ 1 $.  Otherwise, haploid overlap will not lead to the purging of deleterious mutations from the population, and thereby counter the mutation-accumulation induced by $ \mu $.  Indeed, the larger the value of $ \mu $, the greater the rate of mutation-accumulation, and so we expect that $ e^{-\lambda^2 (1 - \alpha)} $ should consequently decrease to purge deleterious mutations sufficiently effectively.  

Thus, as $ \alpha \rightarrow 1 $, we expect $ \lambda^2 \rightarrow \infty $ in order to keep the mean fitness of the diploids produced from haploid fusion sufficiently small to counter mutation-accumulation and thereby localize the population.  However, as $ \lambda^2 \rightarrow \infty $, then the Poisson distribution approaches a Gaussian distribution with a mean of $ \lambda^2 $ and a standard deviation of $ \lambda $.  We may therefore write, in the limit as $ \alpha \rightarrow 1 $, that,
\begin{equation}
\frac{1}{l!} \lambda^{2 l} e^{-\lambda^2} \rightarrow \frac{1}{\lambda \sqrt{2 \pi}} \exp[-\frac{(l - \lambda^2)^2}{2 \lambda^2}]
\end{equation}

For the multiplicative fitness landscape where $ \alpha \rightarrow 1 $, Eq. (44) then becomes,
\begin{eqnarray}
&   &
1 = 2 \sum_{l = 0}^{\infty} \frac{\alpha^l}{\bar{\kappa} + \alpha^l} 
\frac{1}{\lambda \sqrt{2 \pi}} \exp[-\frac{(l - \lambda^2)^2}{2 \lambda^2}]
\nonumber \\
&   &
\mu = \lambda^2 (1 - 2 \sum_{l = 0}^{\infty} \frac{\alpha^{l + 1}}{\bar{\kappa} + \alpha^{l + 1}} 
\frac{1}{\lambda \sqrt{2 \pi}} \exp[-\frac{(l - \lambda^2)^2}{2 \lambda^2}])
\nonumber \\
\end{eqnarray}

Now, let us define a continuous variable $ x $ via $ x = l/\lambda^2 $.  Then we have,
\begin{eqnarray}
&   &
1 = 2 \sum_{l = 0}^{\infty} \frac{1}{\lambda^2} \frac{(\alpha^{\lambda^2})^x}{\bar{\kappa} + (\alpha^{\lambda^2})^x}
\frac{\lambda}{\sqrt{2 \pi}} \exp[-\frac{\lambda^2 (x - 1)^2}{2}]
\nonumber \\
&   &
\mu = \lambda^2 (1 - 2 \sum_{l = 0}^{\infty} \frac{1}{\lambda^2} \frac{\alpha (\alpha^{\lambda^2})^x}
{\bar{\kappa} + \alpha (\alpha^{\lambda^2})^x} 
\frac{\lambda}{\sqrt{2 \pi}} \exp[-\frac{\lambda^2 (x - 1)^2}{2}])
\nonumber \\
\end{eqnarray}

Defining $ \alpha = 1 - s $, it should be noted that,
\begin{equation}
\lim_{\alpha \rightarrow 1} e^{-\lambda^2 (1 - \alpha)} 
= \lim_{s \rightarrow 0} e^{\lambda^2 (-s)} 
= \lim_{s \rightarrow 0} e^{\lambda^2 \ln (1 - s)} 
= \lim_{s \rightarrow 0} (1 - s)^{\lambda^2} 
= \lim_{\alpha \rightarrow 1} \alpha^{\lambda^2}
\end{equation}
and so, when $ \alpha $ is close to $ 1 $, the mean fitness of the diploids produced by the haploid fusion becomes $ \alpha^{\lambda^2} $.  For a given $ \mu $, we expect this to converge to a given quantity in order to allow for the localization of the population at steady-state.

As $ \lambda \rightarrow \infty $, we have that $ (\lambda/\sqrt{2 \pi}) \exp[-\lambda^2 (x - 1)^2/2] \rightarrow \delta(x - 1) $.  So, as $ \alpha \rightarrow 1 $, the above pair of equations may be written as,
\begin{eqnarray}
&   &
1 = 2 \int_{0}^{\infty} \frac{(\alpha^{\lambda^2})^x}{\bar{\kappa} + (\alpha^{\lambda^2})^x} \delta(x - 1) d x =
\frac{2 \alpha^{\lambda^2}}{\bar{\kappa} + \alpha^{\lambda^2}}
\nonumber \\
&   &
\mu = \lambda^2 (1 - 2 \int_{0}^{\infty} \frac{\alpha (\alpha^{\lambda^2})^x} {\bar{\kappa} + \alpha (\alpha^{\lambda^2})^x}
\delta(x - 1) d x) 
\nonumber \\
&   &
= \lambda^2 (1 - 2 \frac{\alpha \alpha^{\lambda^2}}{\bar{\kappa} + \alpha (\alpha^{\lambda^2})})
\end{eqnarray}

The first equation gives us that $ \bar{\kappa} = \alpha^{\lambda^2} $.  Substituting into the second equation, we obtain,
\begin{equation}
\lambda^2 = \mu \frac{1 + \alpha}{1 - \alpha}
\end{equation}
and so,
\begin{equation}
\bar{\kappa} = \alpha^{\mu \frac{1 + \alpha}{1 - \alpha}} = e^{\mu (1 + \alpha) \frac{\ln \alpha}{1 - \alpha}}
\end{equation}

This expression is only valid for $ \alpha $ close $ 1 $.  Again defining $ \alpha = 1 - s $, we then obtain,
\begin{equation}
\lim_{\alpha \rightarrow 1} \bar{\kappa} = 
e^{2 \mu \lim_{s \rightarrow 0} \frac{\ln (1 - s)}{s}} =
e^{-2 \mu}
\end{equation}

\section{Discussion}

\subsection{The basic mechanism for the selective advantage of sexual reproduction}

The basic mechanism explaining the selective advantage of sexual reproduction over asexual reproduction and self-fertilization is as follows:  If a diploid cell has a homologous pair where both genes are non-functional, then, if this cell reproduces either asexually or via the self-fertilization pathway, the daughter cells will also have two non-functional genes in this homologous pair.  The reason for this is that a homologous pair with two non-functional genes will produce four non-functional daughter genes.  If these four genes are the only genes that can produce the corresponding homologous pairs in the daughter cells, as is the case with asexual reproduction and self-fertilization, then the corresponding homologous pairs in the daughter cells will have two non-functional genes.

For sexual reproduction, this is not necessarily the case, since the haploids produced by a diploid cell with two non-functional genes in a given homologous pair may fuse with haploids produced by a diploid cell containing functional copies of the gene in the same homologous pair.  This means that the resulting daughter diploid can have a corresponding homologous pair with one functional and one non-functional copy of the gene (see Figure 8).  This breaks up the association between two defective genes in a given homologous pair, preventing the steady accumulation of non-functional homologous pairs that can occur with the non-sexual pathways.  

The explanation is a bit more involved than the basic mechanism given above, however, since sexual reproduction with the two-chromosomed genome (i.e. no recombination) has a large $ N $ mean fitness that approaches the mean fitness of the non-sexual reproduction strategies.  

In the absence of recombination, a given chromosome cannot reduce the number of defective genes.  Once $ \mu > \ln 2 $, then $ e^{-\mu} < 1/2 $, which means that when a given chromosome replicates and produces two daughter chromosomes, on average less than one of those daughters will be identical to the parent.  Since semiconservative replication effectively destroys the original parent DNA molecule, the result is a steady accumulation of mutations that leads to loss of viability due to genetic drift.  The ability for sexual reproduction to break up associations between defective genes in a homologous pair may lead to a mean fitness that is larger than the mean fitness of the non-sexual strategies for finite $ N $.  However, as $ N $ becomes large this effect steadily disappears, and the result is that sexual reproduction in the absence of recombination has no selective advantage over non-sexual reproduction strategies.

In the case of sexual reproduction with the multi-chromosomed genome, recombination allows for the production of daughter cells with fewer defective genes than were present in the parent.  In the limit of large $ N $, this effect washes out any mutation accumulation effect for any value of $ \mu $.  To understand this, we first note that, in the limit of large $ N $, a given genome will have a number of defective genes that scales as $ \sqrt{N} $.  

To see this, we note that the probability that two haploids, each having $ n $ functional genes, share at least one position where both genes are non-functional, is given by $ 1 - {N - n \choose n}/{N \choose n} $.  Using Stirling's Formula, it may be shown that, in the limit of large $ N $, this probability is $ 1/2 $ when $ n $ is on the order of $ \sqrt{N} $.  As a result, haploid fusion will lead to a loss of fitness, and therefore the purging of deleterious mutations, when the number of non-functional chromosomes in a genome is on the order of $ \sqrt{N} $. 

The defective genes, along with all of the non-defective genes in the genome, segregate themselves among four haploid cells, so that each haploid on average has half the number of defective genes in the original parent (which is a number that still scales as $ \sqrt{N} $).  By the nature of the binomial distribution, the standard deviation for the number of defective genes in a given haploid scales as $ N^{1/4} $.  Therefore, out of the four haploids produced, it may be shown that two will have on the order of $ N^{1/4} $ fewer defective genes than would be expected from a purely symmetric re-distribution of genes, and two will have on the order of $ N^{1/4} $ more defective genes than would be expected from a purely symmetric re-distribution of genes.  Thus, on average, there is no net accumulation of mutations in the population.  Although each replication cycle introduces an average of $ \mu $ mutations per $ N $ template strands from each gene, this effect is washed out by the $ N^{1/4} $ fluctuation in the number of defective genes in the daughter cells due to recombination.  While this effect is not strong enough to prevent a decrease in the mean fitness as $ \mu $ increases, it is strong enough to give a significant advantage to sexual reproduction over other reproduction strategies as $ \alpha \rightarrow 1 $, and to eliminate the error threshold for $ \alpha > 0 $.

This washing out effect is illustrated in Figure 9.

\subsection{Recombination and the evolutionary basis for the meiotic pathway}

An interesting feature of meiosis, the process by which a diploid cell produces four haploids, is that the first diploid division is essentially characterized by $ r_i = 1 $, using the notation of this paper.  The reason for this is that, during the first stage of meiosis, a given chromosome replicates, and the two daughter chromosomes remain paired together.  The two homologous pairs of daughters then line up with one another, during which recombination can occur, after which each pair of daughter chromosomes segregate into distinct cells.  

We offer the following simple explanation for this segregation mechanism:  If the homologous pairs of daughters line up in the first stage of meiosis, then, in the second stage, where haploid production takes place, the homologous pairs no longer need to find each other, since they are already connected.  Thus, this haploid production pathway only requires each homologous pair of chromosomes to line up with one another in the original parent diploid cell.  If the daughters of a given parent were not to co-segregate, then each homologous pair would have to find one another in each of the two daughter diploids, in order to properly form four haploid cells with the haploid complement of genes.  This second pathway requires twice the number of homologous pair alignments, which takes additional time and energy over the first pathway.

Furthermore, during meiosis, crossover between the homologous pairs occurs, leading to an exchange of genes between the homologous pairs, a process known as {\it meiotic recombination}.  Meiotic recombination essentially ensures that, although each chromosome contains numerous genes, the segregation of genes is such that the genes on a given chromosome may be derived from either of the two parent chromosomes.  The result is that meiotic recombination leads to a gene segregation pattern that most closely approximates the segregation pattern for the multi-chromosome, sexual pathway considered in this paper.

\subsection{Sexual reproduction as a stress response in {\it S. cerevisiae}}

It should be noted that the results for the sexual reproduction pathways were obtained in the limit where $ \gamma \rho \rightarrow \infty $, that is, where the time cost for sex may be assumed to be negligible.  For finite values of $ \gamma \rho $ the value of $ \bar{\kappa} $ will be reduced.  This suggests why unicellular organisms such as {\it S. cerevisiae} engage in a sexual stress response.  When conditions are such that the fitness is high, then the relative value of $ \gamma \rho $ is small, i.e., the characteristic time a haploid spends searching for a mate with which to fuse is large compared to the characteristic doubling time, and so the fitness benefit of sex does not outweigh its cost.  However, under stressful conditions, the fitness may drop to values where the characteristic haploid fusion time is small compared to the characteristic doubling time, and so the fitness benefit for sex outweighs the costs.  For more complex, slowly replicating organisms, it is possible that the cost for sex is almost always sufficiently small to keep sex as the optimal strategy.  This, however, is highly species-dependent, since many classes of organisms are able to reproduce both asexually and sexually.

\subsection{Sexual reproduction and the error catastrophe in complex, multicellular organisms}

One of the interesting results of our models is that the sexual reproduction pathway for the multi-chromosomed genome, in contrast to the other reproduction pathways considered in this paper, does not appear to exhibit any kind of error threshold where the mean fitness of the population reaches $ 0 $ at some critical mutation rate and remains there.  For unicellular organisms, such as {\it S. cerevisiae}, where $ \mu $ is on the order of $ 0.01 $, a non-sexual reproduction strategy will not lead to the loss of viability in a population, since this value of $ \mu $ is far below the critical value of $ \ln 2 \approx 0.69 $.  In this case, then, {\it S. cerevisiae} does not need to reproduce sexually in order to survive, though sexual reproduction, when it is not too costly, does provide an additional fitness boost, and so it makes sense for the organism to maintain the pathway in its genome.

However, for more complex, multicellular organisms, the value of $ \mu $ can greatly exceed $ \ln 2 $.  For humans, for example, the value of $ \mu $ per replication cycle is on the order of $ 3 $ (and it is higher if we count $ \mu $ to be the average number of  point mutations by which the gamete genomes of a human differ from the original fertilized egg from which the human was produced).  Although the sexual reproduction pathways considered in this paper were for unicellular organisms, the results in this paper nevertheless suggest that sexual reproduction is necessary in more complex organisms to prevent the steady accumulation of mutations and the loss of viability of the population.  While research explicitly modeling asexual and sexual reproduction pathways in multicellular organisms is necessary, it is nevertheless interesting to note that the production of gametes in multicellular organisms follows a similar meiotic pathway to the one that occurs in {\it S. cerevisiae}.

\subsection{Masking of deleterious genes, sexual reproduction, and diploidy}

One theory for the advantage of sexual reproduction is that it allows for the restoration of a wild-type genome by pairing a defective gene in one haploid with a functional gene in another haploid.  While this ``masking effect" has been discussed above, this paper is not the first to advance it.  This is in fact a relatively old theory to explain the selective advantage for diploidy and for sex.  However, previous mathematical analyses led to the rejection of this theory for the existence of diploidy and sex, while the analysis in this paper shows that this masking effect can indeed occur in diploid, sexually reproducing organisms.

Previous research on sex and diploidy regarded the ability to mask mutations as a function of diploidy only.  The idea was that sex obtained its selective advantage by making use of the masking ability that diploidy presumably confers, thereby providing a selective advantage to the strategy.  The idea that sex itself was not necessary for the masking effect led researchers to first study the hypothesis that diploidy provides a masking effect in asexually reproducing organisms.  However, as is seen from our earlier analysis, with standard asexual reproduction without mitotic recombination ($ r_i = 0 $), at steady-state the mutation-accumulation is such that every homologous pair has at least one non-functional copy of a given gene.  While it is true that these non-functional genes may be masked by a functional copy in the homologous pair, there is no apparent advantage over haploidy in this case.  Furthermore, the haploids produced from such diploid cells would contain a number of defective genes that is proportional to the haploid complement of $ N $ genes, so that haploid fusion would produce diploids with a number of homologous pairs lacking a functional copy of a given gene that scales with $ N $, leading to diploids of essentially $ 0 $ fitness, thereby eliminating any selective advantage for sex.

Indeed, in this paper, we have found that diploidy without sexual reproduction with recombination does not provide any advantage over haploidy, given the fitness landscapes considered in this paper (asexually reproducing haploids would also yield a mean fitness of $ \max \{2 e^{-\mu} - 1, 0\} $).  Diploidy only has an advantage over haploidy when coupled to sexual reproduction with recombination.  The reason for this is that sexual reproduction leads to a $ \sqrt{N} $ scaling in the number of defective genes in the genome, making the masking effect provided by diploidy possible, since the fraction of deleterious genes goes to zero with increasing genome size.  Combined with recombination, which washes out mutation-accumulation effects (something that is only possible if the number of deleterious mutations is much larger than the average number of mutations per replication cycle), the result is the elimination of the error catastrophe and a selective advantage for sexual reproduction over non-sexual forms of reproduction.

It should therefore be apparent that the selective advantage for sexual reproduction identified in this paper shows a very strong connection between diploidy and sexual reproduction.  Without sexual reproduction, diploidy provides no fitness benefit over haploidy with the landscapes considered in this paper.  Conversely, without diploidy, sexual reproduction only provides a selective advantage under relatively restrictive and problematic assumptions.  With diploidy, however, we have shown that sexual reproduction can provide a fitness benefit over other reproduction strategies.  

This analysis suggests that sexual reproduction and diploidy should have evolved together.  However, this seems unlikely, since if these strategies are only advantageous when present together, it appears that the chances that both would randomly evolve simultaneously is negligibly small.  However, because diploidy provides a mechanism for genetic repair via homologous recombination repair, we argue that diploidy does have an important selective advantage that is not connected to sex, at least in more slowly reproducing organisms for which repair of the genome is more important.  

One possibility is that diploidy evolved before sexual reproduction, so that sexually reproducing organisms evolved from asexually reproducing diploid organisms.  Another possibility is that a form of haploid sex evolved first, whereby two haploid organisms temporarily fused to form a diploid organism.  The purpose of this fusion was to allow for homologous recombination repair in each of the haploid genomes.  Once homologous recombination repair was complete, the diploids would divide to form four haploids.  At this point, the purpose of sex would simply be to recover the asexual mean fitness that would exist without double-stranded damage in the haploid genomes.  This hypothesis is consistent with recent experimental work on the multicellular green algae {\it Volvox carteri} (Nedelcu et al. 2004).  However, in time, the benefits of diploidy would have caused it to evolve into the dominant state of the organismal life cycle, making it possible for sexual reproduction to provide a population mean fitness that exceeds that of non-sexual reproduction reproduction strategies in both the haploid and diploid states.

\subsection{Speculations on the evolution of mitotic recombination}

An important issue connected to the evolution of sexual reproduction is the issue of mitotic recombination, since mathematical models with a different set of assumptions than the ones considered here have found that mitotic recombination can often provide an almost identical advantage to sexual reproduction (Mandegar and Otto 2007).  The apparent discrepancy is that here, we do not assume that a homologous gene pair with a single non-functional copy of a gene leads to a fitness penalty, whereas other models do make this assumption.  If the fitness landscape considered in this paper is closer to the fitness landscapes of actual genomes, then our modeling suggests that mitotic recombination is simply not worth the additional time and energy costs involved in finding the homologous pair in the cell nucleus.  

Nevertheless, mitotic recombination does occur on occasion.  The likely explanation is that, while the vast majority of genes in diploid genomes are such that only one functional copy is needed to achieve the wild-type fitness, there may be a few genes where there is a non-negligible fitness penalty for having even one non-functional copy of a gene in a homologous pair.  If this fitness penalty is small, then once again it may not be worth the time and energy to engage in mitotic recombination.  If this fitness penalty is large, then in any event genomes with a non-functional copy of the gene will be purged from the population, so that mitotic recombination may not be necessary.  However, for intermediate values of the fitness penalty, it is possible that mitotic recombination is worth the time and energy costs.

While this discussion on mitotic recombination is speculative at this stage, it should be noted that it is known that certain genes are more prone to mitotic recombination than others.  It is likely that the genes more prone to mitotic recombination are exactly those for which mitotic recombination would provide a fitness benefit.  

\section{Conclusions}

This paper analyzed the evolutionary dynamics associated with three reproduction pathways in unicellular organisms:  (1)  Asexual reproduction, including mitotic recombination.  (2)  Self-fertilization with random mating.  (3) Sexual reproduction with random mating.  In addition, we considered two different forms of genome organization, to study the effects of recombination on the mean fitness for the various reproduction pathways:  We considered a two-chromosomed genome, whereby the haploid complement of genes was all on a single chromosome, and we also considered a multi-chromosomed genome, where each gene defined a separate chromosome, so that the distinct homologous pairs could segregate independently of one another.

We assumed that the purpose of diploidy is to provide genetic redundancy, in particular by allowing for the repair of genetic damage due to various mutagens, radiation, and environmental free radicals.  It was assumed that the fitness of a wild-type organism is $ 1 $, and that the fitness is unaffected as long as the organism has at least one functional copy of every gene.  More generally, we assumed that a genome with $ l $ homologous pairs lacking a functional copy of a given gene has a fitness of  $\kappa_l $, where $ 1 = \kappa_0 > \kappa_1 > \dots > \kappa_{\infty} = 0 $.  

We found, for the asexual, self-fertilization, and sexual, two-chromosomed pathways, that the mean fitness at mutation-selection balance converges to $ \max\{2 e^{-\mu} - 1, 0\} $ as $ N \rightarrow \infty $, where $ \mu $ is the average number of mutations per haploid complement of template gene strands per replication cycle.  This result holds independently of the extent of mitotic recombination or the organization of the genome.  However, for the sexual reproduction pathway with the multi-chromosomed genome, we found, assuming a multiplicative fitness landscape defined by $ \kappa_l = \alpha^{l} $, that the mean fitness at mutation-selection balance exceeds the mean fitness of the other reproduction pathways.  This fitness increase is larger the closer $ \alpha $ is to $ 1 $, while for $ \alpha = 0 $ we do not obtain a selective advantage over the other reproduction pathways.  

It must be emphasized that the results of this paper do not make any assumption regarding population size, nor is it necessary to assume a dynamic fitness landscape (either induced environmentally or due to co-evolutionary dynamics in the case of the Red Queen Hypothesis).  Furthermore, in contrast to the Deterministic Mutation Hypothesis, we do not need to assume that $ \mu > 1 $, nor do we need to assume synergistic (negative) epistasis.  Indeed, we only explicitly considered the multiplicative fitness landscape in this paper, which does not exhibit any epistasis.  However, we conjecture that our results will hold more generally.  In any event, we believe that the multiplicative fitness landscape considered in this paper is a more ``generic" landscape that more closely approximates the fitness landscapes of actual organismal genomes.  Essentially, this landscape is obtained by averaging over the various fitness penalties associated with knocking out individual genes from the genome, and assuming a uniform fitness penalty for each knockout.

Therefore, this paper developed mathematical models that provide a selective advantage for sex under more general and far less restrictive assumptions than previous studies.  Given that the mathematical models developed here are more realistic than previous models, in that they explicitly take into consideration semiconservative replication, diploidy, and suggest an evolutionary basis for meiosis and meiotic recombination, we believe that the work described in this paper points to a much more satisfying and complete resolution of the question of the maintenance of sexual reproduction in diploid organisms, as compared with previous work.

In this vein, we should point out why we believe that the Deterministic Mutation Hypothesis and other explanations for the existence of sex require a number of seemingly overly restrictive assumptions in order to obtain a selective advantage for the sexual reproduction strategy.  The basic reason is that previous models for sexual reproduction ignored the role of diploidy.  Thus, the standard model that was used to analyze sexual reproduction is the following:  Two parent haploids produce a daughter by contributing copies of their genes.  The basic mechanism is that for each gene, the daughter receives a single copy from one of the parents, so that a given parent has a $ 50\% $ chance of contributing a given gene to the daughter (see Figure 10).  While this mechanism in principle allows for the restoration of the wild-type genome from two defective parents, in practice each parent contributes an average of half of their defective genes to the daughter, so that, on average, the daughter has as many defective genes as the parents.  Furthermore, because we are dealing with haploid genomes, once a daughter receives a defective gene, it cannot receive a functional copy of that gene from the other parent and thereby ``cover" the mutation.  Also, in diploid organisms reproducing sexually, we showed that the average number of defective genes per genome scales as $ \sqrt{N} $, which, combined with recombination, leads to fluctuations on the order of $ N^{1/4} $ that wash out any mutation-accumulation effects.  With haploid genomes, the average number of defective genes per genome is a finite number that does not scale with $ N $, and so the fluctuations do not wash out any mutation-accumulation effects.  

These various effects, put together, means that, for sexual reproduction to have a selective advantage in haploid organisms, it is necessary to introduce additional restrictive assumptions that are not necessary if diploidy is taken into account.

Thus, we have argued that in cases where predominantly haploid organisms engage in sexual reproduction (generally as part of a stress response), then this is in order to temporarily form a diploid organism for the purposes of engaging in homologous recombination repair.  Given the previous work on sexual reproduction with haploid organisms, it is likely that sex in this context increases the mean fitness to its value in the absence of double-stranded DNA damage.  As mentioned in the Discussion, the mean fitness can only be increased further with true diploidy and sexual reproduction with recombination.

We should also point out that, in stochastic simulations of the various models we have considered in this paper, we have observed the finite population, Hill-Robertson effect, leading to a reduction in the mean fitness of the population beyond what would be expected in an infinite population model.  It is also true that this effect was smallest for sexual reproduction with the multi-chromosomed genome (the only case for which we provided results from the stochastic simulations),
corroborating previous results by different authors (Keightley and Otto 2006).  However, the extent of the Hill-Robertson effect is strongly dependent on the value of $ \alpha $:  The closer $ \alpha $ is to $ 1 $, the stronger the effect.  This being said, we have found that the Hill-Robertson effect is only appreciable at larger values of $ \mu $, where the cutoff for ``large" decreases with increasing $ \alpha $.  Furthermore, by increasing the population size sufficiently, the Hill-Robertson effect can be essentially eliminated.  For $ \alpha = 0.5 $, we have found good agreement between the infinite population results and stochastic simulations for a population size of $ 20,000 $, where we considered asexual reproduction without mitotic recombination for the multi-chromosomed genome.

In any event, in previous studies using haploid models for sexual reproduction, the difference in fitness between the sexual populations and the asexual populations disappears once the population size is sufficiently large.  In this work, we find that, by considering the role of diploidy, sexual reproduction in the multi-chromosomed genome retains a selective advantage over the other reproduction pathways in the infinite population limit.  This is a significant result, for, as mentioned before, it suggests that sexual reproduction has a selective advantage under far less restrictive conditions than previous models indicate.  Consequently, this result also provides an explanation for the persistence of sexual reproduction in populations that are not sufficiently small for the Hill-Robertson effect (or other finite size effects such as Muller's Ratchet) to be relevant.  Given how small unicellular organisms are, many of which are nevertheless capable of reproducing sexually, and given that there are approximately $ 7 \times 10^9 $ humans on the planet, such populations may in fact be fairly common. 

The results of this paper do not explain why a large variety of sexual and mixed asexual-sexual strategies are observed (e.g. male-female body size, the sex ratio, male parental care versus lack thereof, sperm storage, etc.).  While these complex issues are left for future work, the models presented in this paper nevertheless suggest a basic advantage for sexual reproduction that is at work in slowly reproducing, complex organisms.  The specific form that the sexual strategies take may then depend on other parameters that are connected to the specific environmental niche that the given species inhabit, and the particular survival strategy that is employed.

As a final note, we are aware that many plant species are not diploid, but contain additional copies of their genes (e.g. tetraploid).  Future research on the evolution and maintenance of sex will need to model these organisms, though we suspect that the basic mechanism for the advantage of sex obtained by considering diploidy will persist when considering these more complex genomes as well.

\begin{acknowledgments}

This research was supported by a Start-Up Grant from the United States -- Israel Binational Science Foundation, and by an Alon Fellowship from the Israel Science Foundation.  

\end{acknowledgments}

\begin{appendix}

\section{Derivation of the Evolutionary Dynamics Equations for Asexual Reproduction}

\subsection{Two-chromosomed genome}

The dynamical equations governing the evolution of the asexually replicating, two-chromosomed unicellular population, are given by,
\begin{eqnarray}
&  &
\frac{d n_{\{\sigma_1, \sigma_2\}}}{dt} = 
-\kappa_{\{\sigma_1, \sigma_2\}} n_{\{\sigma_1, \sigma_2\}}
+ \sum_{\{\sigma_1', \sigma_2'\}} \kappa_{\{\sigma_1', \sigma_2'\}} n_{\{\sigma_1', \sigma_2'\}}
\times \nonumber \\
&   &
\sum_{\sigma_{11}'} \sum_{\sigma_{12}'} \sum_{\sigma_{21}'} \sum_{\sigma_{22}'}
p(\sigma_1', \sigma_{11}') p(\sigma_1', \sigma_{12}') 
p(\sigma_2', \sigma_{21}') p(\sigma_2', \sigma_{22}')
\times \nonumber \\
&   &
[r_i (\delta_{\{\sigma_{11}', \sigma_{12}'\}, \{\sigma_1, \sigma_2\}} + 
        \delta_{\{\sigma_{21}', \sigma_{22}'\}, \{\sigma_1, \sigma_2\}}) 
\nonumber \\
&   &
+ \frac{1}{2} (1 - r_i)
       (\delta_{\{\sigma_{11}', \sigma_{21'}\}, \{\sigma_1, \sigma_2\}} +
        \delta_{\{\sigma_{12}', \sigma_{22}'\}, \{\sigma_1, \sigma_2\}}) 
\nonumber \\
&   &
+ \frac{1}{2} (1 - r_i)
       (\delta_{\{\sigma_{11}', \sigma_{22}'\}, \{\sigma_1, \sigma_2\}} +
        \delta_{\{\sigma_{12}', \sigma_{21}'\}, \{\sigma_1, \sigma_2\}})]
\end{eqnarray}
where $ \delta_{\{\sigma_1, \sigma_2\}, \{\sigma_3, \sigma_4\}} = 1 $ if $ \{\sigma_1, \sigma_2\} = \{\sigma_3, \sigma_4\} $, and $ 0 $ otherwise.

The above equation may be expanded into separate terms, which may then be collected and simplified to give,
\begin{eqnarray}
&  &
\frac{d n_{\{\sigma_1, \sigma_2\}}}{dt} =
-\kappa_{\{\sigma_1, \sigma_2\}} n_{\{\sigma_1, \sigma_2\}}
+ r_i \sum_{\{\sigma_1', \sigma_2'\}} \kappa_{\{\sigma_1', \sigma_2'\}} n_{\{\sigma_1', \sigma_2'\}}
\times \nonumber \\
&  &
\sum_{(\sigma_1'', \sigma_2''), \{\sigma_1'', \sigma_2''\} = \{\sigma_1, \sigma_2\}}
[p(\sigma_1', \sigma_1'') p(\sigma_1', \sigma_2'') + p(\sigma_2', \sigma_1'') p(\sigma_2', \sigma_2'')]
\nonumber \\
&  &
+ 2 (1 - r_i) \sum_{\{\sigma_1', \sigma_2'\}} \kappa_{\{\sigma_1', \sigma_2'\}} n_{\{\sigma_1', \sigma_2'\}}
\sum_{(\sigma_1'', \sigma_2''), \{\sigma_1'', \sigma_2''\} = \{\sigma_1, \sigma_2\}}
p(\sigma_1', \sigma_1'') p(\sigma_2', \sigma_2'')
\end{eqnarray}

Converting to the ordered strand-pair representation we have, for $ \sigma_1 \neq \sigma_2 $,
\begin{eqnarray}
&   &
\frac{d n_{(\sigma_1, \sigma_2)}}{dt} = -\kappa_{(\sigma_1, \sigma_2)} n_{(\sigma_1, \sigma_2)}
+ 2 r_i \sum_{\{\sigma_1', \sigma_2'\}, \sigma_1' \neq \sigma_2'} 
\kappa_{(\sigma_1', \sigma_2')} n_{(\sigma_1', \sigma_2')} 
\times \nonumber \\
&  &
[p(\sigma_1', \sigma_1) p(\sigma_1', \sigma_2) + p(\sigma_2', \sigma_1) p(\sigma_2', \sigma_2)] 
\nonumber \\
&  &
+ 2 r_i \sum_{\{\sigma', \sigma'\}} \kappa_{(\sigma', \sigma')} n_{(\sigma', \sigma')}
p(\sigma', \sigma_1) p(\sigma', \sigma_2)
\nonumber \\
&  &
+ 2 (1 - r_i) \sum_{\{\sigma_1', \sigma_2'\}, \sigma_1' \neq \sigma_2'} \kappa_{(\sigma_1', \sigma_2')} n_{(\sigma_1', \sigma_2')}
[p(\sigma_1', \sigma_1) p(\sigma_2', \sigma_2) + p(\sigma_2', \sigma_1) p(\sigma_1', \sigma_2)]
\nonumber \\
&   &
+ 2 (1 - r_i) \sum_{\{\sigma', \sigma'\}} \kappa_{(\sigma', \sigma')} n_{(\sigma', \sigma')} p(\sigma', \sigma_1) p(\sigma', \sigma_2)
\nonumber \\
&   &
= -\kappa_{(\sigma_1, \sigma_2)} n_{(\sigma_1, \sigma_2)} 
+ 2 r_i \sum_{(\sigma_1', \sigma_2')} \kappa_{(\sigma_1', \sigma_2')} n_{(\sigma_1', \sigma_2')} p(\sigma_1', \sigma_1) p(\sigma_1', \sigma_2)
\nonumber \\
&  &
+ 2 (1 - r_i) \sum_{(\sigma_1', \sigma_2')} \kappa_{(\sigma_1', \sigma_2')} n_{(\sigma_1', \sigma_2')}
p(\sigma_1', \sigma_1) p(\sigma_2', \sigma_2)
\nonumber \\
\end{eqnarray}

We also have,
\begin{eqnarray}
&   &
\frac{d n_{(\sigma, \sigma)}}{dt} = 
-\kappa_{(\sigma, \sigma)} n_{(\sigma, \sigma)}
+ 2 r_i \sum_{\{\sigma_1', \sigma_2'\}, \sigma_1' \neq \sigma_2'} \kappa_{(\sigma_1', \sigma_2')} n_{(\sigma_1', \sigma_2')}
\nonumber \\
&   &
[p(\sigma_1', \sigma) p(\sigma_1', \sigma) + p(\sigma_2', \sigma) p(\sigma_2', \sigma)]
\nonumber \\
&   &
+ 2 r_i \sum_{\{\sigma', \sigma'\}} \kappa_{(\sigma', \sigma')} n_{(\sigma', \sigma')} p(\sigma', \sigma) p(\sigma', \sigma)
\nonumber \\
&   &
+ 2 (1 - r_i) \sum_{\{\sigma_1', \sigma_2'\}, \sigma_1' \neq \sigma_2'} \kappa_{(\sigma_1', \sigma_2')} n_{(\sigma_1', \sigma_2')}
[p(\sigma_1', \sigma) p(\sigma_2', \sigma) + p(\sigma_2', \sigma) p(\sigma_1', \sigma)]
\nonumber \\
&   &
+ 2 (1 - r_i) \sum_{\{\sigma', \sigma'\}} \kappa_{(\sigma', \sigma')} n_{(\sigma', \sigma')} p(\sigma', \sigma) p(\sigma', \sigma)
\nonumber \\
&  &
=  -\kappa_{(\sigma, \sigma)} n_{(\sigma, \sigma)}
+ 2 r_i \sum_{(\sigma_1', \sigma_2')} \kappa_{(\sigma_1', \sigma_2')} n_{(\sigma_1', \sigma_2')} p(\sigma_1', \sigma) p(\sigma_2', \sigma)
\nonumber \\
&   &
+ 2 (1 - r_i) \sum_{(\sigma_1', \sigma_2')} \kappa_{(\sigma_1', \sigma_2')} n_{(\sigma_1', \sigma_2')} p(\sigma_1', \sigma) p(\sigma_2', \sigma)
\nonumber \\
\end{eqnarray}
and so, converting from population numbers to population fractions, we obtain,
\begin{eqnarray}
&   &
\frac{d x_{(\sigma_1, \sigma_2)}}{dt} = 
-(\kappa_{(\sigma_1, \sigma_2)} + \bar{\kappa}(t)) x_{(\sigma_1, \sigma_2)} 
\nonumber \\
&  &
+ 2 r_i \sum_{(\sigma_1', \sigma_2')} \kappa_{(\sigma_1', \sigma_2')} x_{(\sigma_1', \sigma_2')} 
p(\sigma_1', \sigma_1) p(\sigma_1', \sigma_2)
\nonumber \\
&  &
+ 2 (1 - r_i) \sum_{(\sigma_1', \sigma_2')} \kappa_{(\sigma_1', \sigma_2')} x_{(\sigma_1', \sigma_2')}
p(\sigma_1', \sigma_1) p(\sigma_2', \sigma_2)
\nonumber \\
\end{eqnarray}
where $ x_{(\sigma_1, \sigma_2)} \equiv n_{(\sigma_1, \sigma_2)}/(n = \sum_{(\sigma_1', \sigma_2')} n_{(\sigma_1', \sigma_2')}) $, and $ \bar{\kappa}(t) = (1/n) (dn/dt) = \sum_{(\sigma_1, \sigma_2)} \kappa_{(\sigma_1, \sigma_2)} x_{(\sigma_1, \sigma_2)} $.

To convert this to a set of equations in terms of the $ z_{l_{10}, l_{01}, l_{00}} $ population fractions, we proceed as follows:  Given a daughter ordered strand-pair $ (\sigma_1, \sigma_2) $ characterized by the parameters $ l_{10}, l_{01}, l_{00} $, and given a parent ordered strand-pair $ (\sigma_1', \sigma_2') $, we let $ l_{i_1 i_2 j_1 j_2} $ denote the number of positions where $ \sigma_1 $ is $ i_1 $, $ \sigma_2 $ is $ i_2 $, $ \sigma_1' $ is $ j_1 $, and $ \sigma_2' $ is $ j_2 $.  We then have,
\begin{eqnarray}
&   &
p(\sigma_1', \sigma_1) = p^{l_{1111} + l_{1110} + l_{1011} + l_{1010}} 
(1 - p)^{l_{0111} + l_{0110} + l_{0011} + l_{0010}}
\delta_{l_{1101} + l_{1100} + l_{1001} + l_{1000}, 0} 
\nonumber \\
&   &
p(\sigma_1', \sigma_2) = p^{l_{1111} + l_{1110} + l_{0111} + l_{0110}} 
(1 - p)^{l_{1011} + l_{1010} + l_{0011} + l_{0010}}
\delta_{l_{1101} + l_{1100} + l_{0101} + l_{0100}, 0}
\nonumber \\
&   &
p(\sigma_2', \sigma_2) = p^{l_{1111} + l_{1101} + l_{0111} + l_{0101}} 
(1 - p)^{l_{1011} + l_{1001} + l_{0011} + l_{0001}}
\delta_{l_{1110} + l_{1100} + l_{0110} + l_{0100}, 0}
\nonumber \\
\end{eqnarray}

Taking into account degeneracies, we then have,
\begin{widetext}
\begin{eqnarray}
&   &
\frac{d z_{l_{10}, l_{01}, l_{00}}}{dt} = -(\kappa_{l_{00}} + \bar{\kappa}(t)) z_{l_{10}, l_{01}, l_{00}}
+ 2 r_i \frac{N!}{l_{10}! l_{01}! l_{00}! (N - l_{10} - l_{01} - l_{00})!}
\times \nonumber \\
&   &
\sum_{l_{1110} = 0}^{N - l_{10} - l_{01} - l_{00}} \sum_{l_{1101} = 0}^{N - l_{10} - l_{01} - l_{00} - l_{1110}}
\sum_{l_{1100} = 0}^{N - l_{10} - l_{01} - l_{00} - l_{1110} - l_{1101}}
\sum_{l_{1010} = 0}^{l_{10}} \sum_{l_{1001} = 0}^{l_{10} - l_{1010}} \sum_{l_{1000} = 0}^{l_{10} - l_{1010} - l_{1001}}
\nonumber \\
&   &
\sum_{l_{0110} = 0}^{l_{01}} \sum_{l_{0101} = 0}^{l_{01} - l_{0110}} \sum_{l_{0100} = 0}^{l_{01} - l_{0110} - l_{0101}}
\sum_{l_{0010} = 0}^{l_{00}} \sum_{l_{0001} = 0}^{l_{00} - l_{0010}} \sum_{l_{0000} = 0}^{l_{00} - l_{0010} - l_{0001}}
\kappa_{l_{1100} + l_{1000} + l_{0100} + l_{0000}} 
\times \nonumber \\
&   &
\frac{z_{l_{1110} + l_{1010} + l_{0110} + l_{0010}, l_{1101} + l_{1001} + l_{0101} + l_{0001}, l_{1100} + l_{1000} + l_{0100} + l_{0000}}}{{N \choose l_{1110} + l_{1010} + l_{0110} + l_{0010}} {N - l_{1110} - l_{1010} - l_{0110} - l_{0010} \choose l_{1101} + l_{1001} + l_{0101} + l_{0001}} {N - l_{1110} - l_{1010} - l_{0110} - l_{0010} - l_{1101} - l_{1001} - l_{0101} - l_{0001} \choose l_{1100} + l_{1000} + l_{0100} + l_{0000}}}
\times \nonumber \\
&   &
{N - l_{10} - l_{01} - l_{00} \choose l_{1110}} {N - l_{10} - l_{01} - l_{00} - l_{1110} \choose l_{1101}}
{N - l_{10} - l_{01} - l_{00} - l_{1110} - l_{1101} \choose l_{1100}}
\times \nonumber \\
&   &
{l_{10} \choose l_{1010}} {l_{10} - l_{1010} \choose l_{1001}} {l_{10} - l_{1010} - l_{1001} \choose l_{1000}}
{l_{01} \choose l_{0110}} {l_{01} - l_{0110} \choose l_{0101}} {l_{01} - l_{0110} - l_{0101} \choose l_{0100}}
\times \nonumber \\
&   &
{l_{00} \choose l_{0010}} {l_{00} - l_{0010} \choose l_{0001}} {l_{00} - l_{0010} - l_{0001} \choose l_{0000}}
\times \nonumber \\
&   &
p^{l_{1111} + l_{1110} + l_{1011} + l_{1010}} (1 - p)^{l_{0111} + l_{0110} + l_{0011} + l_{0010}}
\delta_{l_{1101} + l_{1100} + l_{1001} + l_{1000}, 0} 
\times \nonumber \\
&  &
p^{l_{1111} + l_{1110} + l_{0111} + l_{0110}} (1 - p)^{l_{1011} + l_{1010} + l_{0011} + l_{0010}}
\delta_{l_{1101} + l_{1100} + l_{0101} + l_{0100}, 0}
\nonumber \\
&   &
+ 2 (1 - r_i) \frac{N!}{l_{10}! l_{01}! l_{00}! (N - l_{10} - l_{01} - l_{00})!}
\times \nonumber \\
&   &
\sum_{l_{1110} = 0}^{N - l_{10} - l_{01} - l_{00}} \sum_{l_{1101} = 0}^{N - l_{10} - l_{01} - l_{00} - l_{1110}}
\sum_{l_{1100} = 0}^{N - l_{10} - l_{01} - l_{00} - l_{1110} - l_{1101}}
\sum_{l_{1010} = 0}^{l_{10}} \sum_{l_{1001} = 0}^{l_{10} - l_{1010}} \sum_{l_{1000} = 0}^{l_{10} - l_{1010} - l_{1001}}
\nonumber \\
&   &
\sum_{l_{0110} = 0}^{l_{01}} \sum_{l_{0101} = 0}^{l_{01} - l_{0110}} \sum_{l_{0100} = 0}^{l_{01} - l_{0110} - l_{0101}}
\sum_{l_{0010} = 0}^{l_{00}} \sum_{l_{0001} = 0}^{l_{00} - l_{0010}} \sum_{l_{0000} = 0}^{l_{00} - l_{0010} - l_{0001}}
\kappa_{l_{1100} + l_{1000} + l_{0100} + l_{0000}} 
\times \nonumber \\
&   &
\frac{z_{l_{1110} + l_{1010} + l_{0110} + l_{0010}, l_{1101} + l_{1001} + l_{0101} + l_{0001}, l_{1100} + l_{1000} + l_{0100} + l_{0000}}}{{N \choose l_{1110} + l_{1010} + l_{0110} + l_{0010}} {N - l_{1110} - l_{1010} - l_{0110} - l_{0010} \choose l_{1101} + l_{1001} + l_{0101} + l_{0001}} {N - l_{1110} - l_{1010} - l_{0110} - l_{0010} - l_{1101} - l_{1001} - l_{0101} - l_{0001} \choose l_{1100} + l_{1000} + l_{0100} + l_{0000}}}
\times \nonumber \\
&   &
{N - l_{10} - l_{01} - l_{00} \choose l_{1110}} {N - l_{10} - l_{01} - l_{00} - l_{1110} \choose l_{1101}}
{N - l_{10} - l_{01} - l_{00} - l_{1110} - l_{1101} \choose l_{1100}}
\times \nonumber \\
&   &
{l_{10} \choose l_{1010}} {l_{10} - l_{1010} \choose l_{1001}} {l_{10} - l_{1010} - l_{1001} \choose l_{1000}}
{l_{01} \choose l_{0110}} {l_{01} - l_{0110} \choose l_{0101}} {l_{01} - l_{0110} - l_{0101} \choose l_{0100}}
\times \nonumber \\
&   &
{l_{00} \choose l_{0010}} {l_{00} - l_{0010} \choose l_{0001}} {l_{00} - l_{0010} - l_{0001} \choose l_{0000}}
\times \nonumber \\
&   &
p^{l_{1111} + l_{1110} + l_{1011} + l_{1010}} (1 - p)^{l_{0111} + l_{0110} + l_{0011} + l_{0010}}
\delta_{l_{1101} + l_{1100} + l_{1001} + l_{1000}, 0} 
\times \nonumber \\
&   &
p^{l_{1111} + l_{1101} + l_{0111} + l_{0101}} (1 - p)^{l_{1011} + l_{1001} + l_{0011} + l_{0001}}
\delta_{l_{1110} + l_{1100} + l_{0110} + l_{0100}, 0}
\nonumber
\end{eqnarray}
After some manipulations, we obtain that,
\begin{eqnarray}
&   &
\frac{d z_{l_{10}, l_{01}, l_{00}}}{dt} = -(\kappa_{l_{00}} + \bar{\kappa}(t)) z_{l_{10}, l_{01}, l_{00}}
+ 2 r_i \sum_{l_{1110} = 0}^{N - l_{10} - l_{01} - l_{00}} \sum_{l_{1010} = 0}^{l_{10}}
\sum_{l_{0110} = 0}^{l_{01}} \sum_{l_{0010} = 0}^{l_{00}} \sum_{l_{0001} = 0}^{l_{00} - l_{0010}}
\sum_{l_{0000} = 0}^{l_{00} - l_{0010} - l_{0001}} \kappa_{l_{0000}}
\times \nonumber \\
&   &
z_{l_{1110} + l_{1010} + l_{0110} + l_{0010}, l_{0001}, l_{0000}}
\frac{(l_{1110} + l_{1010} + l_{0110} + l_{0010})!}{l_{1110}! l_{1010}! l_{0110}! l_{0010}!} 
(1 - \epsilon)^{2 l_{1110}} [\epsilon (1 - \epsilon)]^{l_{1010}} [\epsilon (1 - \epsilon)]^{l_{0110}} \epsilon^{2 l_{0010}}
\times \nonumber \\
&   &
\frac{(N - l_{1110} - l_{1010} - l_{0110} - l_{0010} - l_{0001} - l_{0000})!}{(l_{10} - l_{1010})! (l_{01} - l_{0110})! 
(l_{00} - l_{0010} - l_{0001} - l_{0000})! (N - l_{10} - l_{01} - l_{00} - l_{1110})!} 
\times \nonumber \\
&   &
[\epsilon (1 - \epsilon)]^{l_{10} - l_{1010}} 
 [\epsilon (1 - \epsilon)]^{l_{01} - l_{0110}} 
\epsilon^{2 (l_{00} - l_{0010} - l_{0001} - l_{0000})} 
(1 - \epsilon)^{2 (N - l_{10} - l_{01} - l_{00} - l_{1110})} 
\nonumber \\
&   &
+ 2 (1 - r_i) \sum_{l_{1010} = 0}^{l_{10}} \sum_{l_{0101} = 0}^{l_{01}} \sum_{l_{0010} = 0}^{l_{00}} 
\sum_{l_{0001} = 0}^{l_{00} - l_{0010}} \sum_{l_{0000} = 0}^{l_{00} - l_{0010} - l_{0001}}
\kappa_{l_{0000}} z_{l_{1010} + l_{0010}, l_{0101} + l_{0001}, l_{0000}}
\times \nonumber \\
&   &
\frac{(l_{1010} + l_{0010})!}{l_{1010}! l_{0010}!} (1 - \epsilon)^{l_{1010}} \epsilon^{l_{0010}}
\frac{(l_{0101} + l_{0001})!}{l_{0101}! l_{0001}!} (1 - \epsilon)^{l_{0101}} \epsilon^{l_{0001}}
\times \nonumber \\
&   &
\frac{(N - l_{1010} - l_{0101} - l_{0010} - l_{0001} - l_{0000})!}{(l_{10} - l_{1010})! (l_{01} - l_{0101})! (l_{00} - l_{0010} - l_{0001} - l_{0000})! 
(N - l_{10} - l_{01} - l_{00})!}
\times \nonumber \\
&   &
[\epsilon (1 - \epsilon)]^{l_{10} - l_{1010}}  [\epsilon (1 - \epsilon)]^{l_{01} - l_{0101}} 
\epsilon^{2 (l_{00} - l_{0010} - l_{0001} - l_{0000})}
(1 - \epsilon)^{2 (N - l_{10} - l_{01} - l_{00})} 
\nonumber \\
\end{eqnarray}
\end{widetext}
which is equivalent to Eq. (1).

\subsection{Multi-chromosomed genome}

To derive the evolutionary dynamics equations for the multi-chromosomed genomes reproducing asexually, we label each of the daughter cells from a given parent as a ``left" cell and a ``right" cell.  We then first wish to determine the probability that a given daughter cell, either left or right, has a particular genome.  Since the homologous pairs segregate into the daughter cells independently of one another, we may compute the probability of a given segregation pattern for each homologous pair, and then multiply the appropriate probabilities together for a given daughter genome.

For this analysis, we will consider the left daughter cells only, since the arguments are analogous for the right daughter cells.  Then, we wish to compute the probability $ p(rs \rightarrow xy) $, where $ rs, xy = 11, 10, 00 $, which is the probability that a homolgous pair where one gene is of type $ r $ and the other gene is of type $ s $ produces the homologous pair $ xy $ in the left daughter cell.  We handle each case in turn:

\medskip\noindent
\underline{$ 11 \rightarrow 11 $}:  Since each daughter chromosome is the daughter of a $ 1 $ parent, the probability that a given daughter chromosome is $ 1 $ is $ p $, so the probability that both are $ 1 $ is $ p^2 $.

\medskip\noindent
\underline{$ 11 \rightarrow 10 $}:  The probability that a given daughter chromosome is $ 1 $ is $ p $, and the probability that a daughter chromosome is $ 0 $ is $ 1 - p $.  Since it does not matter which daughter is $ 1 $ and which is $ 0 $, we obtain an overall probability of $ 2 p (1 - p) $.

\medskip\noindent
\underline{$ 11 \rightarrow 00 $}:  The probability for this pathway is $ (1 - p)^2 $.

\medskip\noindent
\underline{$ 10 \rightarrow 11 $}:  The $ 0 $ parent always forms two $ 0 $ daughters, while the $ 1 $ parent may form either a $ 11 $, $ 10 $, or a $ 00 $ daughter pair.  In order to form a $ 11 $ daughter cell, the $ 1 $ parent must produce a $ 11 $ daughter pair, which occurs with probability $ p^2 $.  Furthermore, the two $ 1 $ daughters must co-segregate.  Since they are derived from the same parent, this occurs with probability $ r_i $.  Finally, the two co-segregating $ 1 $ daughters must co-segregate into the left cell, which occurs with probability of $ 1/2 $.  The overall probability is then $ r_i p^2/2 $.

\medskip\noindent
\underline{$ 10 \rightarrow 10 $}:  If the $ 1 $ parent forms two $ 1 $ daughters, then the two $ 1 $ daughters cannot co-segregate, for otherwise this would produce a $ 11 $ pair in one cell and a $ 00 $ pair in the other cell.  So, we want each $ 1 $ to co-segregate with a $ 0 $ derived from the other parent gene, which occurs with probability $ 1 - r_i $.  The probability of this particular segregation pattern is $ (1 - r_i) p^2 $.

The $ 1 $ parent forms one $ 1 $ and one $ 0 $ daughter with probability $ 2 p (1 - p) $.  This produces a $ 10 $ pair in one cell, and a $ 00 $ pair in the other cell, so the probability that the left cell receives the $ 10 $ pair is $ 1/2 $, giving an overall probability of $ p (1 - p) $.

Adding the probabilities together, we obtain an overall probability of $ p (1 - r_i p) $.\\

\medskip\noindent
\underline{$ 10 \rightarrow 00 $}:  The probability for this pathway is $ 1 - p (1 - r_i p) - r_i p^2/2 = 1 - p (1 - r_i p + r_i p/2) = 1 - p (1 - r_i p/2) $.

\medskip\noindent
\underline{$ 00 \rightarrow 00 $}:  The probability for this pathway is $ 1 $.

\bigskip
Given a daughter diploid characterized by the parameters $ l_{10}, l_{00} $, and given a parent diploid, let $ l_{i_1 i_2 j_1 j_2} $ denote the number of homologous gene pairs where the daughter is $ i_1, i_2 $ and the parent is $ j_1, j_2 $.  The probability that the parent diploid produces the daughter diploid as the left daughter is,
\begin{eqnarray}
&   &
p^{2 l_{1111}} [2 p (1 - p)]^{l_{1011}} (1 - p)^{2 l_{0011}}
(\frac{r_i}{2} p^2)^{l_{1110}} [p (1 - r_i p)]^{l_{1010}} [1 - p (1 - \frac{r_i}{2} p)]^{l_{0010}}
\delta_{l_{1100} + l_{1000}, 0}
\nonumber \\
\end{eqnarray}

Taking into account degeneracies, we obtain that the evolutionary dynamics equations are then,
\begin{widetext}
\begin{eqnarray}
&   &
\frac{d z_{l_{10}, l_{00}}}{dt} = -(\kappa_{l_{00}} + \bar{\kappa}(t)) z_{l_{10}, l_{00}}
+ 2 \frac{N!}{l_{10}! l_{00}! (N - l_{10} - l_{00})!}
\times \nonumber \\
&   &
\sum_{l_{1110} = 0}^{N - l_{10} - l_{00}} \sum_{l_{1100} = 0}^{N - l_{10} - l_{00} - l_{1110}}
\sum_{l_{1010} = 0}^{l_{10}} \sum_{l_{1000} = 0}^{l_{10} - l_{1010}}
\sum_{l_{0010} = 0}^{l_{00}} \sum_{l_{0000} = 0}^{l_{00} - l_{0010}}
\kappa_{l_{1100} + l_{1000} + l_{0000}} 
\times \nonumber \\
&   &
\frac{z_{l_{1110} + l_{1010} + l_{0010}, l_{1100} + l_{1000} + l_{0000}}}
{{N \choose l_{1110} + l_{1010} + l_{0010}} {N - l_{1110} - l_{1010} - l_{0010} \choose l_{1100} + l_{1000} + l_{0000}}}
\times \nonumber \\
&   &
{N - l_{10} - l_{00} \choose l_{1110}} {N - l_{10} - l_{00} - l_{1110} \choose l_{1100}}
{l_{10} \choose l_{1010}} {l_{10} - l_{1010} \choose l_{1000}}
{l_{00} \choose l_{0010}} {l_{00} - l_{0010} \choose l_{0000}}
\times \nonumber \\
&   &
p^{2 l_{1111}} [2 p (1 - p)]^{l_{1011}} (1 - p)^{2 l_{0011}}
(\frac{r_i}{2} p^2)^{l_{1110}} [p (1 - r_i p)]^{l_{1010}} [1 - p (1 - \frac{r_i}{2} p)]^{l_{0010}}
\delta_{l_{1100} + l_{1000}, 0}
\nonumber 
\end{eqnarray}
\begin{eqnarray}
&   &
= -(\kappa_{l_{00}} + \bar{\kappa}(t)) z_{l_{10}, l_{00}} 
+ 2 \sum_{l_{1110} = 0}^{N - l_{10} - l_{00}} \sum_{l_{1010} = 0}^{l_{10}} \sum_{l_{0010} = 0}^{l_{00}} 
\sum_{l_{0000} = 0}^{l_{00} - l_{0010}} \kappa_{l_{0000}} z_{l_{1110} + l_{1010} + l_{0010}, l_{0000}}
\times \nonumber \\
&   &
\frac{(l_{1110} + l_{1010} + l_{0010})!}{l_{1110}! l_{1010}! l_{0010}!}
[\frac{r_i}{2} (1 - \epsilon)^2]^{l_{1110}} [1 - \epsilon - r_i (1 - \epsilon)^2]^{l_{1010}} [\epsilon + \frac{r_i}{2} (1 - \epsilon)^2]^{l_{0010}}
\times \nonumber \\
&    &
\frac{(N - l_{1110} - l_{1010} - l_{0010} - l_{0000})!}{(l_{10} - l_{1010})! (l_{00} - l_{0010} - l_{0000})! (N - l_{10} - l_{00} - l_{1110})!}
\times \nonumber \\
&   &
[2 \epsilon (1 - \epsilon)]^{l_{10} - l_{1010}} \epsilon^{2 (l_{00} - l_{0010} - l_{0000})} (1 - \epsilon)^{2 (N - l_{10} - l_{00} - l_{1110})} 
\nonumber \\
\end{eqnarray}
\end{widetext}
which is identical to Eq. (15).

\section{Derivation of the Evolutionary Dynamics Equations for Self-Fertilization for the Multi-Chromosomed Genome}

To develop the evolutionary dynamics equations for self-fertilization with random mating, we proceed as follows:  Given a parent diploid cell, we assume that it splits into a left diploid and a right diploid.  The left diploid then splits into two haploids, haploid $ 1 $ on the left and haploid $ 2 $ on the right, while the right diploid also splits into two haploids, haploid $ 3 $ on the left and haploid $ 4 $ on the right.

We then have the following pairings, all with equal probability because of random mating:  (1) $ 1 \leftrightarrow 2, 3 \leftrightarrow 4 $.  (2)  $ 1 \leftrightarrow 3, 2 \leftrightarrow 4 $.  (3)  $ 1 \leftrightarrow 4, 2 \leftrightarrow 3 $.  Each of the three possible pairing schemes have a probability of $ 1/3 $ of occuring.

We may consider each pairing scheme in turn.  Our goal is to determine, for a given parent diploid, what is the probability of obtaining a specific daughter diploid as the left daughter cell.

We consider the various probabilities in order.

\bigskip\noindent
\begin{center}
$ 1 \leftrightarrow 2, 3 \leftrightarrow 4 $
\end{center}

\medskip\noindent
\underline{$ 11 \rightarrow 11 $}:  If a homologous pair in the parent diploid is $ 11 $, then each daughter gene in the final left diploid is the daughter of a $ 1 $ parent.  Since the probability that a given daughter of a $ 1 $ parent is itself $ 1 $ is $ p $, the probability that both daughters are $ 1 $ is $ p^2 $.

\medskip\noindent
\underline{$ 11 \rightarrow 10 $}:  As with the previous case, the probability that a given daughter of a $ 1 $ parent is itself $ 1 $ is  $ p $, while the probability that the daughter is $ 0 $ is $ 1 - p $.  Therefore, the probability that a given daughter of a $ 1 $ parent is $ 1 $ and the other daughter of a $ 1 $ parent is $ 0 $ is $ p (1 - p) $.  Since it does not matter which daughter is $ 1 $ and which is $ 0 $, we obtain a total probability of $ 2 p (1 - p) $.

\medskip\noindent
\underline{$ 11 \rightarrow 00 $}:  The probability of this pathway is $ 1 - p^2 - 2 p(1 - p) = (1 - p)^2 $.

\medskip\noindent
\underline{$ 10 \rightarrow 11 $}:  The probability that a $ 10 $ pair produces two $ 1 $ daughters and two $ 0 $ daughters is $ p^2 $.  Since these two $ 1 $ daughters are from the same $ 1 $ parent, the probability that they co-segregate into the left diploid is $ r_i/2 $, giving a total probability of $ r_i p^2/2 $.

\medskip\noindent
\underline{$ 10 \rightarrow 10 $}:  The probability that a $ 10 $ pair produces $ 2 $ $ 1 $ daughters and $ 2 $ $ 0 $ daughters is $ p^2 $.  Since these two $ 1 $ daughters are from the same $ 1 $ parent, and since the two $ 0 $ daughters are from the same 
$ 0 $ parent, the only way to obtain a $ 10 $ left daughter cell is for the daughter chromosomes of a given parent to not co-segregate.  Since this occurs with probability $ 1 - r_i $, we obtain an overall probability of $ (1 - r_i) p^2 $.

The probability that a $ 10 $ pair produces $ 1 $ $ 1 $ daughter and $ 3 $ $ 0 $ daughters is $ 2 p (1 - p) $.  Since the probability that the $ 1 $ chromosome ends up in the left daughter cell is $ 1/2 $, we obtain an overall probability of $ p (1 - p) $.

The total probability is then $ (1 - r_i) p^2 + p (1 - p) = p (1 - r_i p) $.

\medskip\noindent
\underline{$ 10 \rightarrow 00 $}:  The probability for this pathway is $ 1 - r_i p^2/2 - p(1 - r_i p) = 1 - p (1 - r_i p/2) $.

\medskip\noindent
\underline{$ 00 \rightarrow 00 $}:  Because of the neglect of backmutations, this occurs with probability $ 1 $.

\bigskip\noindent
\begin{center}
$ 1 \leftrightarrow 3, 2 \leftrightarrow 4 $
\end{center}

\medskip\noindent
\underline{$ 11 \rightarrow 11, 10, 00 $}:  Following a similar line of reasoning to the one used above, we obtain an identical corresponding set of transition probabilities.

\medskip\noindent
\underline{$ 10 \rightarrow 11 $}:  The $ 1 $ parent must produce two $ 1 $ daughters with probability $ p^2 $.  These $ 1 $ daughters must segregate into distinct diploids, with a probability of $ 1 - r_i $.  The probability that these $ 1 $ then end up in haploids $ 1 $ and $ 3 $ respectively is $ 1/4 $, for a total probability of $ (1 - r_i) p^2/4 $.

\medskip\noindent
\underline{$ 10 \rightarrow 10 $}:  The $ 1 $ parent produces two $ 1 $ daughters with probability $ p^2 $, while the $ 0 $ parent produces two $ 0 $ daughters with probability $ 1 $.  If the $ 1 $ daughters and the $ 0 $ daughters each co-segregate, which occurs with probability $ r_i $, then the $ 1 $ haploid and the $ 3 $ haploid will together form a $ 10 $ pair.  If the daughters of each parent do not co-segregate, with probability $ 1 - r_i $, then we form two $ 10 $ diploids.  The probability that the $ 1 $ haploid has a $ 1 $ and the $ 3 $ haploid a $ 0 $ is $ 1/4 $, and the probability that the $ 1 $ haploid has a $ 0 $ and the $ 3 $ haploid a $ 1 $ is $ 1/4 $, giving an overall probability of $ p^2 (r_i + (1 - r_i)/2) = (1 + r_i) p^2/2 $.

The $ 1 $ parent produces one $ 1 $ daughter and one $ 0 $ daughter with probability $ 2 p (1 - p) $.  The probability that this 
$ 1 $ daughter ends up in either haploid $ 1 $ or $ 3 $ is $ 1/2 $, for an overall probability of $ p (1 - p) $.

The total probability is then $ p [1 - p + (1 + r_i) p/2] = p [1 - (1 - r_i) p/2] $.

\medskip\noindent
\underline{$ 10 \rightarrow 00 $}:  The probability of this pathway is $ 1 - (1 - r_i) p^2/4 - p (1 - (1 - r_i) p/2) = 1 - p [1 - (1 - r_i) p/4] $.

\medskip\noindent
\underline{$ 00 \rightarrow 00 $}:  The probability for this pathway is simply $ 1 $.

\bigskip\noindent
\begin{center}
$ 1 \leftrightarrow 4 $, $ 2 \leftrightarrow 3 $
\end{center}

\medskip\noindent
This case is symmetric to Case $ 2 $, so all of the probabilities are identical.

Given a diploid parent and a diploid daughter cell, where the daughter is characterized by $ l_{10}, l_{00} $, let $ l_{i_1 i_2 j_1 j_2} $ denote the number of positions where the daughter is $ i_1, i_2 $ and the parent is $ j_1, j_2 $.  The probability that the parent diploid produces the daughter diploid as the left daughter cell is then,
\begin{eqnarray}
&   &
p^{2 l_{1111}} [2 p (1 - p)]^{l_{1011}} (1 - p)^{2 l_{0011}} 
(\frac{r_i}{2} p^2)^{l_{1110}} [p (1 - r_i p)]^{l_{1010}} [1 - p(1 - r_i p/2)]^{l_{0010}}
\delta_{l_{1100} + l_{1000}, 0}, 
\nonumber \\
&   &
\mbox{ for the $ 1 \leftrightarrow 2, 3 \leftrightarrow 4 $ mating pattern.}
\nonumber \\
&   &
p^{2 l_{1111}} [2 p (1 - p)]^{l_{1011}} (1 - p)^{2 l_{0011}} 
(\frac{1 - r_i}{4} p^2)^{l_{1110}} [p (1 - \frac{1 - r_i}{2} p)]^{l_{1010}} [1 - p (1 - \frac{1 - r_i}{4} p)]^{l_{0010}}
\delta_{l_{1100} + l_{1000}, 0}, 
\nonumber \\
&   &
\mbox{for the $ 1 \leftrightarrow 3, 2 \leftrightarrow 4 $ and $ 1 \leftrightarrow 4, 2 \leftrightarrow 3 $ mating patterns.}
\nonumber \\
\end{eqnarray}

Taking into account degeneracies and the probabilities for the various mating patterns, we obtain,
\begin{widetext}
\begin{eqnarray}
&   &
\frac{d z_{l_{10}, l_{00}}}{dt} = -(\kappa_{l_{00}} + \bar{\kappa}(t)) z_{l_{10}, l_{00}} + 2 \frac{N!}{l_{10}! l_{00}! (N - l_{10} - l_{00})!} 
\times \nonumber \\
&   &
\sum_{l_{1110} = 0}^{N - l_{10} - l_{00}} \sum_{l_{1100} = 0}^{N - l_{10} - l_{00} - l_{1110}}
\sum_{l_{1010} = 0}^{l_{10}} \sum_{l_{1000} = 0}^{l_{10} - l_{1010}}
\sum_{l_{0010} = 0}^{l_{00}} \sum_{l_{0000} = 0}^{l_{00} - l_{0010}}
\kappa_{l_{1100} + l_{1000} + l_{0000}} 
\frac{z_{l_{1110} + l_{1010} + l_{0010}, l_{1100} + l_{1000} + l_{0000}}}
{{N \choose l_{1110} + l_{1010} + l_{0010}} {N - l_{1110} - l_{1010} - l_{0010} \choose l_{1100} + l_{1000} + l_{0000}}}
\times \nonumber \\
&   &
{N - l_{10} - l_{00} \choose l_{1110}} {N - l_{10} - l_{00} - l_{1110} \choose l_{1100}} 
{l_{10} \choose l_{1010}} {l_{10} - l_{1010} \choose l_{1000}} {l_{00} \choose l_{0010}} {l_{00} - l_{0010} \choose l_{0000}}
\times \nonumber \\
&   &
\frac{1}{3} [p^{2 l_{1111}} [2 p (1 - p)]^{l_{1011}} (1 - p)^{2 l_{0011}} 
(\frac{r_i}{2} p^2)^{l_{1110}} [p (1 - r_i p)]^{l_{1010}} [1 - p(1 - r_i p/2)]^{l_{0010}}
\delta_{l_{1100} + l_{1000}, 0}
\nonumber \\
&   &
+ 2 p^{2 l_{1111}} [2 p (1 - p)]^{l_{1011}} (1 - p)^{2 l_{0011}} 
(\frac{1 - r_i}{4} p^2)^{l_{1110}} [p (1 - \frac{1 - r_i}{2} p)]^{l_{1010}} [1 - p (1 - \frac{1 - r_i}{4} p)]^{l_{0010}} 
\times \nonumber \\
&   &
\delta_{l_{1100} + l_{1000}, 0}]
\nonumber 
\end{eqnarray}
\begin{eqnarray}
&   &
= -(\kappa_{l_{00}} + \bar{\kappa}(t)) z_{l_{10}, l_{00}} 
+ \frac{2}{3} \sum_{l_{1110} = 0}^{N - l_{10} - l_{00}} \sum_{l_{1010} = 0}^{l_{10}} 
\sum_{l_{0010} = 0}^{l_{00}} \sum_{l_{0000} = 0}^{l_{00} - l_{0010}} 
\kappa_{l_{0000}} z_{l_{1110} + l_{1010} + l_{0010}, l_{0000}}
\times \nonumber \\
&   &
\frac{(l_{1110} + l_{1010} + l_{0010})!}{l_{1110}! l_{1010}! l_{0010}!} 
\times \nonumber \\
&   &
[[\frac{r_i}{2} (1 - \epsilon)^2]^{l_{1110}} [1 - \epsilon - r_i (1 - \epsilon)^2]^{l_{1010}} [\epsilon + \frac{r_i}{2} (1 - \epsilon)^2]^{l_{0010}}
\nonumber \\
&   &
+ 2 [\frac{1 - r_i}{4} (1 - \epsilon)^2]^{l_{1110}} [1 - \epsilon - \frac{1 - r_i}{2} (1 - \epsilon)^2]^{l_{1010}} [\epsilon + \frac{1 - r_i}{4} (1 - \epsilon)^2]^{l_{0010}}]
\times \nonumber \\
&   &
\frac{(N - l_{1110} - l_{1010} - l_{0010} - l_{0000})!}{(l_{10} - l_{1010})! (l_{00} - l_{0010} - l_{0000})! (N - l_{10} - l_{00} - l_{1110})!}
\times \nonumber \\
&   &
[2 \epsilon (1 - \epsilon)]^{l_{10} - l_{1010}}  \epsilon^{2 (l_{00} - l_{0010} - l_{0000})} 
(1 - \epsilon)^{2 (N - l_{10} - l_{00} - l_{1110})}
\nonumber \\
\end{eqnarray}
\end{widetext}
which is identical to Eq. (23).

\section{Derivation of the Evolutionary Dynamics Equations for Sexual Reproduction}

\subsection{Two-chromosomed genome}

For sexual reproduction with random mating, the dynamical equations are,
\begin{eqnarray}
&   &
\frac{d n_{\{\sigma_1, \sigma_2\}}}{dt} =
-\kappa_{\{\sigma_1, \sigma_2\}} n_{\{\sigma_1, \sigma_2\}} +
(\frac{\gamma}{V}) n_{\sigma_1} n_{\sigma_2}, \mbox{ for $ \sigma_1 \neq \sigma_2 $}
\nonumber \\
&   &
\frac{d n_{\{\sigma, \sigma\}}}{dt} =
-\kappa_{\{\sigma, \sigma\}} n_{\{\sigma, \sigma\}} +
\frac{1}{2} (\frac{\gamma}{V}) n_{\sigma}^2
\end{eqnarray}
\begin{eqnarray}
&   &
\frac{d n_{\sigma}}{dt} = -(\frac{\gamma}{V}) n_{\sigma} n_{H} +
\sum_{\{\sigma_1, \sigma_2\}} \kappa_{\{\sigma_1, \sigma_2\}} n_{\{\sigma_1, \sigma_2\}}
\times \nonumber \\
&   &
\sum_{\sigma_{11}} \sum_{\sigma_{12}} \sum_{\sigma_{21}} \sum_{\sigma_{22}}
p(\sigma_1, \sigma_{11}) p(\sigma_1, \sigma_{12}) p(\sigma_2, \sigma_{21}) p(\sigma_2, \sigma_{22})
\times \nonumber \\
&   &
[\delta_{\sigma_{11}, \sigma} + \delta_{\sigma_{12}, \sigma} + \delta_{\sigma_{21}, \sigma} +
\delta_{\sigma_{22}, \sigma}]
\nonumber \\
&   &
= -(\frac{\gamma}{V}) n_{\sigma} n_H + 2 \sum_{\{\sigma_1, \sigma_2\}} \kappa_{\{\sigma_1, \sigma_2\}} n_{\{\sigma_1, \sigma_2\}} 
[p(\sigma_1, \sigma) + p(\sigma_2, \sigma)]
\nonumber \\
&   &
= -(\frac{\gamma}{V}) n_{\sigma} n_H + 4 \sum_{\{\sigma_1, \sigma_2\}, \sigma_1 \neq \sigma_2} \kappa_{(\sigma_1, \sigma_2)} n_{(\sigma_1, \sigma_2)} 
[p(\sigma_1, \sigma) + p(\sigma_2, \sigma)]
\nonumber \\
&  &
+ 4 \sum_{\{\sigma', \sigma'\}} \kappa_{(\sigma', \sigma')} n_{(\sigma', \sigma')} p(\sigma', \sigma)
\nonumber \\
&  &
= -(\frac{\gamma}{V}) n_{\sigma} n_H + 4 \sum_{(\sigma_1, \sigma_2)} \kappa_{(\sigma_1, \sigma_2)} n_{(\sigma_1, \sigma_2)}
p(\sigma_1, \sigma)
\end{eqnarray}

Defining the diploid ordered strand-pair population fractions via $ x_{(\sigma_1, \sigma_2)} = n_{(\sigma_1, \sigma_2)}/n $, and the haploid population fractions via $ x_{\sigma} = n_{\sigma}/(2 n) $, we obtain, after converting from population numbers to population fractions, and using the fact that $ \rho = n/V $, the dynamical equations,
\begin{eqnarray}
&   &
\frac{d x_{(\sigma_1, \sigma_2)}}{dt} = -(\kappa_{(\sigma_1, \sigma_2)} + \bar{\kappa}(t)) x_{(\sigma_1, \sigma_2)}
+ 2 \gamma \rho x_{\sigma_1} x_{\sigma_2}
\nonumber \\
&  &
\frac{d x_{\sigma}}{dt} = -\bar{\kappa}(t) x_{\sigma} - 2 \gamma \rho x_{\sigma} x_H
+ 2 \sum_{(\sigma_1, \sigma_2)} \kappa_{(\sigma_1, \sigma_2)} x_{(\sigma_1, \sigma_2)} p(\sigma_1, \sigma)
\nonumber \\
\end{eqnarray}

To develop the evolutionary dynamics equations in terms of the $ z_{l_{10}, l_{01}, l_{00}} $ and $ z_{l_0} $, we proceed as follows:  Given a haploid with genome $ \sigma $, let $ l_1 $ and $ l_0 $ denote the number of positions where $ \sigma $ is $ 1 $ and $ 0 $, respectively.  Given some $ (\sigma_1, \sigma_2) $, let $ l_{i j_1 j_2} $ denote the number of positions where $ \sigma $ is $ i $, $ \sigma_1 $ is $ j_1 $, and $ \sigma_2 $ is $ j_2 $.  We then have,
\begin{equation}
p(\sigma_1, \sigma) = p^{l_{111} + l_{110}} (1 - p)^{l_{011} + l_{010}} \delta_{l_{101} + l_{100}, 0}
\end{equation}

The evolutionary dynamics equations for the diploid population fractions $ z_{l_{10}, l_{01}, l_{00}} $ are given by,
\begin{widetext}
\begin{eqnarray}
&   &
\frac{d z_{l_{10}, l_{01}, l_{00}}}{dt} = -(\kappa_{l_{00}} + \bar{\kappa}(t)) z_{l_{10}, l_{01}, l_{00}}
+ 2 \gamma \rho \frac{N!}{l_{10}! l_{01}! l_{00}! (N - l_{10} - l_{01} - l_{00})!} \frac{z_{l_{01} + l_{00}}}{{N \choose l_{01} + l_{00}}} \frac{z_{l_{10} + l_{00}}}{{N \choose l_{10} + l_{00}}} 
\nonumber \\
&   &
= -(\kappa_{l_{00}} + \bar{\kappa}(t)) z_{l_{10}, l_{01}, l_{00}}
+ 2 \gamma \rho \frac{(l_{10} + l_{00})! (l_{01} + l_{00})!}{l_{10}! l_{01}! l_{00}!} 
\times \nonumber \\
&   &
\frac{(N - l_{01} - l_{00})!}{(N - l_{10} - l_{01} - l_{00})!} \frac{(N - l_{10} - l_{00})!}{N!} z_{l_{10} + l_{00}} z_{l_{01} + l_{00}}
\nonumber \\
&   &
= -(\kappa_{l_{00}} + \bar{\kappa}(t)) z_{l_{10}, l_{01}, l_{00}}
+ 2 \gamma \rho \frac{(l_{10} + l_{00})! (l_{01} + l_{00})!}{l_{10}! l_{01}! l_{00}!}
\times \nonumber \\
&   &
(\prod_{k = 1}^{l_{10}} \frac{N - l_{10} - l_{01} - l_{00} + k}{N - l_{10} - l_{00} + k})
(\prod_{k = 1}^{l_{00}} \frac{1}{N - l_{00} + k})
z_{l_{10} + l_{00}} z_{l_{01} + l_{00}}
\nonumber \\
\end{eqnarray}
which is identical to the first equation in Eq. (28).

Taking into account the transition probabilities and various degeneracies, then for the haploids, we have,
\begin{eqnarray}
&   &
\frac{d z_{l_0}}{dt} = -\bar{\kappa}(t) z_{l_0} - 2 \gamma \rho z_{l_0} z_H + 2 {N \choose l_0} 
\sum_{l_{110} = 0}^{N - l_0} \sum_{l_{101} = 0}^{N - l_0 - l_{110}} \sum_{l_{100} = 0}^{N - l_0 - l_{110} - l_{101}} 
\sum_{l_{010} = 0}^{l_0} \sum_{l_{001} = 0}^{l_0 - l_{010}} \sum_{l_{000} = 0}^{l_0 - l_{010} - l_{001}}
\kappa_{l_{100} + l_{000}} 
\times \nonumber \\
&   &
\frac{z_{l_{110} + l_{010}, l_{101} + l_{001}, l_{100} + l_{000}}}{{N \choose l_{110} + l_{010}}
{N - l_{110} - l_{010} \choose l_{101} + l_{001}} {N - l_{110} - l_{010} - l_{101} - l_{001} \choose l_{100} + l_{000}}}
\times \nonumber \\
&   &
{N - l_0 \choose l_{110}} {N - l_0 - l_{110} \choose l_{101}} {N - l_0 - l_{110} - l_{101} \choose l_{100}}
{l_0 \choose l_{010}} {l_0 - l_{010} \choose l_{001}} {l_0 - l_{010} - l_{001} \choose l_{000}}
\times \nonumber \\
&   &
p^{l_{111} + l_{110}} (1 - p)^{l_{011} + l_{010}} \delta_{l_{101} + l_{100}, 0}
\nonumber \\
&   &
= -\bar{\kappa}(t) z_{l_0} - 2 \gamma \rho z_H z_{l_0} 
+ 2 \sum_{l_{110} = 0}^{N - l_0} \sum_{l_{010} = 0}^{l_0} \sum_{l_{001} = 0}^{l_0 - l_{010}} \sum_{l_{000} = 0}^{l_0 - l_{010} - l_{001}}
\kappa_{l_{000}} z_{l_{110} + l_{010}, l_{001}, l_{000}} 
\times \nonumber \\
&   &
\frac{(l_{110} + l_{010})!}{l_{110}! l_{010}!}
(1 - \epsilon)^{l_{110}} \epsilon^{l_{010}}
\frac{(N - l_{110} - l_{010} - l_{001} - l_{000})!}{(l_0 - l_{010} - l_{001} - l_{000})! (N - l_0 - l_{110})!} 
\epsilon^{l_0 - l_{010} - l_{001} - l_{000}} 
(1 - \epsilon)^{N - l_0 - l_{110}}
\nonumber \\
\end{eqnarray}
\end{widetext}
which is identical to the second equation in Eq. (28).

\subsection{Multi-chromosomed genome}

To derive the quasispecies equations for sexual replication with random mating for the multi-chromosome case, we proceed as follows:  We assume that a diploid produces four haploids that may be lined up and labelled ``1", ``2", ``3", ``4".  We wish to determine what is the probability that haploid ``1" receives a certain genome from a given parent diploid.  As with the asexual case, since each of the homologous pairs segregate independently of one another, we may consider the probabilities of the various segregation patterns for a given homologous pair.  We consider each case in turn.

\medskip\noindent
\underline{$ 11 \rightarrow 1 $}:  If a given homologous pair in a parent diploid is $ 11 $, then the corresponding gene in the daughter haploid labelled ``1" is the daughter of a $ 1 $ parent, so the probability that this daughter is itself a $ 1 $ is $ p $.  Therefore, the $ 11 \rightarrow 1 $ probability is simply $ p $.

\medskip\noindent
\underline{$ 11 \rightarrow 0 $}:  Following a similar argument to the one given above, we obtain that the $ 11 \rightarrow 0 $ probability is $ 1 - p $.

\medskip\noindent
\underline{$ 10 \rightarrow 1 $}:  If a given homologous pair in a parent diploid is $ 10 $, then since a $ 0 $ parent gene produces two $ 0 $ daughters, the corresponding gene in the daughter haploid labelled ``1" can only be $ 1 $ if it is the daughter of the $ 1 $ parent.  By the symmetry of the chromosome segregation, the probability that the haploid gene is the daughter of the $ 1 $ parent is $ 1/2 $.  Since the probability that a daughter of the $ 1 $ parent is itself a $ 1 $ is $ p $, we obtain an overall probability of $ p/2 $.

\medskip\noindent
\underline{$ 10 \rightarrow 0 $}:  Since the probability of a $ 10 \rightarrow 1 $ pathway is $ p/2 $, the probability of the $ 10 \rightarrow 0 $ pathway is $ 1 - p/2 $.

\medskip\noindent
\underline{$ 00 \rightarrow 0 $}:  The probability of this pathway is $ 1 $.

\bigskip
Suppose a diploid is characterized by the parameters $ l_{10}, l_{00} $.  Suppose that two haploids, with sequences $ \sigma_1 $ and $ \sigma_2 $ fuse.  If $ \sigma_1 \neq \sigma_2 $, then the diploid production rate is given by $ (\gamma/V) n_{\sigma_1} n_{\sigma_2} $, while if $ \sigma_1 = \sigma_2 $, then the diploid production rate is given by $ (1/2) (\gamma/V) n_{\sigma_1} n_{\sigma_2} $.

If we let $ \hat{\sigma} = (\{s_{11}, s_{12}\}, \dots, \{s_{N1}, s_{N2}\}) $ denote the genome of the diploid, where $ \{s_{i1}, s_{i2}\} = \{1, 1\}, \{1, 0\}, \{0, 0\} $, and if we let $ \hat{\sigma}' $ denote the genome formed by the fusion of haploids with genomes $ \sigma_1 $ and $ \sigma_2 $, then we have, 
\begin{eqnarray}
\frac{d n_{\hat{\sigma}}}{dt} 
& = &
-\kappa_{\hat{\sigma}} n_{\hat{\sigma}} + \frac{\gamma}{V} \sum_{\{\sigma_1, \sigma_2\}, \sigma_1 \neq \sigma_2, \hat{\sigma}' = \hat{\sigma}} n_{\sigma_1} n_{\sigma_2} 
+ \frac{1}{2} \frac{\gamma}{V} \sum_{\{\sigma, \sigma\}, \hat{\sigma}' = \hat{\sigma}} n_{\sigma}^2
\nonumber \\
& = &
-\kappa_{\hat{\sigma}} n_{\hat{\sigma}} + \frac{1}{2} \frac{\gamma}{V} \sum_{(\sigma_1, \sigma_2), \hat{\sigma}' = \hat{\sigma}}
n_{\sigma_1} n_{\sigma_2}
\end{eqnarray}

Now, where $ \hat{\sigma} $ is $ \{1, 1\} $, we must have that both $ \sigma_1 $ and $ \sigma_2 $ are $ 1 $.  Where $ \hat{\sigma} $ is $ \{0, 0\} $, we must have that both $ \sigma_1 $ and $ \sigma_2 $ are $ 0 $.  Where $ \hat{\sigma} $ is $ \{1, 0\} $, we must have that $ \sigma_1 $ is $ 1 $ and $ \sigma_2 $ is $ 0 $, or $ \sigma_1 $ is $ 0 $ and $ \sigma_2 $ is $ 1 $.  Let $ l $ denote the number of spots where $ \sigma_1 $ is $ 1 $ and $ \sigma_2 $ is $ 0 $.  Since we want the fusion of $ \sigma_1 $ and $ \sigma_2 $ to produce $ \hat{\sigma} $, then the number of spots where $ \sigma_1 $ is $ 0 $ and $ \sigma_2 $ is $ 1 $ is $ l_{10} - l $.

Taking into account degeneracies, and converting from population numbers to population fractions, we then have,
\begin{widetext}
\begin{eqnarray}
&   &
\frac{d z_{l_{10}, l_{00}}}{dt} = -(\kappa_{l_{00}} + \bar{\kappa}(t)) z_{l_{10}, l_{00}} + 2 \gamma \rho
\frac{N!}{l_{10}! l_{00}! (N - l_{10} - l_{00})!} 
\sum_{l = 0}^{l_{10}} {l_{10} \choose l} \frac{z_{l_{10} - l + l_{00}}}{{N \choose l_{10} - l + l_{00}}} 
\frac{z_{l + l_{00}}}{{N \choose l + l_{00}}}
\nonumber \\
&   &
= -(\kappa_{l_{00}} + \bar{\kappa}(t)) z_{l_{10}, l_{00}} + 2 \gamma \rho
\sum_{l = 0}^{l_{10}} \frac{(l + l_{00})! (l_{10} - l + l_{00})!}{l! (l_{10} - l)! l_{00}!}
\times \nonumber \\
&   &
(\prod_{k = 1}^{l} \frac{N - l_{10} - l_{00} + k}{N - l - l_{00} + k}) (\prod_{k = 1}^{l_{00}} \frac{1}{N - l_{00} + k})
z_{l + l_{00}} z_{l_{10} - l + l_{00}}
\end{eqnarray}
\end{widetext}
which is identical to the first equation in Eq. (41).

To derive the haploid equations, suppose a haploid is characterized by the parameter $ l_0 $.  Given some parent diploid, let $ l_{i j_1 j_2} $ denote the number of positions where the haploid is $ i $ and the diploid is $ j_1, j_2 $.  We then have a total transition probability of,
\begin{equation}
p^{l_{111}} (1 - p)^{l_{011}} (\frac{p}{2})^{l_{110}} (1 - \frac{p}{2})^{l_{010}} \delta_{l_{100}, 0}
\end{equation}
and so, taking into account degeneracies, we obtain,
\begin{widetext}
\begin{eqnarray}
&   &
\frac{d z_{l_0}}{dt} = -\bar{\kappa}(t) z_{l_0} - 2 \gamma \rho z_{l_0} z_H + 2 \frac{N!}{l_0! (N - l_0)!}
\sum_{l_{110} = 0}^{N - l_0} \sum_{l_{100} = 0}^{N - l_0 - l_{110}}
\sum_{l_{010} = 0}^{l_0} \sum_{l_{000} = 0}^{l_0 - l_{010}}
\kappa_{l_{100} + l_{000}} 
\frac{z_{l_{110} + l_{010}, l_{100} + l_{000}}}{{N \choose l_{110} + l_{010}} {N - l_{110} - l_{010} \choose l_{100} + l_{000}}}
\times \nonumber \\
&   &
{N - l_0 \choose l_{110}} {N - l_0 - l_{110} \choose l_{100}} {l_0 \choose l_{010}} {l_0 - l_{010} \choose l_{000}}
p^{l_{111}} (1 - p)^{l_{011}} (\frac{p}{2})^{l_{110}} (1 - \frac{p}{2})^{l_{010}} \delta_{l_{100}, 0}
\nonumber \\
&   &
= -\bar{\kappa}(t) z_{l_0} - 2 \gamma \rho z_H z_{l_0} + 2 \sum_{l_{110} = 0}^{N - l_0} \sum_{l_{010} = 0}^{l_0} \sum_{l_{000} = 0}^{l_0 - l_{010}} \kappa_{l_{000}} z_{l_{110} + l_{010}, l_{000}} \frac{(l_{110} + l_{010})!}{l_{110}! l_{010}!} (\frac{1 - \epsilon}{2})^{l_{110}} (\frac{1 + \epsilon}{2})^{l_{010}}
\times \nonumber \\
&   &
\frac{(N - l_{110} - l_{010} - l_{000})!}{(l_0 - l_{010} - l_{000})! (N - l_0 - l_{110})!} 
\epsilon^{l_0 - l_{010} - l_{000}} (1 - \epsilon)^{N - l_0 - l_{110}} 
\end{eqnarray}
\end{widetext}
which is identical to the second equation in Eq. (41).

\section{Derivation of the Dynamical Equations for $ w_l(\beta_1, \beta_2, t) $ for Asexual Reproduction in the Two-Chromosomed Genome}

For asexual reproduction in the two-chromosomed genome, we have,
\begin{widetext}
\begin{eqnarray}
&   &
\frac{\partial w_l}{\partial t} = -(\kappa_l + \bar{\kappa}(t)) w_l
+ 2 r_i \sum_{l_1 = 0}^{N - l} \sum_{l_2 = 0}^{N - l - l_1}
\sum_{l_1' = 0}^{N - l - l_1 - l_2} \sum_{l_2' = 0}^{l_1} \sum_{l_3' = 0}^{l_2}
\sum_{l_4' = 0}^{l} \sum_{l_5' = 0}^{l - l_4'} \sum_{l_6' = 0}^{l - l_4' - l_5'}
\kappa_{l_6'} z_{l_1' + l_2' + l_3' + l_4', l_5', l_6'}
\times \nonumber \\
&   &
\frac{(l_1' + l_2' + l_3' + l_4')!}{l_1'! l_2'! l_3'! l_4'!} [(1 - \epsilon)^2]^{l_1'} [\beta_1 \epsilon (1 - \epsilon)]^{l_2'} [\beta_2 \epsilon (1 - \epsilon)]^{l_3'} (\epsilon^2)^{l_4'}
\times \nonumber \\
&   &
\frac{(N - l_1' - l_2' - l_3' - l_4' - l_5' - l_6')!}{(l_1 - l_2')! (l_2 - l_3')! (l - l_4' - l_5' - l_6')! (N - l - l_1 - l_2 - l_1')!}
\times \nonumber \\
&   &
[\beta_1 \epsilon (1 - \epsilon)]^{l_1 - l_2'} [\beta_2 \epsilon (1 - \epsilon)]^{l_2 - l_3'} (\epsilon^2)^{l - l_4' - l_5' - l_6'}
[(1 - \epsilon)^2]^{N - l - l_1 - l_2 - l_1'}
\nonumber \\
&   &
+ 2 (1 - r_i) \sum_{l_1 = 0}^{N - l} \sum_{l_2 = 0}^{N - l - l_1} \sum_{l_1' = 0}^{l_1} \sum_{l_2' = 0}^{l_2} \sum_{l_3' = 0}^{l} \sum_{l_4' = 0}^{l - l_3'} \sum_{l_5' = 0}^{l - l_3' - l_4'}
\kappa_{l_5'} z_{l_1' + l_3', l_2' + l_4', l_5'} 
\times \nonumber \\
&   &
\frac{(l_1' + l_3')!}{l_1'! l_3'!} [\beta_1 (1 - \epsilon)]^{l_1'} \epsilon^{l_3'}
\frac{(l_2' + l_4')!}{l_2'! l_4'!} [\beta_2 (1 - \epsilon)]^{l_2'} \epsilon^{l_4'}
\times \nonumber \\
&   &
\frac{(N - l_1' - l_2' - l_3' - l_4' - l_5')!}{(l_1 - l_1')! (l_2 - l_2')! (l - l_3' - l_4' - l_5')! (N - l - l_1 - l_2)!}
\times \nonumber \\
&   &
[\beta_1 \epsilon (1 - \epsilon)]^{l_1 - l_1'} [\beta_2 \epsilon (1 - \epsilon)]^{l_2 - l_2'} (\epsilon^2)^{l - l_3' - l_4' - l_5'}
[(1 - \epsilon)^2]^{N - l - l_1 - l_2}
\nonumber
\end{eqnarray}
\begin{eqnarray}
&   &
= -(\kappa_l + \bar{\kappa}(t)) w_l 
\nonumber \\
&   &
+ 2 r_i \sum_{l_6' = 0}^{l} \sum_{l_5' = 0}^{l - l_6'} \sum_{l_4' = 0}^{l - l_5' - l_6'}
\sum_{l_1' = 0}^{N - l} \sum_{l_2' = 0}^{N - l - l_1'} \sum_{l_3' = 0}^{N - l - l_1' - l_2'} \sum_{l_1 - l_2' = 0}^{N - l - l_1' - l_2' - l_3'}\sum_{l_2 - l_3' = 0}^{N - l - l_1' - l_2' - l_3' - (l_1 - l_2')} 
\kappa_{l_6'} z_{l_1' + l_2' + l_3' + l_4', l_5', l_6'}
\times \nonumber \\
&   &
\frac{(l_1' + l_2' + l_3' + l_4')!}{l_1'! l_2'! l_3'! l_4'!} [(1 - \epsilon)^2]^{l_1'} [\beta_1 \epsilon (1 - \epsilon)]^{l_2'} [\beta_2 \epsilon (1 - \epsilon)]^{l_3'} (\epsilon^2)^{l_4'}
\times \nonumber \\
&   &
\frac{(N - l_1' - l_2' - l_3' - l_4' - l_5' - l_6')!}{(l_1 - l_2')! (l_2 - l_3')! (l - l_4' - l_5' - l_6')! (N - l - l_1 - l_2 - l_1')!}
\times \nonumber \\
&   &
[\beta_1 \epsilon (1 - \epsilon)]^{l_1 - l_2'} [\beta_2 \epsilon (1 - \epsilon)]^{l_2 - l_3'} (\epsilon^2)^{l - l_4' - l_5' - l_6'}
[(1 - \epsilon)^2]^{N - l - l_1 - l_2 - l_1'}
\nonumber \\
&   &
+ 2 (1 - r_i) \sum_{l_5' = 0}^{l} \sum_{l_4' = 0}^{l - l_5'} \sum_{l_3' = 0}^{l - l_4' - l_5'}
\sum_{l_1' = 0}^{N - l} \sum_{l_2' = 0}^{N - l - l_1'} \sum_{l_1 - l_1' = 0}^{N - l - l_1' - l_2'} 
\sum_{l_2 - l_2' = 0}^{N - l - l_1' - l_2' - (l_1 - l_1')}
\kappa_{l_5'} z_{l_1' + l_3', l_2' + l_4', l_5'} 
\times \nonumber \\
&   &
\frac{(l_1' + l_3')!}{l_1'! l_3'!} [\beta_1 (1 - \epsilon)]^{l_1'} \epsilon^{l_3'}
\frac{(l_2' + l_4')!}{l_2'! l_4'!} [\beta_2 (1 - \epsilon)]^{l_2'} \epsilon^{l_4'}
\times \nonumber \\
&   &
\frac{(N - l_1' - l_2' - l_3' - l_4' - l_5')!}{(l_1 - l_1')! (l_2 - l_2')! (l - l_3' - l_4' - l_5')! (N - l - l_1 - l_2)!}
\times \nonumber \\
&   &
[\beta_1 \epsilon (1 - \epsilon)]^{l_1 - l_1'} [\beta_2 \epsilon (1 - \epsilon)]^{l_2 - l_2'} (\epsilon^2)^{l - l_3' - l_4' - l_5'}
[(1 - \epsilon)^2]^{N - l - l_1 - l_2}
\nonumber
\end{eqnarray}
\begin{eqnarray}
&   &
\geq -(\kappa_l + \bar{\kappa}(t)) w_l 
+ 2 r_i \kappa_l \sum_{l_1' = 0}^{N - l} \sum_{l_2' = 0}^{N - l - l_1'} \sum_{l_3' = 0}^{N - l - l_1' - l_2'}
\sum_{l_1 - l_2' = 0}^{N - l - l_1' - l_2' - l_3'} \sum_{l_2 - l_3' = 0}^{N - l - l_1' - l_2' - l_3' - (l_1 - l_2')}
z_{l_1' + l_2' + l_3', 0, l}
\times \nonumber \\
&   &
\frac{(l_1' + l_2' + l_3')!}{l_1'! l_2'! l_3'!} [(1 - \epsilon)^2]^{l_1'} [\beta_1 \epsilon (1 - \epsilon)]^{l_2'} [\beta_2 \epsilon (1 - \epsilon)]^{l_3'} 
\times \nonumber \\
&   &
\frac{(N - l - l_1' - l_2' - l_3')!}{(l_1 - l_2')! (l_2 - l_3')! (N - l - l_1 - l_2 - l_1')!}
[\beta_1 \epsilon (1 - \epsilon)]^{l_1 - l_2'} [\beta_2 \epsilon (1 - \epsilon)]^{l_2 - l_3'}
[(1 - \epsilon)^2]^{N - l - l_1 - l_2 - l_1'}
\nonumber \\
&   &
+ 2 (1 - r_i) \kappa_l \sum_{l_1' = 0}^{N - l} \sum_{l_2' = 0}^{N - l - l_1'} 
\sum_{l_1 - l_1' = 0}^{N - l - l_1' - l_2'} 
\sum_{l_2 - l_2' = 0}^{N - l - l_1' - l_2' - (l_1 - l_1')}
z_{l_1', l_2', l} [\beta_1 (1 - \epsilon)]^{l_1'} [\beta_2 (1 - \epsilon)]^{l_2'}
\times \nonumber \\
&   &
\frac{(N - l - l_1' - l_2')!}{(l_1 - l_1')! (l_2 - l_2')! (N - l - l_1 - l_2)!}
[\beta_1 \epsilon (1 - \epsilon)]^{l_1 - l_1'} [\beta_2 \epsilon (1 - \epsilon)]^{l_2 - l_2'} [(1 - \epsilon)^2]^{N - l - l_1 - l_2}
\nonumber
\end{eqnarray}
\begin{eqnarray}
&   &
= \kappa_l [2 ((\beta_1 + \beta_2) \epsilon (1 - \epsilon) + (1 - \epsilon)^2)^{N - l} 
\times \nonumber \\
&   &
(r_i w_l(1, 0, t) + (1 - r_i)
w_l(\frac{\beta_1 (1 - \epsilon)}{(\beta_1 + \beta_2) \epsilon (1 - \epsilon) + (1 - \epsilon)^2},
\frac{\beta_2 (1 - \epsilon)}{(\beta_1 + \beta_2) \epsilon (1 - \epsilon) + (1 - \epsilon)^2}, t))
- w_l(\beta_1, \beta_2, t)] 
\nonumber \\
&   &
- \bar{\kappa}(t) w_l(\beta_1, \beta_2, t)
\end{eqnarray}
\end{widetext}
where strict equality holds for $ l = 0 $, or if $ z_{l_1, l_2, l_3} = 0 $ for $ l_3 < l $.

\section{Mathematical Details for the Solution of the Sexual Reproduction Pathways in the Limit of Large $ N $}

\subsection{Two-chromosomed genome}

We begin our analysis by deriving the limiting form of the expression,
\begin{widetext}
\begin{equation}
\frac{(l_1 + l_3)! (l_2 + l_3)!}{l_1! l_2! l_3!}
\prod_{k = 1}^{l_1} \frac{N - l_1 - l_2 - l_3 + k}{N - l_1 - l_3 + k}
\prod_{k = 1}^{l_3} \frac{1}{N - l_3 + k}
\end{equation}
\end{widetext}
in the limit of large $ N $, under the assumption that $ l_1, l_2 $ scale as $ \sqrt{N} $, and $ l_3 $ is finite 
as $ N \rightarrow \infty $.

We begin by re-writing the expression as,
\begin{eqnarray}
&   &
\frac{1}{l_3!} \prod_{k = 1}^{l_3} \frac{(l_1 + k) (l_2 + k)}{N - l_3 + k}
\prod_{k = 1}^{l_1} \frac{1 + \frac{k - l_1 - l_2 - l_3}{N}}{1 - \frac{l_1 + l_3 - k}{N}} 
\end{eqnarray}

In the limit of large $ N $, we obtain,
\begin{eqnarray}
&   &
\frac{1}{l_3!} \prod_{k = 1}^{l_3} \frac{(\frac{l_1}{\sqrt{N}} + \frac{k}{\sqrt{N}}) (\frac{l_2}{\sqrt{N}} + \frac{k}{\sqrt{N}})}{1 + \frac{k - l_3}{N}}
\prod_{k = 1}^{l_1} (1 + \frac{k - l_1 - l_2 - l_3}{N}) (1 + \frac{l_1 + l_3 - k}{N})
\nonumber \\
&   &
\rightarrow
\frac{1}{l_3!} (\frac{l_1 l_2}{N})^{l_3} (1 - \frac{l_2}{N})^{l_1}
\rightarrow
\frac{1}{l_3!} (\frac{l_1 l_2}{N})^{l_3} [(1 - \frac{l_2}{N})^{-\frac{N}{l_2}}]^{-\frac{l_1 l_2}{N}}
\rightarrow
\frac{1}{l_3!} (\frac{l_1 l_2}{N})^{l_3} e^{-\frac{l_1 l_2}{N}}
\end{eqnarray}

And so, as is given in Eq. (35), we have,
\begin{widetext}
\begin{equation}
\tilde{z}_l = 2 e^{-\mu} \sum_{k = 0}^{l} \frac{\mu^{k}}{k!} \tilde{z}_{l - k}
\sum_{l_4 = 0}^{l - k} \frac{\kappa_{l_4}}{\bar{\kappa} + \kappa_{l_4}}
\sum_{l_1 = 0}^{N - l}
\frac{1}{l_4!} (\frac{l_1 (l - l_4 - k)}{N})^{l_4} e^{-\frac{l_1 (l - l_4 - k)}{N}} \tilde{z}_{l_1 + l_4}
\end{equation}
\end{widetext}

Now, in the limit of large $ N $, we have observed from simulations that the $ \tilde{z}_l $ converge to a Gaussian distribution with a mean that is proportional to $ \sqrt{N} $ and a standard deviation that is proportional to $ N^{1/4} $.  While this observation is not a proof, we may nevertheless make an ansatz that the $ \tilde{z}_l $ do indeed converge to a Gaussian in the limit of large $ N $, and see if this allows us to solve for the steady-state of this reproduction pathway.  If this ansatz leads to a self-consistent set of equations that may be used to solve for the steady-state in the limit of large $ N $, then we may assume that it is a correct assumption.

If the mean of the Gaussian scales as $ \sqrt{N} $, then we may write that the mean of the Gaussian is given by $ \lambda \sqrt{N} $.  If the standard deviation of the Gaussian scales as $ N^{1/4} $, then we may write that the standard deviation is $ \gamma N^{1/4} $.  As a result, we may transform from a discrete representation in terms of the $ \tilde{z}_l $ into a continuous representation, denoted by $ p(x) $, where $ x = l/\sqrt{N} $.  Conservation of probability implies that $ \tilde{z}_l = p(x)/\sqrt{N} \Rightarrow p(x) = \sqrt{N} \tilde{z}_l $.

In these re-scaled coordinates, the Gaussian has a mean of $ \lambda $ and a standard deviation of $ \gamma N^{-1/4} $.  As a result, we obtain that,
\begin{equation}
p(x) = \frac{N^{1/4}}{\gamma \sqrt{2 \pi}} e^{-\frac{ \sqrt{N} (x - \lambda)^2}{2 \gamma^2}}
\end{equation}

We then have,
\begin{eqnarray}
&   &
p(x) = 2 e^{-\mu} \sum_{l_4 = 0}^{x \sqrt{N}} \frac{1}{l_4!} \frac{\kappa_{l_4}}{\bar{\kappa} + \kappa_{l_4}}
\sum_{k = 0}^{x \sqrt{N} - l_4} \frac{\mu^{k}}{k!} p(x - \frac{k}{\sqrt{N}})
\times \nonumber \\
&   &
\sum_{l_1 = 0}^{N - x \sqrt{N}}
\frac{1}{\sqrt{N}} (\frac{l_1}{\sqrt{N}} (x - \frac{l_4 + k}{\sqrt{N}}))^{l_4}
e^{-\frac{l_1}{\sqrt{N}} (x - \frac{l_4 + k}{\sqrt{N}})} 
p(\frac{l_1}{\sqrt{N}} + \frac{l_4}{\sqrt{N}})
\end{eqnarray}

Defining $ x_1 = l_1/\sqrt{N} $ we have, in the limit of large $ N $, that,
\begin{eqnarray}
&   &
p(x) = 2 e^{-\mu} \sum_{l_4 = 0}^{\infty} \frac{1}{l_4!} \frac{\kappa_{l_4}}{\bar{\kappa} + \kappa_{l_4}}
\sum_{k = 0}^{\infty} \frac{\mu^{k}}{k!} p(x - \frac{k}{\sqrt{N}})
\times \nonumber \\
&   &
\int_{0}^{\infty} d x_1 (x_1 x)^{l_4} (1 - \frac{l_4 + k}{\sqrt{N} x})^{l_4}
e^{-x_1 x} e^{x_1 \frac{l_4 + k}{\sqrt{N}}}
p(x_1 + \frac{l_4}{\sqrt{N}})
\end{eqnarray}

In the limit of large $ N $, we can evaluate the integral out to order $ 1/\sqrt{N} $.  The idea is that the integrand is a product of two functions of $ x_1 $, where one of the functions is a Gaussian that converges to a $ \delta $-function centered at $ \lambda $.  The integral is then evaluated to order $ 1/\sqrt{N} $ by expanding the other function out to second order in $ x_1 - \lambda $, and integrating under the narrow Gaussian envelope.  In the Taylor expansion, we may ignore any terms containing an $ x_1^n/\sqrt{N} $ where $ n \geq 1 $, since such terms either vanish or contribute a term of order at least $ 1/N $.  

Following these guidelines, the integral becomes,
\begin{widetext}
\begin{eqnarray}
&   &
\lambda^{l_4} (1 + \lambda \frac{l_4 + k}{\sqrt{N}}) x^{l_4} (1 - \frac{l_4 (l_4 + k)}{\sqrt{N} x})  e^{-\lambda x} 
\int_{0}^{\infty} d x_1 
[1 + l_4 \frac{x_1 - \lambda}{\lambda} + \frac{l_4 (l_4 - 1)}{2} (\frac{x_1 - \lambda}{\lambda})^2] 
\times \nonumber \\
&   &
[1 - x (x_1 - \lambda) + \frac{1}{2} x^2 (x_1 - \lambda)^2]
\frac{N^{1/4}}{\gamma \sqrt{2 \pi}} \exp[-\frac{\sqrt{N}}{2 \gamma^2} (x_1 - \lambda + \frac{l_4}{\sqrt{N}})^2]
\end{eqnarray}

Defining $ x_1' = x_1 - \lambda $, this becomes,
\begin{eqnarray}
&   &
\lambda^{l_4} (1 + \lambda \frac{l_4 + k}{\sqrt{N}}) x^{l_4} (1 - \frac{l_4 (l_4 + k)}{\sqrt{N} x})  e^{-\lambda x}
\int_{-\infty}^{\infty} d x_1'
[1 + \frac{l_4}{\lambda} x_1' + \frac{l_4 (l_4 - 1)}{2 \lambda^2} x_1'^2]
\times \nonumber \\
&   &
[1 - x x_1' + \frac{1}{2} x^2 x_1'^2] \exp[-x_1' \frac{l_4}{\gamma^2}] \exp[-\frac{l_4^2}{2 \sqrt{N} \gamma^2}]
\frac{N^{1/4}}{\gamma \sqrt{2 \pi}}
\exp[-\frac{\sqrt{N}}{2 \gamma^2} x_1'^2] 
\nonumber \\
&   &
= \lambda^{l_4} (1 + \lambda \frac{l_4 + k}{\sqrt{N}}) (1 - \frac{l_4^2}{2 \sqrt{N} \gamma^2}) 
x^{l_4} (1 - \frac{l_4 (l_4 + k)}{\sqrt{N} x})  e^{-\lambda x}
\times \nonumber \\
&   &
\int_{-\infty}^{\infty} d x_1'
[1 + (\frac{l_4}{\lambda} - x) x_1' + (\frac{1}{2} x^2 - \frac{l_4}{\lambda} x + \frac{l_4 (l_4 - 1)}{2 \lambda^2}) x_1'^2]
\times \nonumber \\
&   &
[1 - \frac{l_4}{\gamma^2} x_1' + \frac{l_4^2}{2 \gamma^4} x_1'^2]
\frac{N^{1/4}}{\gamma \sqrt{2 \pi}}
\exp[-\frac{\sqrt{N}}{2 \gamma^2} x_1'^2] 
\nonumber \\
&   &
= \lambda^{l_4} [1 + \frac{1}{\sqrt{N}} (\lambda (l_4 + k) - \frac{l_4^2}{2 \gamma^2} - \frac{l_4 (l_4 + k)}{x})]
x^{l_4} e^{-\lambda x}
\times \nonumber \\
&   &
\int_{-\infty}^{\infty} d x_1'
[1 + (\frac{l_4^2}{2 \gamma^4} - \frac{l_4}{\gamma^2} (\frac{l_4}{\lambda} - x) +
\frac{1}{2} x^2 - \frac{l_4}{\lambda} x + \frac{l_4 (l_4 - 1)}{2 \lambda^2}) x_1'^2]
\frac{N^{1/4}}{\gamma \sqrt{2 \pi}}
\exp[-\frac{\sqrt{N}}{2 \gamma^2} x_1'^2] 
\nonumber \\
&   &
= \lambda^{l_4} x^{l_4} e^{-\lambda x} 
[1 + \frac{1}{\sqrt{N}} (\lambda (l_4 + k) - \frac{l_4 (l_4 + k)}{x}
- l_4 (\frac{l_4}{\lambda} - x) +
\frac{1}{2} \gamma^2 x^2 - \frac{l_4 \gamma^2}{\lambda} x + \frac{l_4 (l_4 - 1) \gamma^2}{2 \lambda^2})]
\nonumber \\
\end{eqnarray}
\end{widetext}

Now, instead of working with $ p(x) $ directly, we work with its Laplace Transform, $ P(s) \equiv \int_{0}^{\infty}
p(x) e^{-s x} d x $.  By Taylor expanding out to second-order in $ x - \lambda $, we have that, to first-order in $ 1/\sqrt{N} $, the Laplace Transform of $ p(x) $ is given by,
\begin{eqnarray}
&   &
P(s) = e^{-s \lambda} \int_{0}^{\infty}  d x (1 - s (x - \lambda) + \frac{1}{2} s^2 (x - \lambda)^2) 
\frac{N^{1/4}}{\gamma \sqrt{2 \pi}}
\exp[-\frac{\sqrt{N}}{2 \gamma^2} (x - \lambda)^2]
\nonumber \\
&   &
= e^{-s \lambda} \int_{-\infty}^{\infty} [1 - s x' + \frac{1}{2} s^2 x'^2]  
\frac{N^{1/4}}{\gamma \sqrt{2 \pi}}
\exp[-\frac{\sqrt{N}}{2 \gamma^2} x'^2]
\nonumber \\
&   &
= e^{-s \lambda} (1 + \frac{\gamma^2}{2 \sqrt{N}} s^2)
\end{eqnarray}

We also have,
\begin{widetext}
\begin{eqnarray}
&   &
P(s) = 2 e^{-\mu} \sum_{l_4 = 0}^{\infty} \frac{1}{l_4!} \frac{\lambda^{l_4} \kappa_{l_4}}{\bar{\kappa} + \kappa_{l_4}}
\sum_{k = 0}^{\infty} \frac{\mu^{k}}{k!} 
\int_{0}^{\infty} d x e^{-s x} 
x^{l_4} e^{-\lambda x}
\times \nonumber \\
&    &
[1 + \frac{1}{\sqrt{N}} (\lambda (l_4 + k) - \frac{l_4 (l_4 + k)}{x} - l_4 (\frac{l_4}{\lambda} - x) 
+ \frac{1}{2} \gamma^2 x^2 - \frac{l_4 \gamma^2}{\lambda} x + \frac{l_4 (l_4 - 1) \gamma^2}{2 \lambda^2})]
\times \nonumber \\
&   &
\frac{N^{1/4}}{\gamma \sqrt{2 \pi}} \exp[-\frac{\sqrt{N}}{2 \gamma^2} (x - \lambda - \frac{k}{\sqrt{N}})^2]
\nonumber \\
&   &
= 2 e^{-\mu} e^{-s \lambda} e^{-\lambda^2} \sum_{l_4 = 0}^{\infty} \frac{1}{l_4!} \frac{\lambda^{2 l_4} \kappa_{l_4}}{\bar{\kappa} + \kappa_{l_4}}
\sum_{k = 0}^{\infty} \frac{\mu^{k}}{k!}
\times \nonumber \\
&   &
\int_{-\infty}^{\infty} d x' (1 - s x' + \frac{1}{2} s^2 x'^2) (1 + \frac{l_4}{\lambda} x' + \frac{l_4 (l_4 - 1)}{2 \lambda^2} x'^2)
(1 - \lambda x' + \frac{1}{2} \lambda^2 x'^2) 
\times \nonumber \\
&   &
[1 + \frac{1}{\sqrt{N}} (\lambda (l_4 + k) - \frac{l_4 (l_4 + k)}{\lambda} - l_4 (\frac{l_4}{\lambda} - \lambda) 
+ \frac{1}{2} \gamma^2 \lambda^2 - l_4 \gamma^2 + \frac{l_4 (l_4 - 1) \gamma^2}{2 \lambda^2})]
\times \nonumber \\
&   &
\exp[-\frac{k^2}{2 \sqrt{N} \gamma^2}] \exp[\frac{k x'}{\gamma^2}] 
\frac{N^{1/4}}{\gamma \sqrt{2 \pi}} \exp[-\frac{\sqrt{N}}{2 \gamma^2} x'^2] 
\nonumber \\
&   &
= 2 e^{-\mu} e^{-s \lambda} e^{-\lambda^2} \sum_{l_4 = 0}^{\infty} \frac{1}{l_4!} \frac{\lambda^{2 l_4} \kappa_{l_4}}{\bar{\kappa} + \kappa_{l_4}}
\sum_{k = 0}^{\infty} \frac{\mu^{k}}{k!}
\times \nonumber \\
&   &
[1 + \frac{1}{\sqrt{N}} (\lambda (l_4 + k) - \frac{l_4 (l_4 + k)}{\lambda} - l_4 (\frac{l_4}{\lambda} - \lambda) 
+ \frac{1}{2} \gamma^2 \lambda^2 - l_4 \gamma^2 + \frac{l_4 (l_4 - 1) \gamma^2}{2 \lambda^2} - \frac{k^2}{2 \gamma^2})]
\times \nonumber \\
&   &
\int_{-\infty}^{\infty} d x' [1 + (\frac{l_4}{\lambda} - s) x' + (\frac{1}{2} s^2 - \frac{l_4 s}{\lambda} + \frac{l_4 (l_4 - 1)}{2 \lambda^2}) x'^2]
\times \nonumber \\
&   &
[1 + (\frac{k}{\gamma^2} - \lambda) x' + (\frac{1}{2} \lambda^2 - \frac{k \lambda}{\gamma^2} + \frac{k^2}{2 \gamma^4}) x'^2] 
\frac{N^{1/4}}{\gamma \sqrt{2 \pi}} \exp[-\frac{\sqrt{N}}{2 \gamma^2} x'^2] 
\nonumber
\end{eqnarray}
\begin{eqnarray}
&   &
= 2 e^{-\mu} e^{-s \lambda} e^{-\lambda^2} \sum_{l_4 = 0}^{\infty} \frac{1}{l_4!} \frac{\lambda^{2 l_4} \kappa_{l_4}}{\bar{\kappa} + \kappa_{l_4}}
\sum_{k = 0}^{\infty} \frac{\mu^{k}}{k!}
\times \nonumber \\
&   &
[1 + \frac{1}{\sqrt{N}} (\lambda (l_4 + k) - \frac{l_4 (l_4 + k)}{\lambda} - l_4 (\frac{l_4}{\lambda} - \lambda) 
+ \frac{1}{2} \gamma^2 \lambda^2 - l_4 \gamma^2 + \frac{l_4 (l_4 - 1) \gamma^2}{2 \lambda^2} - \frac{k^2}{2 \gamma^2})]
\times \nonumber \\
&   &
\int_{-\infty}^{\infty} d x'
[1 + (\frac{1}{2} s^2 - \frac{l_4 s}{\lambda} + \frac{l_4 (l_4 - 1)}{2 \lambda^2} + (\frac{l_4}{\lambda} - s) (\frac{k}{\gamma^2} - \lambda) + \frac{1}{2} \lambda^2 - \frac{k \lambda}{\gamma^2} + \frac{k^2}{2 \gamma^4}) x'^2]
\frac{N^{1/4}}{\gamma \sqrt{2 \pi}} \exp[-\frac{\sqrt{N}}{2 \gamma^2} x'^2] 
\nonumber \\
&   &
= 2 e^{-\mu} e^{-s \lambda} e^{-\lambda^2} \sum_{l_4 = 0}^{\infty} \frac{1}{l_4!} 
\frac{\lambda^{2 l_4} \kappa_{l_4}}{\bar{\kappa} + \kappa_{l_4}}
\sum_{k = 0}^{\infty} \frac{\mu^{k}}{k!}
\times \nonumber \\
&   &
[1 + \frac{1}{\sqrt{N}}
((\lambda - \frac{l_4}{\lambda}) (2 l_4 + k) + \gamma^2 (\lambda^2 - l_4) 
+ \frac{l_4 (l_4 - 1) \gamma^2}{\lambda^2} + \frac{l_4}{\lambda} (k - \lambda \gamma^2) - k \lambda
\nonumber \\
&   &
+ (\lambda \gamma^2 - k - \frac{l_4 \gamma^2}{\lambda}) s + \frac{1}{2} \gamma^2 s^2)]
\nonumber \\
&   &
= 2 e^{-s \lambda} e^{-\lambda^2} \sum_{l_4 = 0}^{\infty} \frac{1}{l_4!} 
\frac{\lambda^{2 l_4} \kappa_{l_4}}{\bar{\kappa} + \kappa_{l_4}}
\times \nonumber \\
&   &
[1 + \frac{1}{\sqrt{N}}
((\lambda - \frac{l_4}{\lambda}) (2 l_4 + \mu) + \gamma^2 (\lambda^2 - l_4) 
+ \frac{l_4 (l_4 - 1) \gamma^2}{\lambda^2} + \frac{l_4}{\lambda} (\mu - \lambda \gamma^2) - \lambda \mu
\nonumber \\
&   &
+ (\lambda \gamma^2 - \mu - \frac{l_4 \gamma^2}{\lambda}) s + \frac{1}{2} \gamma^2 s^2)]
\end{eqnarray}
\end{widetext}

Matching powers of $ s $ between the two expressions for $ P(s) $ gives,
\begin{eqnarray}
&   &
1 = 2 e^{-\lambda^2} \sum_{l_4 = 0}^{\infty} \frac{1}{l_4!} 
\frac{\lambda^{2 l_4} \kappa_{l_4}}{\bar{\kappa} + \kappa_{l_4}}
[1 + \frac{1}{\sqrt{N}}
((\lambda - \frac{l_4}{\lambda}) (2 l_4 + \mu) + \gamma^2 (\lambda^2 - l_4) 
\nonumber \\
&   &
+ \frac{l_4 (l_4 - 1) \gamma^2}{\lambda^2} + \frac{l_4}{\lambda} (\mu - \lambda \gamma^2) - \lambda \mu)]
\nonumber \\
&   &
0 = 2 e^{-\lambda^2} \sum_{l_4 = 0}^{\infty} \frac{1}{l_4!} 
\frac{\lambda^{2 l_4} \kappa_{l_4}}{\bar{\kappa} + \kappa_{l_4}}
[\lambda \gamma^2 - \mu - \frac{l_4 \gamma^2}{\lambda}]
\nonumber \\
&   &
1 = 2 e^{-\lambda^2} \sum_{l_4 = 0}^{\infty} \frac{1}{l_4!} 
\frac{\lambda^{2 l_4} \kappa_{l_4}}{\bar{\kappa} + \kappa_{l_4}}
\end{eqnarray}

As $ N \rightarrow \infty $, the $ 1/\sqrt{N} $ term in the first equality becomes negligible, and so the first and the third equalities become identical to one another.  As a result, we obtain the pair of equations,
\begin{eqnarray}
&   &
1 = 2 e^{-\lambda^2} \sum_{l_4 = 0}^{\infty} \frac{1}{l_4!} 
\frac{\lambda^{2 l_4} \kappa_{l_4}}{\bar{\kappa} + \kappa_{l_4}}
\nonumber \\
&   &
\mu = \lambda \gamma^2 (1 - 2 e^{-\lambda^2}
\sum_{l_4 = 0}^{\infty} \frac{1}{l_4!} 
\frac{\lambda^{2 l_4} \kappa_{l_4 + 1}}{\bar{\kappa} + \kappa_{l_4 + 1}})
\nonumber \\
\end{eqnarray}

Note that these equations are not sufficient by themselves to solve for $ \bar{\kappa}, \lambda, \gamma $.  However, they show that the assumption of a Gaussian profile leads to a self-consistent set of equations that allows us to solve for the steady-state of the system in the limit of large $ N $.  This validates the analysis carried out in the main text, which leads us to the large $ N $ result of $ \bar{\kappa} = \max\{2 e^{-\mu} - 1, 0\} $.

\subsection{Multi-chromosomed genome}

We may transform Eq. (47) into its continuous analogue as follows.  We first note that the binomial probability distribution
$ 2^{-N} {N \choose n} $ approaches a Gaussian in the limit of large $ N $, with a mean of $ N/2 $ and a variance $ \sigma^2 =
N/4 $.  Since a normalized Gaussian is given by,
\begin{equation}
\frac{1}{\sigma \sqrt{2 \pi}} \exp[-\frac{(x - \bar{x})^2}{2 \sigma^2}]
\end{equation}
we obtain, in the limit of large $ N $,
that,
\begin{equation}
2^{-N} {N \choose n} \rightarrow \sqrt{\frac{2}{\pi N}} \exp[-2 \frac{(n - \frac{N}{2})^2}{N}]
\end{equation}

Therefore, we have that,
\begin{eqnarray}
&   &
\tilde{z}_l = 2 e^{-\mu} \sum_{l_3 = 0}^{l} \frac{1}{l_3!} \frac{\kappa_{l_3}}{\bar{\kappa} + \kappa_{l_3}} \sum_{k = 0}^{l - l_3} \frac{\mu^{k}}{k!}
\sum_{l_1 = 0}^{N - l} \sqrt{\frac{2}{\pi (l_1 + l - l_3 - k)}} \exp[-\frac{(l_1 - l + l_3 + k)^2}{2 (l_1 + l - l_3 - k)}]
\times \nonumber \\
&   &
\sum_{l_4 = 0}^{l_1 + l - l_3 - k} (\frac{l_4 (l_1 + l - l_4 - l_3 - k)}{N})^{l_3}
e^{-\frac{l_4 (l_1 + l - l_4 - l_3 - k)}{N}} \tilde{z}_{l_4 + l_3} \tilde{z}_{l_1 + l - l_4 - k}
\end{eqnarray}

Defining $ x = l/\sqrt{N} $, $ x_1 = l_1/\sqrt{N} $, $ x_4 = l_4/\sqrt{N} $, we obtain,
\begin{eqnarray}
&   &
p(x) = 2 e^{-\mu} \sum_{l_3 = 0}^{x \sqrt{N}} \frac{1}{l_3!} \frac{\kappa_{l_3}}{\bar{\kappa} + \kappa_{l_3}}
\sum_{k = 0}^{x \sqrt{N} - l_3} \frac{\mu^{k}}{k!}
\times \nonumber \\
&   &
\sum_{l_1 = 0}^{N - l} \frac{1}{\sqrt{N}} N^{1/4}
\sqrt{\frac{2}{\pi (x_1 + x)}} [1 - \frac{l_3 + k}{\sqrt{N} (x_1 + x)}]^{-1/2}
\times \nonumber \\
&   &
\exp[-\frac{\sqrt{N} (x_1 - x)^2 + 2 (l_3 + k) (x_1 - x) + \frac{(l_3 + k)^2}{\sqrt{N}}}{2 (x_1 + x) (1 - \frac{l_3 + k}{\sqrt{N} (x_1 + x)})}]
\times \nonumber \\
&   &
\sum_{l_4 = 0}^{\sqrt{N} (x_1 + x - \frac{l_3 + k}{\sqrt{N}})}
\frac{1}{\sqrt{N}} (x_4 (x_1 + x - x_4 - \frac{l_3 + k}{\sqrt{N}}))^{l_3}
\times \nonumber \\
&   &
e^{-x_4 (x_1 + x - x_4 - \frac{l_3 + k}{\sqrt{N}})} p(x_4 + \frac{l_3}{\sqrt{N}}) p(x_1 + x - x_4 - \frac{k}{\sqrt{N}})
\nonumber \\
\end{eqnarray}
and so, for very large $ N $, keeping terms up to order $ 1/\sqrt{N} $, we obtain,
\begin{widetext}
\begin{eqnarray}
&   &
p(x) = 2 e^{-\mu} \sum_{l_3 = 0}^{\infty} \frac{1}{l_3!} \frac{\kappa_{l_3}}{\bar{\kappa} + \kappa_{l_3}}
\sum_{k = 0}^{\infty} \frac{\mu^{k}}{k!}
\int_{0}^{\infty} d x_1 (1 + \frac{l_3 + k}{2 \sqrt{N} (x_1 + x)})
N^{1/4} \sqrt{\frac{2}{\pi (x_1 + x)}}
\times \nonumber \\
&   &
\exp[-(\frac{\sqrt{N} (x_1 - x)^2}{2 (x_1 + x)} + \frac{(l_3 + k) (x_1 - x)}{x_1 + x} + \frac{(l_3 + k)^2}{2 \sqrt{N} (x_1 + x)}) 
(1 + \frac{l_3 + k}{\sqrt{N} (x_1 + x)} + \frac{(l_3 + k)^2}{N (x_1 + x)^2})]
\times \nonumber \\
&   &
\int_{0}^{x_1 + x} d x_4 [x_4 (x_1 + x - x_4)]^{l_3} (1 - \frac{l_3 + k}{\sqrt{N} (x_1 + x - x_4)})^{l_3}
e^{-x_4 (x_1 + x - x_4)} e^{x_4 \frac{l_3 + k}{\sqrt{N}}}
\times \nonumber \\
&   &
p(x_4 + \frac{l_3}{\sqrt{N}}) p(x_1 + x - x_4 - \frac{k}{\sqrt{N}})
\nonumber \\
\end{eqnarray}
\end{widetext}

Instead of working with $ p(x) $ directly, we will work with its Laplace Transform.  To this end, we define $ P(s) = \int_{0}^{\infty} p(x) e^{-s x} dx $.  As with the two-chromosomed genome, our strategy will be to take the Laplace Transform of both sides of the above equation, and expand out to order $ 1/\sqrt{N} $.  This will provide a set of equalities that must be satisfied in the limit of large $ N $, which will allow us to solve for $ \bar{\kappa} $ in the $ N \rightarrow \infty $ limit.

To begin, since, in the large $ N $ limit, we are assuming that the $ \tilde{z}_l $ converge to a Gaussian distribution with a mean that scales as $ \sqrt{N} $ and a standard deviation that scales as $ N^{1/4} $, we may let $ \lambda \sqrt{N} $ denote the mean of the distribution and $ \gamma N^{1/4} $ denote the standard deviation.  If we switch from the $ l $ to the $ x $ representation, then the Gaussian distribution has a mean of $ \lambda $ and a standard deviation of $ \gamma N^{-1/4} $, so that we obtain,
\begin{equation}
p(x) = \frac{N^{1/4}}{\gamma \sqrt{2 \pi}} \exp[-\sqrt{N} \frac{(x - \lambda)^2}{2 \gamma^2}]
\end{equation}

As with the two-chromosomed genome, we have,
\begin{equation}
P(s) = e^{-s \lambda} (1 + \frac{\gamma^2}{2 \sqrt{N}} s^2)
\end{equation}

However, from Eq. (E18) we also have that,
\begin{widetext}
\begin{eqnarray}
&   &
P(s) = 2 e^{-\mu} \sum_{l_3 = 0}^{\infty} \frac{1}{l_3!} \frac{\kappa_{l_3}}{\bar{\kappa} + \kappa_{l_3}}
\sum_{k = 0}^{\infty} \frac{\mu^{k}}{k!}
\int_{0}^{\infty} dx e^{-s x} 
\times \nonumber \\
&   &
\int_{0}^{\infty} d x_1
(1 + \frac{l_3 + k}{2 \sqrt{N} (x_1 + x)}) 
N^{1/4} \sqrt{\frac{2}{\pi (x_1 + x)}} 
\exp[-\frac{\sqrt{N} (x_1 - x)^2}{2 (x_1 + x)}]
\times \nonumber \\
&   &
\exp[-\frac{(l_3 + k) (x_1 - x)}{x_1 + x}] 
\exp[-\frac{(l_3 + k)^2}{2 \sqrt{N} (x_1 + x)}]
\exp[-\frac{(l_3 + k) (x_1 - x)^2}{2 (x_1 + x)^2}]
\times \nonumber \\
&   &
\exp[-\frac{(l_3 + k)^2 (x_1 - x)}{\sqrt{N} (x_1 + x)^2}]
\exp[-\frac{(l_3 + k)^2 (x_1 - x)^2}{2 \sqrt{N} (x_1 + x)^3}]
\times \nonumber \\
&   &
\int_{0}^{x_1 + x} d x_4 [x_4 (x_1 + x - x_4)]^{l_3} (1 - \frac{l_3 (l_3 + k)}{\sqrt{N} (x_1 + x - x_4)})
e^{-x_4 (x_1 + x - x_4)} (1 + x_4 \frac{l_3 + k}{\sqrt{N}})
\times \nonumber \\
&   &
\frac{N^{1/4}}{\gamma \sqrt{2 \pi}} \exp[-\frac{\sqrt{N}}{2 \gamma^2} (x_4 - \lambda + \frac{l_3}{\sqrt{N}})^2]
\frac{N^{1/4}}{\gamma \sqrt{2 \pi}} \exp[-\frac{\sqrt{N}}{2 \gamma^2} (x_1 + x - x_4 - \lambda - \frac{k}{\sqrt{N}})^2]
\end{eqnarray}

Now, define $ y = x_1 + x $, so that,
\begin{eqnarray}
&   &
P(s) = 2 e^{-\mu} \sum_{l_3 = 0}^{\infty} \frac{1}{l_3!} \frac{\kappa_{l_3}}{\bar{\kappa} + \kappa_{l_3}}
\sum_{k = 0}^{\infty} \frac{\mu^{k}}{k!}
\int_{0}^{\infty} dy 
\times \nonumber \\
&   &
\int_{0}^{y} d x e^{-s (x - \frac{y}{2})} e^{-s \frac{y}{2}} (1 + \frac{l_3 + k}{2 \sqrt{N} y}) N^{1/4} \sqrt{\frac{2}{\pi y}}
\exp[-\frac{2 \sqrt{N} (x - \frac{y}{2})^2}{y}]
\times \nonumber \\
&   &
\exp[\frac{2 (l_3 + k) (x - \frac{y}{2})}{y}] 
\exp[-\frac{(l_3 + k)^2}{2 \sqrt{N} y}]
\exp[-\frac{2 (l_3 + k) (x - \frac{y}{2})^2}{y^2}]
\times \nonumber \\
&   &
\exp[\frac{2 (l_3 + k)^2 (x - \frac{y}{2})}{\sqrt{N} y^2}]
\exp[-\frac{2 (l_3 + k)^2 (x - \frac{y}{2})^2}{\sqrt{N} y^3}]
\times \nonumber \\
&   &
\int_{0}^{y} d x_4 [x_4 (y - x_4)]^{l_3} (1 - \frac{l_3 (l_3 + k)}{\sqrt{N} (y - x_4)}) (1 + x_4 \frac{l_3 + k}{\sqrt{N}})
e^{-x_4 (y - x_4)}
\times \nonumber \\
&   &
\frac{N^{1/4}}{\gamma \sqrt{2 \pi}} \exp[-\frac{\sqrt{N}}{2 \gamma^2} ((x_4 - \lambda)^2 + 2 \frac{l_3 (x_4 - \lambda)}{\sqrt{N}}
+ \frac{l_3^2}{N})]
\times \nonumber \\
&   &
\frac{N^{1/4}}{\gamma \sqrt{2 \pi}} \exp[-\frac{\sqrt{N}}{2 \gamma^2} ((y - x_4 - \lambda)^2 - 2 \frac{k (y - x_4 - \lambda)}{\sqrt{N}}
+ \frac{k^2}{N})]
\end{eqnarray}

Defining $ x' = x - y/2 $ we obtain, in the limit of large $ N $, that,
\begin{eqnarray}
&    &
P(s) = 2 e^{-\mu} \sum_{l_3 = 0}^{\infty} \frac{1}{l_3!} \frac{\kappa_{l_3}}{\bar{\kappa} + \kappa_{l_3}}
\sum_{k = 0}^{\infty} \frac{\mu^{k}}{k!}
\int_{0}^{\infty} dy e^{-s \frac{y}{2}} 
(1 - \frac{(l_3 + k)(l_3 + k - 1)}{2 \sqrt{N} y}) 
\times \nonumber \\
&   &
\int_{-\infty}^{\infty} d x' 
(1 - s x' + \frac{1}{2} s^2 x'^2)
(1 + \frac{2 (l_3 + k)}{y} x' + \frac{2 (l_3 + k)^2}{y^2} x'^2)
(1 - \frac{2 (l_3 + k)}{y^2} x'^2)
\times \nonumber \\
&   &
N^{1/4} \sqrt{\frac{2}{\pi y}}
\exp[-\frac{2 \sqrt{N} x'^2}{y}]
\times \nonumber \\
&   &
\int_{0}^{y} d x_4 [x_4 (y - x_4)]^{l_3} (1 - \frac{l_3 (l_3 + k)}{\sqrt{N} (y - x_4)}) (1 + x_4 \frac{l_3 + k}{\sqrt{N}})
e^{-x_4 (y - x_4)}
\times \nonumber \\
&   &
\frac{N^{1/4}}{\gamma \sqrt{2 \pi}} \exp[-\frac{\sqrt{N}}{2 \gamma^2} ((x_4 - \lambda)^2 + 2 \frac{l_3 (x_4 - \lambda)}{\sqrt{N}}
+ \frac{l_3^2}{N})]
\times \nonumber \\
&   &
\frac{N^{1/4}}{\gamma \sqrt{2 \pi}} \exp[-\frac{\sqrt{N}}{2 \gamma^2} ((y - x_4 - \lambda)^2 - 2 \frac{k (y - x_4 - \lambda)}{\sqrt{N}}
+ \frac{k^2}{N})]
\nonumber \\
\end{eqnarray}

To evaluate the $ x' $-integral to order $ 1/\sqrt{N} $, we note that only terms up to $ x'^2 $ will give a contribution that is up to order $ 1/\sqrt{N} $, and terms of order $ x' $ will integrate out to $ 0 $.  The integral is then,
\begin{eqnarray}
&    &
\int_{-\infty}^{\infty} d x' 
(1 + (\frac{2 (l_3 + k)^2}{y^2} - \frac{2 (l_3 + k)}{y} s + \frac{1}{2} s^2 - \frac{2 (l_3 + k)}{y^2}) x'^2)
N^{1/4} \sqrt{\frac{2}{\pi y}} \exp[-\frac{2 \sqrt{N} x'^2}{y}]
\nonumber \\
&   &
=
1 + \frac{1}{\sqrt{N}} (\frac{y}{8} s^2 - \frac{(l_3 + k)}{2} s + \frac{(l_3 + k) (l_3 + k - 1)}{2 y})
\end{eqnarray}
and so,
\begin{eqnarray}
&   &
P(s) = 2 e^{-\mu} \sum_{l_3 = 0}^{\infty} \frac{1}{l_3!} \frac{\kappa_{l_3}}{\bar{\kappa} + \kappa_{l_3}}
\sum_{k = 0}^{\infty} \frac{\mu^{k}}{k!}
\int_{0}^{\infty} d y e^{-s \frac{y}{2}} 
(1 - \frac{(l_3 + k) (l_3 + k - 1)}{2 \sqrt{N} y})
\times \nonumber \\
&   &
(1 + \frac{1}{\sqrt{N}} (\frac{y}{8} s^2 - \frac{l_3 + k}{2} s + \frac{(l_3 + k) (l_3 + k - 1)}{2 y}))
\times \nonumber \\
&   &
\int_{0}^{y} d x_4 [x_4 (y - x_4)]^{l_3} (1 - \frac{l_3 (l_3 + k)}{\sqrt{N} (y - x_4)}) (1 + x_4 \frac{l_3 + k}{\sqrt{N}})
e^{-x_4 (y - x_4)}
\times \nonumber \\
&   &
\frac{N^{1/4}}{\gamma \sqrt{2 \pi}} \exp[-\frac{\sqrt{N}}{2 \gamma^2} ((x_4 - \lambda)^2 + 2 \frac{l_3 (x_4 - \lambda)}{\sqrt{N}}
+ \frac{l_3^2}{N})]
\times \nonumber \\
&   &
\frac{N^{1/4}}{\gamma \sqrt{2 \pi}} \exp[-\frac{\sqrt{N}}{2 \gamma^2} ((y - x_4 - \lambda)^2 - 2 \frac{k (y - x_4 - \lambda)}{\sqrt{N}}
+ \frac{k^2}{N})]
\nonumber \\
\end{eqnarray}

Defining $ x_5 = y - x_4 $, we obtain that the double integral over $ y $ and $ x_4 $ is,
\begin{eqnarray}
&   &
\int_{0}^{\infty} \int_{0}^{\infty} d x_4 d x_5 e^{-s \frac{x_4 + x_5}{2}} 
(1 + \frac{1}{\sqrt{N}} (\frac{x_4 + x_5}{8} s^2 - \frac{l_3 + k}{2} s + x_4 (l_3 + k) -
\frac{l_3 (l_3 + k)}{x_5}))
\times \nonumber \\
&   &
(x_4 x_5)^{l_3} e^{-x_4 x_5} 
\times \nonumber \\
&   &
\frac{N^{1/4}}{\gamma \sqrt{2 \pi}} \exp[-\frac{\sqrt{N}}{2 \gamma^2} ((x_4 - \lambda)^2 + 2 \frac{l_3 (x_4 - \lambda)}{\sqrt{N}}
+ \frac{l_3^2}{N})]
\times \nonumber \\
&   &
\frac{N^{1/4}}{\gamma \sqrt{2 \pi}} \exp[-\frac{\sqrt{N}}{2 \gamma^2} ((x_5 - \lambda)^2 - 2 \frac{k (x_5 - \lambda)}{\sqrt{N}}
+ \frac{k^2}{N})]
\end{eqnarray}
\end{widetext}

Now, defining $ x_4' = x_4 - \lambda $, and $ x_5' = x_5 - \lambda $, we have, in the limit of large $ N $, that the integral from $ -\lambda $ to $ \infty $ may be taken to be an integral from $ -\infty $ to $ \infty $, because of the narrowness of the Gaussian distribution.  Also, any term containing an $ x_4 $ or $ x_5 $ that is coupled to a $ 1/\sqrt{N} $ factor may have the $ x_4 $ and $ x_5 $ replaced with $ \lambda $, since for an integral involving $ x_4' $ or $ x_5' $ to survive, it must be on the order of at least $ x_4'^2 $ or $ x_5'^2 $.  Since such integrals produce a $ 1/\sqrt{N} $ factor or higher, the overall term is of order at least $ 1/N $, which is beyond the order of the expansion we are seeking.

We therefore have that the integral is given by,
\begin{widetext}
\begin{eqnarray}
&   &
e^{-s \lambda} e^{-\lambda^2} \lambda^{2 l_3}
(1 + \frac{1}{\sqrt{N}} (\frac{\lambda}{4} s^2 - \frac{l_3 + k}{2} s + \lambda (l_3 + k) - \frac{l_3 (l_3 + k)}{\lambda}))
\times \nonumber \\
&   &
\int_{-\infty}^{\infty} \int_{-\infty}^{\infty} d x_4' d x_5' e^{-s \frac{x_4' + x_5'}{2}}
(1 + \frac{x_4'}{\lambda})^{l_3} (1 + \frac{x_5'}{\lambda})^{l_3} e^{-\lambda x_4'} e^{-\lambda x_5'} e^{-x_4' x_5'}
\times \nonumber \\
&   &
\exp[-\frac{l_3 x_4'}{\gamma^2}] 
\exp[-\frac{l_3^2}{2 \sqrt{N} \gamma^2}]
\exp[\frac{k x_5'}{\gamma^2}]
\exp[-\frac{k^2}{2 \sqrt{N} \gamma^2}]
\times \nonumber \\
&   &
\frac{N^{1/4}}{\gamma \sqrt{2 \pi}} 
\exp[-\frac{\sqrt{N}}{2 \gamma^2} x_4'^2] 
\frac{N^{1/4}}{\gamma \sqrt{2 \pi}} 
\exp[-\frac{\sqrt{N}}{2 \gamma^2} x_5'^2]
\end{eqnarray}
\end{widetext}

To evaluate this integral, we expand the functions that do not converge to $ \delta $-functions in a Taylor series.  Since we are only interested in terms up to order $ 1/\sqrt{N} $, we only expand out to order $ x_4'^2 $ or $ x_5'^2 $.  Furthermore, we may neglect any cross terms of $ x_4' $ and $ x_5' $.  The reason for this is that for such terms to survive, the $ x_4' $ term must be coupled to at least another $ x_4' $ term, and similarly for the $ x_5' $ term.  This produces an integral which is of order at least $ 1/'\sqrt{N} \times 1/\sqrt{N} = 1/N $, and so may be neglected.

The integral is then,
\begin{widetext}
\begin{eqnarray}
&   &
(1 - \frac{l_3^2 + k^2}{2 \sqrt{N} \gamma^2}) 
\int_{-\infty}^{\infty} \int_{-\infty}^{\infty} d x_4' d x_5' (1 - \frac{1}{2} (x_4' + x_5') s + \frac{1}{8} (x_4'^2 + x_5'^2) s^2)
\times \nonumber \\
&   &
(1 + \frac{l_3}{\lambda} x_4' + \frac{l_3 (l_3 - 1)}{2 \lambda^2} x_4'^2)
(1 - \lambda x_4' + \frac{1}{2} \lambda^2 x_4'^2)
(1 - \frac{l_3}{\gamma^2} x_4' + \frac{l_3^2}{2 \gamma^4} x_4'^2)
\times \nonumber \\
&   &
(1 + \frac{l_3}{\lambda} x_5' + \frac{l_3 (l_3 - 1)}{2 \lambda^2} x_5'^2)
(1 - \lambda x_5' + \frac{1}{2} \lambda^2 x_5'^2)
(1 + \frac{k}{\gamma^2} x_5' + \frac{k^2}{2 \gamma^4} x_5'^2)
\times \nonumber \\
&   &
\frac{N^{1/4}}{\gamma \sqrt{2 \pi}} 
\exp[-\frac{\sqrt{N}}{2 \gamma^2} x_4'^2] 
\frac{N^{1/4}}{\gamma \sqrt{2 \pi}} 
\exp[-\frac{\sqrt{N}}{2 \gamma^2} x_5'^2]
\nonumber
\end{eqnarray}
\begin{eqnarray}
&   &
= (1 - \frac{l_3^2 + k^2}{2 \sqrt{N} \gamma^2})
\int_{-\infty}^{\infty} \int_{-\infty}^{\infty} d x_4' d x_5' (1 - \frac{1}{2} (x_4' + x_5') s + \frac{1}{8} (x_4'^2 + x_5'^2) s^2)
\times \nonumber \\
&   &
(1 + (\frac{l_3}{\lambda} - \lambda - \frac{l_3}{\gamma^2}) x_4'
+ (\frac{l_3^2}{2 \gamma^4} - \frac{l_3^2}{\lambda \gamma^2} + \frac{l_3 \lambda}{\gamma^2} +
\frac{1}{2} \lambda^2 - l_3 + \frac{l_3 (l_3 - 1)}{2 \lambda^2}) x_4'^2)
\times \nonumber \\
&   &
(1 + (\frac{l_3}{\lambda} - \lambda + \frac{k}{\gamma^2}) x_5' 
+ (\frac{k^2}{2 \gamma^4} + \frac{l_3 k}{\lambda \gamma^2} - \frac{k \lambda}{\gamma^2} +
\frac{1}{2} \lambda^2 - l_3 + \frac{l_3 (l_3 - 1)}{2 \lambda^2}) x_5'^2)
\times \nonumber \\
&   &
\frac{N^{1/4}}{\gamma \sqrt{2 \pi}} 
\exp[-\frac{\sqrt{N}}{2 \gamma^2} x_4'^2] 
\frac{N^{1/4}}{\gamma \sqrt{2 \pi}} 
\exp[-\frac{\sqrt{N}}{2 \gamma^2} x_5'^2]
\nonumber \\
&   &
= (1 - \frac{l_3^2 + k^2}{2 \sqrt{N} \gamma^2})
\int_{-\infty}^{\infty} \int_{-\infty}^{\infty} d x_4' d x_5' 
(1 - \frac{1}{2} s x_4' + \frac{1}{8} s^2 x_4'^2 - \frac{1}{2} s x_5' + \frac{1}{8} s^2 x_5'^2)
\times \nonumber \\
&   &
(1 + (\frac{l_3}{\lambda} - \lambda - \frac{l_3}{\gamma^2}) x_4'
+ (\frac{l_3^2}{2 \gamma^4} - \frac{l_3^2}{\lambda \gamma^2} + \frac{l_3 \lambda}{\gamma^2} +
\frac{1}{2} \lambda^2 - l_3 + \frac{l_3 (l_3 - 1)}{2 \lambda^2}) x_4'^2
\nonumber \\
&   &
+ (\frac{l_3}{\lambda} - \lambda + \frac{k}{\gamma^2}) x_5' 
+ (\frac{k^2}{2 \gamma^4} + \frac{l_3 k}{\lambda \gamma^2} - \frac{k \lambda}{\gamma^2} +
\frac{1}{2} \lambda^2 - l_3 + \frac{l_3 (l_3 - 1)}{2 \lambda^2}) x_5'^2)
\times \nonumber \\
&   &
\frac{N^{1/4}}{\gamma \sqrt{2 \pi}} 
\exp[-\frac{\sqrt{N}}{2 \gamma^2} x_4'^2] 
\frac{N^{1/4}}{\gamma \sqrt{2 \pi}} 
\exp[-\frac{\sqrt{N}}{2 \gamma^2} x_5'^2]
\nonumber \\
&   &
= (1 - \frac{l_3^2 + k^2}{2 \sqrt{N} \gamma^2})
\int_{-\infty}^{\infty} \int_{-\infty}^{\infty} d x_4' d x_5'
\times \nonumber \\
&   &
(1 + (\frac{l_3^2}{2 \gamma^4} - \frac{l_3^2}{\lambda \gamma^2} + \frac{l_3 \lambda}{\gamma^2} +
\frac{1}{2} \lambda^2 - l_3 
+ \frac{l_3 (l_3 - 1)}{2 \lambda^2} - \frac{1}{2} s (\frac{l_3}{\lambda} - \lambda
- \frac{l_3}{\gamma^2}) + \frac{1}{8} s^2) x_4'^2
\nonumber \\
&   &
+ (\frac{k^2}{2 \gamma^4} + \frac{l_3 k}{\lambda \gamma^2} - \frac{k \lambda}{\gamma^2} +
\frac{1}{2} \lambda^2 - l_3 + \frac{l_3 (l_3 - 1)}{2 \lambda^2}
- \frac{1}{2} s (\frac{l_3}{\lambda} - \lambda + \frac{k}{\gamma^2}) +
\frac{1}{8} s^2) x_5'^2)
\times \nonumber \\
&   &
\frac{N^{1/4}}{\gamma \sqrt{2 \pi}} 
\exp[-\frac{\sqrt{N}}{2 \gamma^2} x_4'^2] 
\frac{N^{1/4}}{\gamma \sqrt{2 \pi}} 
\exp[-\frac{\sqrt{N}}{2 \gamma^2} x_5'^2]
\nonumber \\
&   &
= 1 + \frac{1}{\sqrt{N}} 
[\frac{l_3 k - l_3^2}{\lambda} + (l_3 - k) \lambda +
\lambda^2 \gamma^2 - 2 l_3 \gamma^2 + \frac{l_3 (l_3 - 1) \gamma^2}{\lambda^2}]
\nonumber \\
&   &
+ \frac{1}{\sqrt{N}} s [\gamma^2 (\lambda - \frac{l_3}{\lambda}) + \frac{1}{2} (l_3 - k)]
+ \frac{1}{\sqrt{N}} \frac{\gamma^2}{4} s^2
\nonumber \\
\end{eqnarray}

Going back to Eq. (E27), we then have that the overall integral is given by,
\begin{eqnarray}
&   &
e^{-s \lambda} e^{-\lambda^2} \lambda^{2 l_3} 
(1 + \frac{1}{\sqrt{N}} [2 l_3 (\lambda - \frac{l_3}{\lambda}) + \gamma^2 (\lambda^2 - 2 l_3) + \frac{l_3 (l_3 - 1) \gamma^2}{\lambda^2}]
\nonumber \\
&   &
+ \frac{1}{\sqrt{N}} [\gamma^2 (\lambda - \frac{l_3}{\lambda}) - k] s 
+ \frac{1}{\sqrt{N}} \frac{\lambda + \gamma^2}{4} s^2)
\nonumber \\
\end{eqnarray}

Now, noting that $ \sum_{k = 0}^{\infty} \mu^{k}/k! = e^{\mu} $ and $ \sum_{k = 0}^{\infty} k \mu^{k}/k! = \mu e^{\mu} $, we obtain,
\begin{eqnarray}
&   &
P(s) = 2 e^{-s \lambda} e^{-\lambda^2} \sum_{l_3 = 0}^{\infty} \frac{1}{l_3!} \frac{\lambda^{2 l_3} \kappa_{l_3}}{\bar{\kappa} + \kappa_{l_3}} [1 + \frac{1}{\sqrt{N}} (2 l_3 (\lambda - \frac{l_3}{\lambda}) + \gamma^2 (\lambda^2 - 2 l_3) + \frac{l_3 (l_3 - 1) \gamma^2}{\lambda^2})]
\nonumber \\
&   &
+ \frac{s}{\sqrt{N}} 2 e^{-s \lambda} e^{-\lambda^2} 
\sum_{l_3 = 0}^{\infty} \frac{1}{l_3!} \frac{\lambda^{2 l_3} \kappa_{l_3}}{\bar{\kappa} + \kappa_{l_3}}
[\gamma^2 (\lambda - \frac{l_3}{\lambda}) - \mu] 
\nonumber \\
&   &
+ \frac{s^2}{\sqrt{N}} \frac{\lambda + \gamma^2}{4} 2 e^{-s \lambda} e^{-\lambda^2} 
\sum_{l_3 = 0}^{\infty} \frac{1}{l_3!} \frac{\lambda^{2 l_3} \kappa_{l_3}}{\bar{\kappa} + \kappa_{l_3}}
\end{eqnarray}
\end{widetext}

However, given that, to first order in $ 1/\sqrt{N} $, we have $ P(s) = e^{-s \lambda} (1 + \gamma^2/(2 \sqrt{N}) s^2) $, matching powers of $ s $ gives us that,
\begin{eqnarray}
&   &
1 = 2 e^{-\lambda^2} \sum_{l_3 = 0}^{\infty} \frac{1}{l_3!} \frac{\lambda^{2 l_3} \kappa_{l_3}}{\bar{\kappa} + \kappa_{l_3}} 
\times \nonumber \\
&   &
[1 + \frac{1}{\sqrt{N}} (2 l_3 (\lambda - \frac{l_3}{\lambda}) + \gamma^2 (\lambda^2 - 2 l_3) + \frac{l_3 (l_3 - 1) \gamma^2}{\lambda^2})]
\nonumber \\
&   &
0 = \sum_{l_3 = 0}^{\infty} \frac{1}{l_3!} \frac{\lambda^{2 l_3} \kappa_{l_3}}{\bar{\kappa} + \kappa_{l_3}}
[\gamma^2 (\lambda - \frac{l_3}{\lambda}) - \mu] 
\nonumber \\
&   &
\gamma^2 = (\lambda + \gamma^2) e^{-\lambda^2} 
\sum_{l_3 = 0}^{\infty} \frac{1}{l_3!} \frac{\lambda^{2 l_3} \kappa_{l_3}}{\bar{\kappa} + \kappa_{l_3}}
\end{eqnarray}

In the limit of large $ N $, the $ 1/\sqrt{N} $ factor in the first equality becomes negligible, and so we obtain,
\begin{eqnarray}
&   &
1 = 2 e^{-\lambda^2} \sum_{l_3 = 0}^{\infty} \frac{1}{l_3!} \frac{\lambda^{2 l_3} \kappa_{l_3}}{\bar{\kappa} + \kappa_{l_3}} 
\nonumber \\
&   &
0 = \sum_{l_3 = 0}^{\infty} \frac{1}{l_3!} \frac{\lambda^{2 l_3} \kappa_{l_3}}{\bar{\kappa} + \kappa_{l_3}}
[\gamma^2 (\lambda - \frac{l_3}{\lambda}) - \mu] 
\nonumber \\
&   &
\gamma^2 = \frac{\lambda + \gamma^2}{2} 
\end{eqnarray}

The last equality implies that $ \gamma^2 = \lambda $, and so, we have that, in the limit of large $ N $, the mean fitness $ \bar{\kappa} $ may be obtained by solving the pair of equations,
\begin{eqnarray}
&   &
1 = 2 e^{-\lambda^2} \sum_{l_3 = 0}^{\infty} \frac{1}{l_3!} \frac{\lambda^{2 l_3} \kappa_{l_3}}{\bar{\kappa} + \kappa_{l_3}} 
\nonumber \\
&   &
\mu = \lambda^2 (1 - 2 e^{-\lambda^2} \sum_{l_3 = 0}^{\infty} \frac{1}{l_3!} \frac{\lambda^{2 l_3} 
\kappa_{l_3 + 1}}{\bar{\kappa} + \kappa_{l_3 + 1}})
\end{eqnarray}

As a final calculation for this subsection, we compute, in the limit of large $ N $, the probability that the fusion of two haploids produces a diploid with $ l $ homologous gene pairs lacking a functional copy of the given gene.  So, suppose two haploids with $ n $ defective genes overlap.  To determine the probability that the overlap produces a diploid with exactly $ l $ homologous gene pairs lacking a functional copy of the given gene, we note that, given a haploid, there are $ {n \choose l} $ ways of placing defective genes in the other haploid so that the diploid has $ l $ homologous gene pairs lacking a functional copy of the given gene.  The remaining $ n - l $ defective genes in the other haploid must be in the $ N - n $ slots where the first haploid has a functional copy of the gene.  Since there are $ {N - n \choose n - l} $ ways of placing these genes, we obtain that there are a total of $ {n \choose l} {N - n \choose n - l} $ distinct haploid sequences which can fuse with the given haploid to produce a diploid with $ l $ homologous gene pairs lacking a functional copy of the given gene.  Since there are a total of $ {N \choose n} $ distinct haploids having $ n $ defective genes, the probability that haploid fusion will lead to a diploid that has exactly $ l $ homologous pairs lacking a functional copy of the given gene is,
\begin{eqnarray}
&    &
\frac{{n \choose l}{N - n \choose n - l}}{{N \choose n}} 
= \frac{1}{l!} \prod_{k = 1}^{l} \frac{(n - l + k)^2}{N - l + k} \prod_{k = 1}^{n - l} \frac{1 + \frac{l + k - 2 n}{N}}{1 - \frac{n - k}{N}}
\nonumber \\
\end{eqnarray}

In the limit of large $ N $, with $ n \rightarrow \lambda \sqrt{N} $, the above expression becomes,
\begin{eqnarray}
&   &
\frac{1}{l!} \lambda^{2 l} (1 - \frac{n - l}{N})^{n - l} =
\frac{1}{l!} \lambda^{2 l} [(1 - \frac{n - l}{N})^{-\frac{N}{n - l}}]^{-\frac{(n - l)^2}{N}}
\rightarrow
\frac{1}{l!} \lambda^{2 l} e^{-\lambda^2}
\end{eqnarray}

\end{appendix}

 \bigskip\noindent
{\bf References}

\medskip\noindent
Agrawal AF (2006) Evolution of sex:  Why do organisms shuffle their genotypes?  Curr. Biol. 16: R696-R704

\medskip\noindent
Barton NH, Otto SP (2005) Evolution of recombination due to random drift.  Genetics 169:2353-2370

\medskip\noindent
Bell G (1982) The masterpiece of nature:  The evolution and genetics of sexuality.  Croom Helm, London. 

\medskip\noindent
Bernstein H, Byerly HC, Hopf FA, Michod RE (1984) Origin of sex. J. Theor. Biol. 110:323-351

\medskip\noindent
Bruggeman J, Debets AJM, Wijngaarden PJ, de Visser JAGM, Hoekstra RF (2003) Sex slows down the accumulation of deleterious mutations in the homothallic fungus {\it Aspergillus nidulans}.  Genetics 164:479-485

\medskip\noindent
Bull JJ, Meyers LA, Lachmann M (2005) Quasispecies made simple.  PLoS Comput. Biol. 1:e61

\medskip\noindent
Bull JJ, Wilke CO (2008) Lethal Mutagenesis of Bacteria.  Genetics.  180:  1061-1070
\medskip\noindent
Crow JF, Kimura M (1965) Evolution in sexual and asexual populations. Am. Nat. 99:439-450

\medskip\noindent
De Massy B, Baudat F, Nicolas A (1994) Initiation of recombination in {\it Saccharmocyes cerevisiae} haploid meiosis.  Proc. Natl. Acad. Sci. USA 91:11929-11933

\medskip\noindent
Eigen M (1971) Self-organisation of matter and the evolution of biological macromolecules. Naturwissenschaften 58:  465-523

\medskip\noindent
Hamilton WD, Axelrod R, Tanese R (1990) Sexual reproduction as an adaptation to resist parasites (a review).  Proc. Natl. Acad. Sci. USA 87:3566-3573

\medskip\noindent
Herskowitz I (1988) Life cycle of the budding yeast {\it Saccharomyces cerevisiae}.  Microbiol. Rev. 52:536-553

\medskip\noindent
Howard RS, Lively CM (1994) Parasitism, mutation accumulation, and the maintenance of sex.  Nature (London) 367:554-557

\medskip\noindent
Hurst LD, Peck JR (1996) Recent advances in understanding of the evolution and maintenance of sex.  Trends Evol. Ecol. 11:46-52

\medskip\noindent
Kamp C, Bornholdt S (2002) Co-evolution of quasispecies:  B-cell mutation rates maximize viral error catastrophes.  Phys. Rev. Lett. 88:068104

\medskip\noindent
Keightley PD, Otto SP (2006) Interference among deleterious mutations favours sex and recombination in finite populations.  Nature (London) 443:89-92

\medskip\noindent
Kondrashov AS (1988) Deleterious mutations and the evolution of sexual reproduction.  Nature (London) 336: 435-440

\medskip\noindent
Kondrashov AS, Crow JF (1991) Haploidy or diploidy:  Which is better?  Nature (London) 351:314-315

\medskip\noindent
Mable BK, Otto SP (1998) The evolution of life cycles with haploid and diploid phases.  BioEssays 20:435-462

\medskip\noindent
Mandegar MA, Otto SP (2007) Mitotic recombination counteracts the benefits of genetic segregation.  Proc. Roy. Soc. B Biol. Sci. 274:1301-1307

\medskip\noindent
Maynard-Smith J (1978) The evolution of sex.  Cambridge University Press, Cambridge, UK

\medskip\noindent
Michod RE (1995)  Eros and evolution:  A natural philosophy of sex.  Addison-Wesley, New York

\medskip\noindent
Muller JH (1964) The relation of recombination to mutational advance. Mutat. Res. 1:2-9

\medskip\noindent
Nedelcu AM, Marcu O, Michod RE (2004) Sex as a response to oxidative stress:  A two-fold increase in cellular reactive oxygen species activates sex genes.  Proc. R. Soc. B. Biol. Sci. 271:1591-1596

\medskip\noindent
Paland S, Lynch M (2006) Transitions to asexuality result in excess amino-acid substitutions.  Science 311:990-992

\medskip\noindent
Perrot V, Richerd S, Valero M (1991) Transition from haploidy to diploidy.  Nature (London) 351:315-317

\medskip\noindent
Roeder GS (1995) Sex and the single cell:  Meiosis in yeast.  Proc. Natl. Acad. Sci. USA 92:10450-10456

\medskip\noindent
Tannenbaum E, Fontanari JF (2008) A quasispecies approach to the evolution of sexual replication in unicellular organisms.  Theor. Biosci. 127:53-65

\medskip\noindent
Tannenbaum E, Shakhnovich EI (2005) Semiconservative replication, genetic repair, and many-gened genomes:  Extending the quasispecies paradigm to living systems.  Phys. Life Rev. 2:290-317

\medskip\noindent
de Visser JAGM, Elena SF (2007) The evolution of sex:  Empirical insights into the roles of epistasis and drift.  Nat. Genet. 8:139-149

\medskip\noindent
Wilf HS (2006) Generatingfunctionology.  A.K. Peters, Ltd., Wellesley, MA
\medskip\noindent
Wilke CO (2005) Quasispecies theory in the context of population genetics.  BMC Evol. Biol. 5:44

\medskip\noindent
Williams GC (1975) Sex and evolution.  Princeton University Press, Princeton

\medskip\noindent
Zeldovich KB, Chen P, Shakhnovich EI (2007) Protein stability imposes limits on organism complexity and speed of molecular evolution.  Proc. Natl. Acad. Sci. USA 104:16152-16157

\end{document}